\documentclass[12pt]{article}
\usepackage[margin=1in]{geometry}
\usepackage[slantedGreek]{mathpazo}
\usepackage{graphicx,amsmath,amsfonts,amssymb,dsfont}
\usepackage{setspace}
\usepackage{float}
\usepackage{natbib}
\usepackage{algorithm}
\usepackage{algpseudocode}
\usepackage{amsthm}
\usepackage{mathtools}
\usepackage{stackrel}
\usepackage{tikz}
\usetikzlibrary{arrows.meta, positioning, shapes, arrows, calc}
\usepackage[bottom]{footmisc}
\usepackage{url}
\usepackage{caption}
\usepackage{array}

\graphicspath{{fig/}}
\def\boxit#1{\vbox{\hrule\hbox{\vrule\kern6pt
          \vbox{\kern6pt#1\kern6pt}\kern6pt\vrule}\hrule}}

\title{Generative Modeling: A Review}

\author{
\makebox[.4\linewidth]{Maria Nareklishvili\footnote{Maria Nareklishvili is a Postdoctoral Fellow at Stanford University, Graduate School of Business, email: marnar@stanford.edu}}\\
	Graduate School of Business\\
	Stanford University \\
    \and
	\makebox[.4\linewidth]{Nick Polson\footnote{Corresponding author; Nick Polson is Professor of Econometrics and Statistics at Chicago Booth, email: ngp@chicagobooth.edu}}\\
	Booth School of Business\\
	University of Chicago\\
	\and
	\makebox[.4\linewidth]{Vadim Sokolov\footnote{Vadim Sokolov is an Assistant Professor in Operations Research at George Mason University, email: vsokolov@gmu.edu}}\\
	Department of Industrial Engineering\\
	George Mason University
}

\begin{document}

\date{First Draft: September 10, 2024\\
This Draft: June 26, 2026}

\maketitle

\begin{abstract}
\noindent We organize the generative-modeling literature around three classes of generators, corresponding to three distinct inferential tasks: estimating counterfactual outcome distributions in causal inference, recovering posteriors from simulated parameter--outcome pairs, and forming predictive outcome distributions. The unifying representation relies on the noise outsourcing theorem of Kallenberg, which expresses a conditional distribution as a deterministic function of its inputs and an independent noise variable. Within this organization we develop generative Bayesian computation, a method in the parameter--outcome class: a quantile neural network, trained on simulated pairs under the pinball loss, that targets the posterior of the parameter directly, without invertible architectures or density evaluation, and that serves equally as a predictive generator once the roles of parameter and outcome are exchanged. We illustrate the framework on an agent-based Ebola transmission application, where generative Bayesian computation recovers accurate posteriors at substantially lower cost than rejection-based simulation inference, while avoiding the density-evaluation and invertibility constraints of competing generators.
\end{abstract}


\onehalfspacing

\section{Introduction}

Generative models provide a common language for several apparently distinct objectives in statistical inference: estimating counterfactual outcome distributions in causal inference, computing posterior distributions in Bayesian inference, and quantifying uncertainty in prediction. We organize this review around three generator classes that correspond to these three tasks. (1) Counterfactual outcome generators: recover the outcome map $Y = G(c, Z)$ with $Z \sim \mathrm{Unif}(0,1)$ and conditioning input $c$; potential outcomes in causal inference represent a special case $Y(a) = G(\tilde{S}(X), a, Z)$ with treatment state $a \in \{0,1\}$, sufficient dimension reduction \(\tilde{S}(X)\), from which causal estimands $\Delta = \mathcal{T}(G)$ are recovered. (2) Parameter--outcome generators: recover the inverse posterior map $(\theta \mid Y = y) \stackrel{d}{=} G(y, Z)$ from simulated parameter--outcome pairs, as an alternative to Markov chain Monte Carlo \citep{gelfand2000gibbs, smith2007boa} or rejection-based approximate Bayesian computation \citep{beaumont2002approximate, sisson2018overview, marin2012approximate}. (3) Predictive generators: recover the conditional outcome map $Y = G(X, Z)$ for prediction including its predictive uncertainty. 

The three generator classes share a common representation theorem, the noise outsourcing theorem of \citet{kallenberg1997foundations}, which identifies each conditional distribution with a deterministic function of inputs and reference noise. The classes differ in what is targeted, in the data used to estimate the generator, and in the identification assumptions required: parameter generation requires a specified prior distribution and forward simulator, with consistency governed by the usual Bayesian posterior-concentration conditions; the counterfactual outcome generation additionally requires the Neyman-Rubin potential outcomes framework and the classical causal assumptions; and predictive inference requires only that training and target data share the same conditional distribution. 

Alongside the review, we present generative Bayesian computation (GBC), which casts inference as a quantile regression estimated on simulated data. The key idea is to simulate the respective data set from the joint distribution: parameter--outcome pairs for parameter generation, input--output pairs for predictive inference, and conditional outcome triples for counterfactual outcome generation. A deep neural network is estimated to the relevant pair or triple type. Once fitted, the function estimator produces draws or quantiles at new inputs by direct evaluation, without further simulation or likelihood computation. 

Formally, for parameter generation, we draw the training pairs $\{(\theta_i, Y_i)\}_{i=1}^N$ by forward simulation: one draws $\theta_i \sim p(\theta)$ from the prior distribution and then simulates $Y_i \sim p(Y \mid \theta_i)$ from the forward model. The goal is to learn a deterministic function $G$ such that
\[
\theta = G(Y, Z), \quad Z \sim p(Z),
\]
where $Z$ is an auxiliary random variable drawn from a fixed reference distribution, for example uniform or standard normal, independent of the data. The function $G$ transforms this reference noise into posterior draws conditional on $Y$. When $Z \sim \mathrm{Unif}(0,1)$ and $\theta$ is scalar, this reduces to learning the inverse conditional quantile function $\theta = F^{-1}_{\theta \mid Y}(Z)$. The existence of such a representation is guaranteed by the noise outsourcing theorem \citep{kallenberg1997foundations}, discussed in Section~\ref{sec:framework}. Applying the same theorem to the pair $(c, Y)$ yields the outcome map $Y = G(c, Z)$ for outcome generation; the conditioning input $c$ may be empty (unconditional generation),  include a covariate vector (conditional generation), or include a treatment indicator $A$ in which case the construction specializes to the counterfactual outcome map under standard assumptions in causal inference. Applying the theorem to the input--output pair $(X, Y)$ yields the conditional outcome map $Y = G(X, Z)$ for predictive inference.


The number of simulated pairs $N$ is the central control parameter of the framework, the analog of the reference-table size in approximate Bayesian computation, and is chosen by the researcher rather than fixed by the observed data; when simulation is computationally cheap, values of $N$ in the range $10^4$--$10^6$ are feasible; when simulation is costly, $N$ is a design choice constrained by the computational budget. Larger $N$ improves approximation accuracy at the cost of simulation effort, and the statistical properties of the resulting estimators are governed by $N$ via approximation theory rather than by classical fixed-sample asymptotics. This simulation-based training strategy is applicable whenever a forward model is available, even when the likelihood function cannot be evaluated in closed form.

A natural objective for training is to estimate the conditional expectation $\hat{\theta}(Y_i) = \mathbb{E}[\theta \mid Y_i]$, which minimizes mean squared error and can be viewed as a nonparametric regression problem:
\[
\theta_i = f(Y_i) + \varepsilon_i,
\]
where $\varepsilon_i$ is a mean-zero error term. This formulation targets the posterior mean $\mathbb{E}[\theta \mid Y_i]$; recovering the full posterior distribution requires the quantile-network extension below. Deep neural networks have emerged as effective approximators for $f$ in high-dimensional and nonparametric regimes \citep{lecun2015deep, jiang2017learning, polson2018posterior, montanelli2020error, schmidt-hieber2020nonparametric}. To recover the full conditional distribution $p(\theta \mid Y)$ rather than just the posterior mean, one extends this framework by introducing an auxiliary random input $Z \sim \mathrm{Unif}(0,1)$ and learning a function $G(S(Y), \chi(Z))$, where $\chi$ is a learnable embedding of the reference noise (for example, the cosine quantile embedding of Section~\ref{sec:gbc}); this maps the data-dependent summary $S(Y)$ together with the embedded latent input into parameter space. Because the training sample size $N$ is chosen by the researcher and can be made arbitrarily large, the statistical properties of the resulting estimators (including generalization error, convergence rates, and interpolation behavior) are governed by approximation theory rather than by fixed-sample asymptotics. Convergence rates for quantile neural networks are established by \citet{shen2021deep} and \citet{padilla2022quantile}, building on the deep nonparametric regression rates of \citet{schmidt-hieber2020nonparametric}; the interpolation and double-descent literature \citep{belkin2019does, bach2024highdimensional} provides additional asymptotic context. This framework is related to bootstrap methods \citep{efron1982jackknife, efron1992bootstrap, efron1994introduction, diciccio1996bootstrap}, which approximate sampling distributions by repeatedly re-estimating the model on resampled data. Generative modeling differs in that it jointly simulates parameter--outcome pairs and fits the model only once, substantially reducing computation.

Generative Bayesian computation connects to several established lines of work: latent-variable models and auto-encoders \citep{albert2022learning, akesson2021convolutional}, simulation-based or implicit models \citep{diggle1984monte, baker2022analyzing, schultz2022bayesian}, and indirect inference, which constructs estimators by aligning features of observed and simulated data \citep{pastorello2003iterative, stroud2003nonlinear, drovandi2011approximate, drovandi2015bayesian}. Each tradition developed with a distinct original purpose (data compression, density estimation, or parameter calibration), but all share the common structure of learning mappings between data and latent representations. We position these alternatives by asking which generator class they naturally serve and what changes when the same architecture is used in a different class.

The contributions of this paper are: (1)~a three-class organization of generative inference into counterfactual outcome distributions, simulated parameter--outcome generators, and predictive generators, which we use as the spine of the methodological review; (2)~a common representation theorem, the  noise outsourcing theorem of \cite{kallenberg1997foundations}, which expresses each class as a deterministic function of inputs and independent reference noise (the outcome map $Y = G(c, Z)$, the inverse posterior map $\theta = G(Y, Z)$, and the conditional outcome map $Y = G(X, Z)$), and identifies the choices of training pair, conditioning structure, and loss function that distinguish them (3)~a comparison of generative architectures by generator class, emphasizing training objects, outputs, and limitations; (4)~proposed generative Bayesian computation framework, which is a quantile neural network trained by the pinball loss that requires no invertible architecture and no density evaluation, and that also serves as a predictive generator when the parameter and outcome roles are exchanged; and (5)~three applications aligned with the three classes:  Ebola inverse inference for the parameter-generation class; Ebola predictive emulation for the predictive class, and a synthetic program-evaluation example for counterfactual outcome distributions.

The remainder of the paper is organized as follows. Section~2 defines the three generator classes and states the common representation theorem. Section~3 develops the theoretical framework, establishing the noise outsourcing theorem for parameter generation and stating the analogous identification results for outcome generation and predictive inference. Section~4 reviews the generative architectures. Section~5 develops generative Bayesian computation. Section~6 details the application. Section~7 concludes with a decision guide and open challenges.

\section{Three Generator Classes}
\label{sec:tasks}

We distinguish three classes of generators that share a common representation, a deterministic function of inputs and reference noise, but differ in their conditioning variables and target applications. The noise outsourcing theorem of \cite{kallenberg1997foundations} provides the common foundation for all three.


\label{sec:tasks_cat1} The first class targets a new outcome variable given a (possibly empty) conditioning input $c$. The outcome map
\[
Y = G(c, Z), \qquad Z \sim \mathrm{Unif}(0,1),
\]
subsumes three common use cases: unconditional sample generation ($c = \emptyset$, the canonical task of generative adversarial networks and diffusion models for synthetic data), conditional generation ($c$ representing a covariate or class label), and counterfactual generation ($c$ includes a treatment indicator). Applied to the pair $(c, Y)$, the noise outsourcing theorem identifies $G$ pointwise as the generalized inverse of the conditional outcome distribution, $G(c, q) = F^{-1}_{Y \mid c}(q)$ for $q \in (0,1)$ \citep{parzen2004quantile}.

When $c$ includes a binary treatment state $a \in \{0,1\}$ alongside a sufficient covariate reduction $\tilde{S}(X)$, the outcome map reduces to the \emph{counterfactual generator}
\[
G : (0,1) \times \mathcal{S} \times \{0,1\} \to \mathcal{Y}, \qquad q \mapsto G(q, \tilde{s}, a),
\]
with $G(q, \tilde{s}, a) = F^{-1}_{Y \mid \tilde{S}(X) = \tilde{s},\, A = a}(q)$, non-decreasing and left-continuous in $q$. Under the standard causal assumptions (consistency $Y = Y(A)$, no interference, unconfoundedness $Y(a) \perp A \mid \tilde{S}(X)$, and positivity $0 < P(A = a \mid \tilde{S}(X) = \tilde{s}) < 1$ for $a \in \{0,1\}$), one obtains the identification equality $F_{Y(a) \mid \tilde{S}(X) = \tilde{s}} = F_{Y \mid \tilde{S}(X) = \tilde{s},\, A = a}$ for each $a$, so that $G(q, \tilde{s}, a) = F^{-1}_{Y(a) \mid \tilde{S}(X) = \tilde{s}}(q)$. The noise outsourcing theorem then recovers the potential outcomes $Y(a) = G(U, \tilde{S}(X), a)$, and any scalar causal estimand is a functional $\Delta = \mathcal{T}(G)$ of the generator. Conditional and average treatment effects are
\[
\Delta(\tilde{s}) = \int_0^1 \bigl[G(q, \tilde{s}, 1) - G(q, \tilde{s}, 0)\bigr]\, dq, \qquad \Delta = \int \Delta(\tilde{s})\, dF_{\tilde{S}(X)}(\tilde{s}).
\]
The unconditional and non-causal conditional distributions are recovered by setting $c = \emptyset$ or by taking $c$ as a generic covariate vector without necessarily invoking commonly imposed identification assumptions in causal inference.

The second class targets the \emph{Bayesian posterior} $p(\theta \mid Y)$. Applying the noise outsourcing theorem of \cite{kallenberg1997foundations} to the pair $(\theta, Y)$ yields a measurable $G$ such that
\[
(\theta \mid Y = y) \;\stackrel{d}{=}\; G(y, Z), \qquad Z \sim \mathrm{Unif}(0,1).
\]
When $\theta$ is scalar, $G(y, \tau) = F^{-1}_{\theta \mid y}(\tau)$ is the inverse conditional cumulative distribution function. In the multivariate case, $F^{-1}$ is not unique without an ordering convention; we adopt the autoregressive Rosenblatt-style factorization of Section~\ref{sec:gbc}, while alternative vector-quantile constructions based on the Brenier optimal-transport map provide a coordinate-free generalization \citep{carlier2016vector}. Section~\ref{sec:framework} develops this representation and its sufficient-statistic and quantile-based forms; Section~\ref{sec:gbc} presents generative Bayesian computation, a parameter-generation specialization built on the pinball loss.

The third class targets the \emph{predictive distribution} $p(Y \mid X)$. Applying the noise outsourcing theorem to the pair $(X, Y)$ yields
\[
Y \mid X = x \;\stackrel{d}{=}\; G(x, Z), \qquad Z \sim \mathrm{Unif}(0,1),
\]
where $G$ plays the role of a forward emulator. Quantile networks trained with the pinball loss are well suited to this predictive class; Section~\ref{sec:ebola} demonstrates this construction in the Ebola application by predicting 56-week infection trajectories from model parameters with full predictive uncertainty.


\emph{Notation.} For parameter generation, $\theta$ denotes the Bayesian parameter, $Y$ the data, and $Z$ auxiliary noise; $\tau \in (0,1)$ denotes a quantile level. For predictive inference, $X$ denotes covariates, $Y$ the outcome, and $Z$ auxiliary noise. For outcome generation, $c$ denotes a possibly empty conditioning input and $Z$ auxiliary noise; the counterfactual outcome generation adds a binary treatment state $a \in \{0,1\}$, a treatment indicator $A$, and a sufficient covariate reduction $\tilde{S}(X)$, with $U \sim \mathrm{Unif}(0,1)$ for the latent rank index and $q \in (0,1)$ for a fixed quantile level. The symbols $q$ and $\tau$ play the same mathematical role (a fixed level in $(0,1)$); we use $\tau$ in the parameter-generation sections following the convention of the generative Bayesian computation and $q$ in the sections describing counterfactual outcome generation following the convention in causal inference. We reserve $\theta$ for Bayesian model parameters; causal estimands are denoted by $\Delta$. 

\section{Generative Framework}
\label{sec:framework}

This section formalizes the generative approach to posterior inference. The central idea is to represent the posterior distribution $p(\theta \mid Y)$ through a deterministic function of the observed data and an independent source of randomness, rather than through density evaluation or iterative sampling. Throughout, we use $N$ for the number of simulated training pairs, $n$ for the number of observations in a statistical model, and $\tau$ for quantile levels. The noise outsourcing theorem of Section~\ref{sec:noise_outsourcing} gives the parameter-generation mechanism; the analogous identification results for outcome generation and predictive inference are stated in Section~\ref{sec:framework_outcome_predictive}.

Consider a Bayesian model defined by a prior $p(\theta)$ and a forward model $p(Y \mid \theta)$. We introduce an auxiliary latent variable $Z \sim p(Z)$ drawn from a known reference distribution (for example, $Z \sim \mathrm{Unif}(0,1)$) independently of both $\theta$ and $Y$. The goal is to find a measurable function $G$ such that
\begin{align}
\theta = G(Y, Z) , \quad Z \sim p(Z),
\end{align}
and the conditional distribution of $G(Y, Z)$ given $Y = y$ matches the posterior $p(\theta \mid Y = y)$ for all $y$. The existence of such a function is guaranteed under mild conditions by the noise outsourcing theorem, described below. Once $G$ is learned from simulated training pairs $\{(\theta_i, Y_i)\}_{i=1}^N$, approximate posterior samples for any new observation $y$ are obtained by drawing $Z_1, \ldots, Z_M \sim p(Z)$ and evaluating $G(y, Z_j)$ for $j = 1, \ldots, M$, where $M$ is the desired number of posterior draws, without further simulation or likelihood evaluation.

\subsection{Noise Outsourcing Theorem}
\label{sec:noise_outsourcing}

Let $(Y, \theta)$ be random variables defined on a Borel space $(\mathcal{Y} \times \Theta)$. The noise outsourcing theorem \citep[Theorem~5.10]{kallenberg1997foundations} guarantees the existence of a measurable function $G: \mathcal{Y} \times [0,1] \to \Theta$ and a random variable $Z \sim \mathrm{Unif}(0,1)$, independent of $Y$, such that
\[
(Y, \theta) \stackrel{a.s.}{=} (Y, G(Y, Z)).
\]
Equivalently, for any $y \in \mathcal{Y}$, the conditional distribution of $G(y, Z)$ given $Y = y$ equals the posterior $p(\theta \mid Y = y)$:
\[
(\theta \mid Y = y) \;\stackrel{d}{=}\; G(y, Z), \quad Z \sim \mathrm{Unif}(0,1).
\]
This result, also known as the functional representation lemma \citep{kallenberg1997foundations}, provides a unified view of conditional distribution estimation and nonparametric regression: finding the posterior $p(\theta \mid Y)$ is equivalent to finding a deterministic function $G$ of the data and independent noise. The function $G$ is measurable but not necessarily differentiable; when $\theta$ is scalar, it coincides with the inverse conditional cumulative distribution function, $G(y, \tau) = F^{-1}_{\theta \mid y}(\tau)$. The theorem requires that $(\mathcal{Y} \times \Theta)$ be a standard Borel space, which holds for Euclidean spaces and most spaces encountered in statistical applications. It does not require continuity of the posterior or existence of a density; however, when the posterior is discrete or singular, the resulting function $G$ may be discontinuous in $\tau$, which can complicate approximation by smooth neural networks.

The same theorem is invoked across all three generator classes, so what it actually delivers must be stated precisely, both to prevent reading more into the ``single foundation'' than is there and to fix the perimeter for the identification and approximation arguments that follow. The theorem is a statement of representation and existence: it guarantees that a generator $G$ exists and characterizes it pointwise. It supplies no estimation procedure: it does not specify how to learn $G$ from a finite simulated data set. It supplies no rates: approximation quality is a distinct question, governed by the smoothness of $G$ and the network theory of Section~\ref{sec:gbc}. Additionally, it does not guarantee identification: in the counterfactual class it is the causal assumptions (consistency, unconfoundedness, positivity, the stable unit treatment value assumption) that license interpreting $G$ as a potential-outcome map, not the theorem. The unification across the three classes is therefore representational: one object, applied to three pairs. It is not an inferential unification, and the identification and approximation content of each class must be argued on its own terms.

\subsection{Sufficient Statistics and Dimensionality Reduction}
\label{}

When the observed data $Y$ is high-dimensional, the generator $G$ can be estimated more efficiently by conditioning on a low-dimensional summary $S(Y)$ rather than on $Y$ directly. A statistic $S(Y)$ is sufficient in the Bayesian sense \citep{kolmogorov1942definition} if, for any prior $p(\theta)$, the posterior satisfies $p(\theta \in B \mid Y) = p(\theta \in B \mid S(Y))$ for all measurable $B \subseteq \Theta$. When such a sufficient statistic exists and satisfies $Y \perp \theta \mid S(Y)$, the noise outsourcing theorem simplifies to
\[
\theta \mid Y \stackrel{d}{=} G(S(Y), Z).
\]
In parametric models, sufficient statistics are often available in closed form. For forward models with intractable likelihoods, sufficient statistics must be approximated. Approaches include deep neural networks \citep{jiang2017learning}, autoencoders, partial least squares \citep{polson2021deep, brillinger2012generalized, bhadra2021merging}, and kernel-based embeddings \citep{park2016k2abc}. The choice of summary statistic is critical: \cite{beaumont2002approximate} and \cite{nunes2010optimal} discuss optimal selection in exponential families, while \cite{jiang2018approximate} and \cite{bernton2019approximate} propose smoothing-based extensions for approximate Bayesian computation. \cite{jiang2017learning} show that the posterior mean $\mathbb{E}[\theta \mid Y]$ is an optimal summary under quadratic loss and provide asymptotic guarantees for deep network estimators of $S(Y)$.

\subsection{Quantile-Based Posterior Estimation}
\label{sec:quantile_estimation}

Given training pairs $\{(\theta_i, Y_i)\}_{i=1}^N$ simulated from the joint distribution (the analog of an approximate Bayesian computation reference table, with $N$ chosen by the researcher), the posterior can be estimated through its quantile function. We adopt the quantile-function viewpoint of \citet{parzen2004quantile}, who shows that quantile methods are direct alternatives to density estimation and provides three properties we use throughout: (i)~the conditional quantile $Q_{\theta \mid Y}(\tau, y) = F^{-1}_{\theta \mid y}(\tau)$ is non-decreasing and left-continuous in $\tau$; (ii)~it satisfies the identity $\theta \stackrel{d}{=} Q_{\theta \mid Y}(Z, Y)$ when $Z \sim \mathrm{Unif}(0,1)$, providing a sampler that requires only one uniform draw and one network evaluation per posterior sample; and (iii)~quantile composition, $Q_{g(\theta) \mid Y}(\tau, y) = g(Q_{\theta \mid Y}(\tau, y))$ for non-decreasing left-continuous $g$, makes quantile-based posteriors invariant to monotone reparameterizations and is used in Appendix~\ref{app_bayes_rule} to derive a quantile-form Bayes rule. A deep neural network $\hat{Q}(\tau, y; w)$ is estimated to approximate this quantile function by minimizing the pinball (quantile) loss:
\[
\mathcal{L}(w) = \frac{1}{NK} \sum_{i=1}^N \sum_{k=1}^K \rho_{\tau_k}\bigl(\theta_i - \hat{Q}(\tau_k, Y_i; w)\bigr), \quad \rho_\tau(u) = u(\tau - \mathds{1}_{u < 0}),
\]
where $\rho_\tau$ is the asymmetric check function \citep{koenker1978regression}. By training over a grid of quantile levels or by treating $\tau$ as an additional input to the network \citep{dabney2018implicit, ostrovski2018autoregressive}, one can approximate the full posterior distribution. Unlike approximate Bayesian computation, which relies on rejection sampling and becomes inefficient in high-dimensional settings, this approach evaluates the trained network directly at the observed data. The quality of the approximation depends on the smoothness of the posterior map and the generalization ability of the network; see \cite{padilla2022quantile} and \cite{moon2021learning} for convergence rates of quantile neural networks. Appendix~\ref{app_bayes_risk} characterizes Bayes risk bounds for such estimators.

\subsection{Mechanisms for Outcome Generation and Predictive Inference}
\label{sec:framework_outcome_predictive}

The noise outsourcing theorem of \citet{kallenberg1997foundations}, stated in Section~\ref{sec:noise_outsourcing}, extends to outcome generation and predictive inference. Applied to the pair $(c, Y)$, where $c$ is a (possibly empty) conditioning input, the theorem yields a Borel-measurable $G$ such that
\[
Y \mid c \;\stackrel{d}{=}\; G(c, Z), \qquad Z \sim \mathrm{Unif}(0,1),
\]
identified pointwise via the generalized inverse $G(c, q) = F^{-1}_{Y \mid c}(q)$, following \citet{parzen2004quantile}. Three distinct cases recover the canonical applications: (i) unconditional sample generation when $c = \emptyset$, where $Y \stackrel{d}{=} G(Z)$ produces synthetic samples from the marginal data distribution; (ii) conditional generation when $c$ is a covariate or class label, providing $Y \mid c$ for synthetic-data problems; and (iii) counterfactual generation when $c$ includes a treatment indicator $A \in \{0,1\}$ alongside a sufficient covariate reduction $\tilde{S}(X)$, yielding the counterfactual outcome map $Y(a) = G(U, \tilde{S}(X), a)$ under commonly invoked assumptions in causal inference (consistency, unconfoundedness given $\tilde{S}(X)$, positivity, and the stable unit treatment value assumption).  Under these assumptions, any scalar causal estimand is identified as a functional $\Delta = \mathcal{T}(G)$ of the generator, including the conditional and average treatment effects.


The conditional outcome map in predictive inference admits the same noise-outsourcing representation with $(X, Y)$ in place of $(c, Y)$:
\[
Y \mid X = x \;\stackrel{d}{=}\; G(x, Z), \qquad Z \sim \mathrm{Unif}(0,1),
\]
where $G(x, \tau) = F^{-1}_{Y \mid X = x}(\tau)$ is the conditional quantile function. Quantile neural networks trained with the pinball loss (Section~\ref{sec:quantile_estimation}) are the natural estimators for predictive inference: the same estimation procedure as in parameter generation, with the roles of $\theta$ and $Y$ exchanged. One conceptual caution: in parameter generation the training pairs are drawn from the product of prior and simulator and $Z$ encodes \emph{epistemic} posterior uncertainty, whereas in prediction the pairs come from observed (or simulated) data and $Z$ encodes \emph{aleatoric} outcome variability. The symmetry is architectural; the source of the training distribution, and the interpretation of $Z$, differ. 



\section{Generator Classes and Associated Methods}
\label{sec:architectures}

The three generator classes share a common representation, noise outsourcing theorem of \cite{kallenberg1997foundations} applied to different variable pairs, but differ in what they condition on, which identification assumptions they require, and which statistical question they answer.  The architecture of the method itself does not determine class membership; the training object and conditioning structure do, hence, any architecture that can represent a conditional distribution can in principle serve any of the three classes once its loss, conditioning variables, and output interpretation are changed accordingly. For example, the unconditional adversarial or diffusion generator, built to match a data distribution, is therefore primarily an outcome generator; conditioning it on covariates redirects it to prediction, and conditioning it on a treatment indicator under causal assumptions redirects it to counterfactual outcome generation. Approximate Bayesian computation and fiducial inference invert a forward data-generating equation and are for that reason tied to the parameter--outcome class, with no natural counterfactual or predictive analogue. A single architecture can thus serve several classes: a method may estimate one target without retraining (its primary class) and reach another only after re-conditioning or a change of loss (a secondary role). The subsections below provide a detailed formulation of each architecture under the class it primarily serves. Appendix~\ref{sec:arch_cross} compares the methods in greater detail, with tables displaying training object, output, and limitations of each class; the primary/secondary class membership of the generative architectures; and their computational properties.

\subsection{Generators for Counterfactual Outcome Distributions}

The first generator class targets both observed or counterfactual outcome distributions. The generator of potential outcomes is defined as
\[
Y(a) = G(U, \tilde{S}(X), a), \qquad U \sim \mathrm{Unif}(0,1), \quad a \in \{0,1\},
\]
where \(\tilde{S}(X)\) is a sufficient covariate reduction and \(a\) indexes the treatment state. In non-causal applications the same class reduces to conditional or unconditional outcome generation, \(Y = G(c,Z)\), where \(c\) is empty, a class label, or a vector of covariates. The statistical target is the conditional distribution of outcomes, not a posterior over model parameters.

Generative adversarial networks \citep{goodfellow2014generative, goodfellow2020generative} and diffusion models are natural members of this class. An adversarial generator learns to transform noise and conditioning variables into synthetic outcomes by matching the generated and observed distributions through a discriminator. This is useful for rich outcome spaces, but the adversarial objective does not by itself guarantee calibrated counterfactual intervals or valid causal identification. Diffusion models, in their discrete-time \citep{sohl-dickstein2015deep, ho2020denoising} and continuous-time score-based \citep{song2021scorebased} formulations, primarily generate outcomes: they learn a reverse denoising process that maps Gaussian noise into draws from the outcome distribution. Their advantage is stable generation of multimodal distributions; their cost is many denoising steps per draw, although consistency-model variants reduce this cost \citep{song2023consistency}. In the causal setting, GANITE \citep{yoon2018ganite} adapts the adversarial construction to individualized treatment effects, and double machine learning \citep{chernozhukov2018double} provides an orthogonalized alternative for average effects.

Conditional variational autoencoders and conditional normalizing flows can also be used for outcome generation. A conditional variational autoencoder represents unobserved heterogeneity with a latent variable and decodes it into outcomes given covariates or treatment state. This gives a flexible generator, but the Gaussian latent approximation may understate tail behavior or multimodality. A conditional flow provides exact density evaluation when the map is invertible and dimensions match, but the bijection requirement can be restrictive in causal problems where the outcome dimension, covariate dimension, and latent heterogeneity need not align.

Quantile generators provide a particularly direct formulation for counterfactual distributions. Estimating \(G(q,\tilde{s},a)=F^{-1}_{Y\mid \tilde{S}(X)=\tilde{s},A=a}(q)\) by the pinball loss yields the full conditional quantile process. Average effects, conditional effects, and quantile treatment effects \citep{chernozhukov2005iv} are then functionals of the same fitted generator. This approach separates the statistical identification assumptions from the numerical architecture: unconfoundedness and positivity identify the counterfactual distribution, while the generator estimates its conditional quantile map.

We now give the detailed formulation of the architectures whose primary home is this class: generative adversarial networks, diffusion models, and the decoder of the variational autoencoder. Each is presented in its own established notation, with the symbol $\theta$ noted explicitly whenever it denotes network parameters rather than the Bayesian parameter of interest.

\subsubsection*{Generative adversarial networks} 
The generative adversarial network of \citet{goodfellow2020generative} learns an implicit outcome distribution by training two networks in competition; here $\theta_g$ and $\theta_d$ are network parameters, not the Bayesian parameter of interest. In its original form the generator $G_{\theta_g}(Z)$ maps noise $Z \sim p(Z)$ to a synthetic outcome and matches the marginal data distribution, so the canonical generative adversarial network is an \emph{unconditional outcome generator}; conditioning the generator as $G_{\theta_g}(Z, X)$ redirects it to a predictive target when $X$ is a covariate vector, and supplying a treatment indicator $a$ as in $G_{\theta_g}(Z, X, a)$ redirects it to a counterfactual outcome prediction. A discriminator $D_{\theta_d}$ classifies samples as real or generated. The networks are trained by maximizing the value function 
\begin{equation}
J(\theta_d,\theta_g) = \mathbb{E}_{Y}[\log D_{\theta_d}(Y)] + \mathbb{E}_{Z}[\log(1- D_{\theta_d}(G_{\theta_g}(Z)))],
\end{equation}
whose first term is estimated by an empirical average over observed training samples. Under a zero-sum formulation the parameters solve the minimax problem 
\(
\min_{\theta_g}\max_{\theta_d} J(\theta_d,\theta_g),
\) so the generator learns to produce samples the discriminator cannot distinguish from real data. A GAN produces a synthetic outcome $\tilde Y = G_{\theta_g}(Z)$ (Figure~\ref{fig_gan}), and its target can be written through the Jensen--Shannon divergence between the observed law $p_{\mathrm{obs}}$ and the generated law $p_G$,
\[
\mathrm{JS}(p_{\mathrm{obs}}, p_G) = \log 2 + \tfrac{1}{2} \sup_{D : \mathcal{Y} \to [0,1]} \bigl( \mathbb{E}_{Y \sim p_{\mathrm{obs}}} [\log D(Y)] + \mathbb{E}_{Y \sim p_G} [\log(1 - D(Y))] \bigr).
\]
The discriminator thus plays the role that the $\epsilon$-neighborhood plays in approximate Bayesian computation: it supplies the criterion for deciding whether the generated distribution is close to the observed one.

\begin{figure}[H]
    \centering
    \resizebox{0.85\textwidth}{!}{
\begin{tikzpicture}[
    node distance=3cm,
    box/.style={rectangle, draw, rounded corners=8pt, minimum width=3cm, minimum height=1.2cm, fill=blue!8, align=center},
    gen/.style={rectangle, draw, rounded corners=8pt, minimum width=3.5cm, minimum height=3.5cm, fill=green!8, align=center},
    disc/.style={rectangle, draw, rounded corners=8pt, minimum width=3cm, minimum height=2.5cm, fill=red!8, align=center},
    arrow/.style={->, thick, draw=gray!60},
    every node/.style={font=\small}
]
\node[box] (z) at (0,0.5) {$Z$ (Noise)};
\node[gen] (g) at (5,0.5) {Generator \\[0.2cm] $G_{\theta_g}(Z)$};
\node[box] (fake) at (9.5,2) {Generated outcome $\tilde{Y}$};
\node[box] (real) at (9.5,-1) {Real outcome $Y$};
\node[disc] (d) at (13.5,0.5) {Discriminator \\[0.2cm] $D_{\theta_d}$};
\node[box] (out) at (17,0.5) {Real/Fake};
\draw[arrow] (z) -- (g);
\draw[arrow] (g) -- (fake);
\draw[arrow] (fake) -- (d);
\draw[arrow] (real) -- (d);
\draw[arrow] (d) -- node[above, font=\small] {} (out);
\node[align=center, font=\small] at (5,-1.9) {Generate};
\node[align=center, font=\small] at (13.5,-1.9) {Classify};
\node[align=center] at (9,-3.6) {
    \footnotesize
    $\min\limits_{\theta_g}\max\limits_{\theta_d} J(\theta_d,\theta_g)$ \\[0.3cm]
    $J(\theta_d,\theta_g) = \mathbb{E}_{Y}[\log D_{\theta_d}(Y)] + \mathbb{E}_{Z}[\log(1-D_{\theta_d}(G_{\theta_g}(Z)))]$
};
\end{tikzpicture}
}
\caption{Generative adversarial network. The generator maps noise $Z$ to a synthetic outcome $\tilde{Y}$; the discriminator scores outcomes as real ($Y$) or generated ($\tilde{Y}$), and the two networks are trained against the minimax objective.}
\label{fig_gan}
\end{figure}

Consider a study that records patient outcomes $Y$ under treatments $A$ together with characteristics $X$ such as age and biomarkers. A conditional GAN \citep{mirza2014conditional} generates synthetic outcomes $\tilde{Y} = G_{\theta_g}(Z, X, a)$ by transforming noise $Z \sim \mathcal{N}(0, I)$ and patient features into plausible responses, while the discriminator evaluates whether an outcome is consistent with the patient profile and treatment assignment. Conditioned on a treatment indicator, the generator instantiates the counterfactual outcome map $G(U, X, a)$ of Section~\ref{sec:tasks_cat1} (with the reference noise \(Z\) playing the role of the latent rank \(U\)), and treatment-effect functionals $\Delta = \mathcal{T}(G)$ such as the conditional average treatment effect follow from sampling different noise vectors for a fixed patient. \citet{athey2022smiles} provide a complementary application: GAN-generated profile images on a micro-lending platform isolate the causal effect of modifiable style features (smiling, framing) from fixed characteristics on funding outcomes.

At the minimax optimum $p_G = p_{\mathrm{obs}}$, the generator reproduces the data distribution exactly. Unlike normalizing flows, which minimize a forward Kullback--Leibler divergence by maximum likelihood, and variational autoencoders, which optimize an evidence lower bound, the Jensen--Shannon objective requires no density evaluation and no parametric assumption on $p_{\mathrm{obs}}$, so GANs apply to implicit models where only sampling is possible. The cost is training instability: the minimax problem can cycle rather than converge \citep{salimans2016improved}, and the generator can suffer \emph{mode collapse}, concentrating mass on a subset of the modes of $p_{\mathrm{obs}}$, because the Jensen--Shannon penalty discourages mass outside the support of $p_{\mathrm{obs}}$ but does not penalize missing modes. This makes GANs less reliable than diffusion models for covering a multimodal outcome distribution, despite their faster sampling. A conditional GAN extends to predictive inference, but because it produces samples rather than a density, predictive intervals must be read off the generated sample and coverage assessed empirically; the discriminator yields no calibrated parameter posterior, so the GAN is not a natural parameter--outcome estimator.

\subsubsection*{Diffusion models}
Diffusion models are transport-based generators built from a forward noising process, a learned reverse denoising process, and a sampling procedure \citep{ho2020denoising, song2021scorebased}; this paragraph writes $Y_t$ for the outcome sample at diffusion step $t$ and $\phi$ for network parameters. The forward process corrupts a data point $Y_0 \sim q(Y_0)$ with Gaussian noise over $T$ steps,
\[
q(Y_{1:T} \mid Y_0) = \prod_{t=1}^T q(Y_t \mid Y_{t-1}), \qquad q(Y_t \mid Y_{t-1}) = \mathcal{N}(\sqrt{1-\beta_t}\, Y_{t-1},\, \beta_t I),
\]
with variance schedule $\beta_t$, so that $q(Y_T)$ approaches an isotropic Gaussian for large $T$. The product $\prod_{t=1}^T q(Y_t \mid Y_{t-1})$ is the joint density of the entire noising trajectory $Y_1, \dots, Y_T$ conditional on the noise-free (uncorrupted) data point $Y_0$, factorized into one-step transitions by the Markov property of the forward process. The reverse process learns a denoising model that inverts the corruption,
\[
p_\phi(Y_{0:T}) = p(Y_T) \prod_{t=1}^T p_\phi(Y_{t-1} \mid Y_t), \qquad p(Y_T) = \mathcal{N}(0, I).
\]
Training maximizes a variational bound on the data log-likelihood that reduces to denoising score matching: with closed-form marginal $q(Y_t \mid Y_0) = \mathcal{N}(\sqrt{\bar{\alpha}_t}\, Y_0,\, (1-\bar{\alpha}_t)I)$ and $\bar{\alpha}_t = \prod_{s \le t}(1-\beta_s)$, the network $s_\phi$ minimizes
\[
\min_\phi \sum_{t=1}^T \lambda_t \,\mathbb{E}\bigl[\lVert s_\phi(Y_t, t) - \nabla_{Y_t} \log q(Y_t \mid Y_0) \rVert^2\bigr],
\]
so $s_\phi(Y_t,t)$ approximates the score $\nabla_{Y_t}\log q_t(Y_t)$ of the noised data distribution, the same object used in score-based estimation \citep{hyvarinen2005estimation}. \citet{song2021scorebased} show that the forward and reverse processes solve stochastic differential equations whose reverse direction depends on this score; sampling starts from $Y_T \sim \mathcal{N}(0,I)$ and integrates the reverse process to produce $\hat{Y}_0$. Because each step is a local correction rather than a global bijection, diffusion models handle well-separated modes better than normalizing flows, at the cost of hundreds of network evaluations per draw, which consistency-model variants reduce \citep{song2023consistency}. The multimodal energy landscape of protein-conformation generation is a canonical setting where this mode coverage makes diffusion preferable to flows. The same construction targets a posterior when the diffusing variable is the parameter $\theta$ rather than the outcome: conditioning the score on outcome $Y$, \citet{geffner2023compositional} decompose $\nabla_\theta \log p_t(\theta \mid Y) = \nabla_\theta \log p_t(\theta) + \nabla_\theta \log p_t(Y \mid \theta)$ for compositional approximation and \citet{sharrock2024sequential} train on simulated pairs with sequential refinement, while classifier-free guidance \citep{ho2022classifier} conditions on covariates for prediction.

\subsubsection*{Variational autoencoders} 

A variational autoencoder \citep{kingma2014stochastic, kingma2019introduction} is a latent-variable generative model that learns both a low-dimensional representation of high-dimensional data and a mechanism for generating new samples. We present it in the paper's convention, where $Y \in \mathcal{Y}$ denotes an outcome, $Z \in \mathcal{Z}$ a latent variable,  and $(w, \phi)$ the decoder and encoder network parameters, respectively (reserving $\theta$ for the Bayesian parameter of interest). The variational autoencoder mirrors the Bayesian setup with the latent $Z \leftrightarrow \theta$ playing the role of the parameter: the encoder approximates the posterior $p(\theta \mid Y)$ while the decoder corresponds to the forward model $p(Y \mid \theta)$; the difference is that the variational autoencoder assumes a fixed latent prior $p(Z) = \mathcal{N}(0, I)$ and optimizes a variational bound, whereas in Bayesian inference the prior $p(\theta)$ is part of the model.

\begin{figure}[H]
\centering
\resizebox{0.85\textwidth}{!}{%
\begin{tikzpicture}[
    box/.style={rectangle,draw,minimum width=2.5cm,minimum height=1.2cm,rounded corners},
    data/.style={rectangle,draw,minimum width=2cm,minimum height=2cm,rounded corners,align=center},
    arrow/.style={->,thick},
    node distance=4cm
]
    \node[data] (input1) at (-6,1) {$y_1$\\(Face 1)};
    \node[data] (input2) at (-6,-1) {$y_2$\\(Face 2)};
    \node at (-6,-2.5) {$\vdots$};
    \node[data] (inputn) at (-6,-4) {$y_n$\\(Face n)};

    \node[box,fill=blue!10,minimum width=3cm] (encoder) at (-2,-1.5) {Encoder $(\mu, \sigma^2)$};
    \node[align=center] at (-2,0) {$f_\phi := f_1 \circ \cdots \circ f_L$};
    \node[align=center] at (-2,-2.8) {$Y \mapsto Z$};

    \node[draw,circle,minimum size=4cm,align=center] (latent) at (2,-1.5) {Latent Space\\$Z = \mu + \sigma\odot \epsilon$\\$ \epsilon \sim \mathcal{N}(0, I)$};

    \node[box,fill=red!10,minimum width=3cm] (decoder) at (6,-1.5) {Decoder};
    \node[align=center] at (6,0) {$g_w := g_1  \circ \cdots \circ g_{L'}$};
    \node[align=center] at (6,-2.8) {$Z \mapsto \hat{Y}$};

    \node[data] (out1) at (9,1) {$\hat{y}_1$\\(New \\ Face 1)};
    \node[data] (out2) at (9,-1) {$\hat{y}_2$\\(New \\ Face 2)};
    \node at (9,-2.5) {$\vdots$};
    \node[data] (outn) at (9,-4) {$\hat{y}_n$\\(New \\ Face n)};

    \draw[arrow] (input1) -- (encoder);
    \draw[arrow] (input2) -- (encoder);
    \draw[arrow] (inputn) -- (encoder);

    \draw[arrow] (encoder) -- (latent);
    \draw[arrow] (latent) -- (decoder);

    \draw[arrow] (decoder) -- (out1);
    \draw[arrow] (decoder) -- (out2);
    \draw[arrow] (decoder) -- (outn);

    \node[align=center] at (-6,2.5) {Training\\Data};
    \node[align=center] at (9,2.5) {Generated\\Samples};

    \node[text width=3cm, align=center] at (2,1.5)
        {Compressed\\Outcomes};

\end{tikzpicture}
}
\caption{Variational autoencoder. The encoder ($f_\phi$ parameterizing \(q_\phi\)) and decoder ($g_w$ parameterizing \(p_w\)) are composite maps between outcome space ($Y$) and latent space ($Z$). The encoder compresses an enumerated training sample $y_1,\dots,y_n$ into latent representations, from which the decoder generates new examples $\hat{y}_1,\dots,\hat{y}_n$.}
\label{fig:vae_overview}
\end{figure}

This generative model has two parts: a latent prior $Z \sim p(Z) = \mathcal{N}(0, I)$ representing mean-zero independent unobservables, and a conditional outcome model $Y \mid Z \sim p_w(Y \mid Z)$ in which $w$ indexes a flexible function such as a neural network. The encoder targets the latent posterior $p_w(Z \mid Y) = p_w(Y \mid Z)\,p(Z)/p_w(Y)$, whose marginal likelihood \(p_w(Y)\) is in general intractable. The variational autoencoder therefore introduces a Gaussian variational approximation $q_\phi(Z \mid Y) = \mathcal{N}(Z \mid \mu, \operatorname{diag}(\sigma^2))$ with $\mu = \mu_\phi(Y)$ and $\sigma = \sigma_\phi(Y)$ output by an encoder network, and uses the reparameterization $Z = \mu + \sigma \odot \epsilon$, $\epsilon \sim \mathcal{N}(0, I)$, so that sampling is a differentiable function of $(\mu, \sigma)$ and the decoder reconstruction $\hat{Y} = g_w(Z)$ can be optimized by gradient descent (see Figure \ref{fig:vae_overview}). Training maximizes the evidence lower bound\footnotemark
\begin{align}
\mathcal{L}(w,\phi, y)
= \mathbb{E}_{z\sim q_\phi(\cdot\mid y)}\bigl[\log p_w(y\mid z)\bigr]
   - \mathrm{KL}\!\bigl(q_\phi(z\mid y)\,\|\,p(z)\bigr),
\end{align}
with full objective $\mathbb{E}_{Y\sim p_{\mathrm{obs}}}[\mathcal{L}(w,\phi, Y)]$,
estimated by its average over the training sample.

\footnotetext{For a sample $y \sim p_{\mathrm{obs}}$ and any variational
density $q_\phi(z\mid y)$ that integrates to one, the marginal likelihood
admits the identity
\begin{align*}
p_w(y)
= \int \frac{p_w(y,z)}{q_\phi(z\mid y)}\,q_\phi(z\mid y)\,dz
= \mathbb{E}_{z\sim q_\phi(\cdot\mid y)}\!\Bigl[\tfrac{p_w(y,z)}{q_\phi(z\mid y)}\Bigr].
\end{align*}
Applying Jensen's inequality to the concave $\log$,
\begin{align*}
\log p_w(y)
= \log \mathbb{E}_{z\sim q_\phi(\cdot\mid y)}\!\Bigl[\tfrac{p_w(y,z)}{q_\phi(z\mid y)}\Bigr]
\ge \mathbb{E}_{z\sim q_\phi(\cdot\mid y)}\!\Bigl[\log\tfrac{p_w(y,z)}{q_\phi(z\mid y)}\Bigr]
=: \mathcal{L}(w,\phi, y).
\end{align*}
Because $p_w(y,z)=p_w(y\mid z)\,p(z)$, taking expectations under
$q_\phi(z\mid y)$ yields the bound above, where $\mathrm{KL}(\cdot\|\cdot)$
is the Kullback--Leibler divergence. Maximizing it tightens the gap to
$\log p_w(y)$ and drives $q_\phi(z\mid y)$ toward the true latent posterior
$p_w(z\mid y)$.}

We note that the decoder $p_w(Y \mid Z)$ is the outcome generator: drawing $Z \sim p(Z)$ and decoding yields synthetic draws from the marginal data distribution, and a conditional variational autoencoder that appends a treatment indicator instantiates the counterfactual outcome map $G(U, X, a)$ when the Gaussian latent is an adequate model for the potential-outcome distribution. The encoder is instead an approximate posterior, placing the same architecture in the parameter--outcome class; because the variational family is Gaussian, multimodal or heavy-tailed posteriors are systematically underdispersed regardless of network capacity, a limitation not shared by the likelihood-free methods of that class. A conditional variational autoencoder that takes covariates as input produces a predictive distribution $p(Y \mid X)$ with calibrated intervals only when the Gaussian latent structure matches the predictive law, in contrast to quantile networks, which make no such assumption. Read against classical structure, the variational autoencoder is a nonlinear factor-analysis model: the Gaussian latent prior is a modeling convention rather than the inferential prior, and the decoder is a nonlinear analogue of the factor-loading map.

\subsection{Generators from Simulated Parameter--Outcome Pairs}

The second generator class targets Bayesian inverse inference. The data object is a simulated reference table
\[
\{(\theta_i,Y_i)\}_{i=1}^N, \qquad \theta_i \sim p(\theta), \quad Y_i \sim p(Y\mid \theta_i),
\]
and the target is a generator for the posterior distribution,
\[
(\theta\mid Y=y) \stackrel{d}{=} G(y,Z), \qquad Z \sim \mathrm{Unif}(0,1).
\]
This is the class most directly associated with simulation-based inference \citep{cranmer2020frontier}. The generator consumes an observed outcome or summary statistic and returns posterior draws, posterior quantiles, or a posterior density approximation.

The two classical members are approximate Bayesian computation and fiducial inference. Approximate Bayesian computation generates parameter--outcome pairs from the prior and forward simulator, then keeps parameters whose simulated outcomes are close to the observed outcome. It requires no likelihood evaluation and is broadly applicable, but its acceptance rate deteriorates quickly with the dimension of the summary statistic. Fiducial inference also inverts a forward data-generating equation, but it treats the resulting distribution over parameters as induced by the structural equation rather than by a prior. The connection to the present framework is structural and worth stating directly: the noise outsourcing representation $\theta = G(Y, Z)$ \emph{is} an inverted data-generating equation, and modern generalized fiducial inference \citep{hannig2016generalized} and the theory of confidence distributions \citep{xie2013confidence} study exactly such inversions and the conditions under which the resulting distribution has calibrated frequentist coverage. In regular problems the fiducial distribution can coincide with the Bayesian posterior under Jeffreys' prior; outside such settings it requires additional conventions. A generator trained on $(\theta, Y)$ pairs and a generalized fiducial construction can therefore be read as two routes to the same object, and the fiducial literature is a natural source of coverage guarantees for the parameter--outcome class. Approximate Bayesian computation and fiducial inference are tied to this class because the simulated table $\{(\theta_i, Y_i)\}$ is their natural training object; normalizing flows are bidirectional and serve this class through the same change-of-variables framework used for outcome and predictive generation.

The modern simulation-based inference literature replaces rejection with a trained network. Neural posterior estimation fits a conditional density (typically a normalizing flow) directly to the posterior \citep{papamakarios2016fast, greenberg2019automatic}; the BayesFlow framework \citep{radev2020bayesflow} pairs an invertible flow with a summary network learned jointly from the simulator and is widely used for amortized Bayesian inference. These approaches yield exact density evaluation at the price of an invertible architecture and a tractable Jacobian. Neural likelihood estimation instead models $p(Y \mid \theta)$ \citep{papamakarios2019sequential}, and neural ratio estimation targets the likelihood-to-evidence ratio by classification \citep{hermans2020likelihood, durkan2020contrastive}; both then require a downstream sampler. Conditional diffusion models learn a posterior score and generate parameter draws through iterative denoising \citep{geffner2023compositional, sharrock2024sequential}, handling multimodality at substantial prediction-time cost. Software for this family is mature \citep{tejerocantero2020sbi}, and standardized benchmarks exist \citep{lueckmann2021benchmarking}.

Generative Bayesian computation belongs to this same parameter--outcome class but uses a quantile rather than density parameterization. For scalar \(\theta\), it estimates \(F^{-1}_{\theta\mid Y=y}(\tau)\) directly with the pinball loss; for vector parameters, it uses an autoregressive conditional quantile factorization. It therefore requires neither likelihood evaluation nor an invertible map. The price is that it returns posterior draws and quantiles rather than a closed-form posterior density, so calibration checks such as coverage diagnostics remain essential (Section~\ref{sec:app_ebola_inverse}). The detailed formulation of the classical members and the bidirectional flow follows.

\subsubsection*{Approximate Bayesian computation} 

Approximate Bayesian computation conducts likelihood-free inference when data can be simulated from a structural model \citep{beaumont2002approximate, marin2012approximate, sisson2018overview}. It simulates a reference table $\{(\theta_i, Y_i^\star)\}_{i=1}^N$ with $Y_i^\star \sim p(Y \mid \theta_i)$ and compares each simulated output to the data through a low-dimensional summary statistic $S(\cdot) \in \mathbb{R}^k$, $k \ll \dim(Y)$. A common posterior approximation is
\begin{equation}
p^\epsilon_{\mathrm{ABC}}(\theta \mid Y = y) = \frac{1}{m_\epsilon(y)} \int K_\epsilon(S(y^\star) - S(y))\, p(y^\star \mid \theta)\, p(\theta)\, dy^\star,
\end{equation}
where $K_\epsilon$ is a kernel with bandwidth $\epsilon$ and $m_\epsilon(y)$ a normalizing constant. As $\epsilon \to 0$ with $S$ sufficient, $p^\epsilon_{\mathrm{ABC}}(\theta \mid Y) \to p(\theta \mid Y)$; with a uniform kernel the approximation reduces to the conditional law of $\theta$ given $\|S(Y^\star) - S(Y)\| < \epsilon$, so that $p^\epsilon_{\mathrm{ABC}}(\theta \mid Y) \propto p(\theta)\, p(\|S(Y^\star) - S(Y)\| < \epsilon \mid \theta)$. Figure~\ref{fig_abc} summarizes the construction. The table size $N$ is the central control parameter: a larger table sustains a smaller bandwidth at a fixed acceptance rate, the same role $N$ plays in the generative methods of Section~\ref{sec:gbc}. A central application is population genetics, where likelihood-based inference for demographic processes such as migration is intractable but genetic data can be simulated under competing scenarios and compared through summary statistics such as allele frequencies.

\begin{figure}[H]
\centering
\resizebox{0.85\textwidth}{!}{%
\begin{tikzpicture}[
    node distance=3cm,
    box/.style={rectangle, draw, rounded corners=8pt, minimum width=2.2cm, minimum height=1.2cm, fill=blue!8},
    proc/.style={rectangle, draw, rounded corners=8pt, minimum width=3.5cm, minimum height=1.8cm, fill=green!8},
    data/.style={rectangle, draw, rounded corners=8pt, minimum width=3.2cm, minimum height=1.2cm, fill=red!8},
    arrow/.style={->, thick, draw=gray!60},
    every node/.style={font=\small}
]
\node[box] (prior) at (0,1.5) {Prior $p(\theta)$};
\node[proc] (sim) at (3.5,1.5) {\newline $Y^\star \sim p(Y\mid\theta)$};
\node[box] (stat) at (6.7,1.5) {$S(Y^\star)$};
\node[data] (yobs) at (0,-1.5) {Outcome $Y$};
\node[box] (sobs) at (6.7,-1.5) {$S(Y)$};
\node[proc] (comp) at (10.5,0) {\newline $\|S(Y^\star) - S(Y)\| < \epsilon$};
\node[data] (post) at (14.5,0) {\newline $p_{\mathrm{ABC}}^{\epsilon}(\theta\mid Y)$};
\draw[arrow] (prior) -- (sim);
\draw[arrow] (sim) -- (stat);
\draw[arrow] (yobs) -- (sobs);
\draw[arrow] (stat) -- (comp);
\draw[arrow] (sobs) -- (comp);
\draw[arrow] (comp) -- (post);
\node[align=left, font=\footnotesize] at (3,-3) {
    $\epsilon \to 0$: Better approximation\\
    $\epsilon$ large: Higher acceptance rate
};
\end{tikzpicture}
}
\caption{Approximate Bayesian computation. Parameters drawn from the prior are pushed through the simulator; a parameter is retained when its simulated summary statistic falls within $\epsilon$ of the observed summary statistic.}
\label{fig_abc}
\end{figure}

\citet{jiang2017learning} prove that the posterior mean $\mathbb{E}[\theta \mid Y]$ is the optimal summary statistic under squared-error loss and learn $S$ with deep networks, the result underlying the sufficient-statistic discussion of Section~\ref{}. The cost of the method is that the acceptance rate falls as $\epsilon^{\dim(S)}$, so it becomes prohibitive when the summary dimension is large, and when $S$ is not sufficient the approximation is biased for every $\epsilon$. Approximate Bayesian computation has no differentiable training objective, in contrast to the pinball loss of generative Bayesian computation, the flow log-likelihood, and the evidence lower bound; neural variants \citep{papamakarios2016fast, cranmer2020frontier} replace the rejection step with a trained density estimator, exhibiting the method as the conceptual ancestor of simulation-based inference.

\subsubsection*{Fiducial inference}
Fiducial inference constructs a data-dependent distribution over parameters without a prior \citep{hannig2016generalized}. Given a structural model $Y = G_{\mathrm{fwd}}(U, \theta)$, in which $G_{\mathrm{fwd}}$ maps latent noise and parameter to data (the opposite direction from the inverse posterior map $\theta = G(Y,Z)$) and $U \sim F_U$ is known, invertibility in $\theta$ for fixed $U$ yields a fiducial map $\theta = Q_{Y}(U) := G_{\mathrm{fwd}}^{-1}(Y, U)$, exactly the parameter--outcome inverse map of Section~\ref{sec:framework}; propagating the randomness of $U$ induces a distribution over $\theta$. For the additive model $Y = \theta + U$ with $U \sim \mathcal{N}(0,1)$ the fiducial distribution is $\theta \mid Y = y \sim \mathcal{N}(y, 1)$. For a sample $Y_1,\dots,Y_n \sim \mathcal{N}(\mu, \sigma^2)$, the sufficient statistics satisfy $\bar{Y} = \mu + \sigma V_1$ and $S^2 = \sigma^2 V_2$, where $S^2 = \tfrac{1}{n-1}\sum_{i=1}^n (Y_i - \bar{Y})^2$ and the independent latents are $V_1 \sim \mathcal{N}(0, 1/n)$ and $V_2 \sim \mathrm{Gamma}(\tfrac{n-1}{2}, \tfrac{n-1}{2})$; inverting these relations ($\mu = \bar{Y} - \sigma V_1$, $\sigma = S/\sqrt{V_2}$) gives the fiducial distributions for $\mu$ and $\sigma$. A fiducial rejection sampler, for observed data $Y = y$ and draws $i = 1,\dots,N$, samples $U_i^\star \sim F_U$, solves $\theta_i^\star = \arg\min_\theta \|y - G_{\mathrm{fwd}}(U_i^\star, \theta)\|$, and accepts $\theta_i^\star$ when $\|y - G_{\mathrm{fwd}}(U_i^\star, \theta_i^\star)\| \le \epsilon$, a procedure structurally identical to approximate Bayesian computation: both simulate auxiliary randomness, invert the forward model, and accept on a distance criterion. The distinction is conceptual, since fiducial inference is prior-free while approximate Bayesian computation targets the Bayesian posterior; in regular models the fiducial distribution coincides with the posterior under Jeffreys' prior, the Jacobian of $G_{\mathrm{fwd}}^{-1}$ encoding the Fisher information, and outside such models the generalized fiducial framework of \citet{hannig2016generalized} supplies the additional conventions. Quantile-based generative Bayesian computation, by contrast, requires neither a structural equation nor invertibility, approximating $F^{-1}_{\theta \mid y}(\tau)$ from simulated pairs directly.

\subsubsection*{Normalizing flows and independent component analysis} 

Normalizing flows and independent component analysis (ICA) share a common idea, modeling observed data as an invertible transformation of independent latent variables, with ICA the linear special case. In ICA \citep{mackay1999maximum}, the data $X \in \mathbb{R}^K$ is a linear mixture $X = A Z$ of independent latent sources $Z \in \mathbb{R}^K$, and the goal is to recover the matrix $W = A^{-1}$ so that $Z = W X$ (Figure~\ref{fig_ica}). The sources are mutually independent, $p_Z(z) = \prod_{k=1}^K p_k(z_{k})$, with each $p_k$ non-Gaussian, since a Gaussian $Z$ would make the model rotationally invariant and hence non-identifiable. Writing the joint density as $p(x, z \mid A) = \delta(x - A z)\, p_Z(z)$ and marginalizing the latent variable gives the closed-form likelihood
\[
p_X(x \mid A) = \int p(x, z\mid A)\, dz =  \int \delta(x - A z)\, p_Z(z)\, dz = \frac{1}{|\det A|} \prod_{k=1}^K p_k\bigl((A^{-1} x)_k\bigr),
\]
using the Dirac scaling identity $\int \delta(x - A z) f(z)\, dz = |\det A|^{-1} f(A^{-1} x)$, valid for invertible $A$. ICA is thus a latent-variable model with a deterministic linear decoder $X = A Z$ and an independent latent prior.

\begin{figure}[H]
\centering
\resizebox{0.6\textwidth}{!}{%
\begin{tikzpicture}[
    latent/.style={circle, draw=black!50, thick, minimum size=1.3cm, fill=blue!8, font=\small},
    obs/.style={circle, draw=black!50, thick, minimum size=1.3cm, fill=red!8, font=\small},
    mat/.style={rectangle, draw=black!50, thick, rounded corners=8pt,
                minimum width=2.4cm, minimum height=1.3cm, fill=green!8, font=\small},
    arrow/.style={->, thick, draw=gray!60},
    rowlabel/.style={font=\small, text=black, anchor=east}
]
\node[rowlabel] at (-0.4, 1.5) {mixing};
\node[latent] (z1) at (0.8, 1.5)  {$Z$};
\node[mat]    (A)  at (4,   1.5)  {$A$};
\node[obs]    (x1) at (7.2, 1.5)  {$X$};
\draw[arrow] (z1) -- (A);
\draw[arrow] (A)  -- (x1);
\node[rowlabel] at (-0.4, -0.5) {unmixing};
\node[obs]    (x2) at (0.8, -0.5) {$X$};
\node[mat]    (W)  at (4,   -0.5) {$W = A^{-1}$};
\node[latent] (z2) at (7.2, -0.5) {$\hat{Z}$};
\draw[arrow] (x2) -- (W);
\draw[arrow] (W)  -- (z2);
\node[font=\footnotesize, align=left] at (1.8, -1.8)
    {$p_Z(z) = \prod_k p_k(z_{k})$};
\node[font=\footnotesize, align=center] at (6.4, -1.8)
    {$X = AZ$ };
\end{tikzpicture}
}
\caption{Linear independent component analysis. The mixing matrix $A$ maps independent latent sources $Z$ to observed data $X$; inference recovers the unmixing matrix $W = A^{-1}$.}
\label{fig_ica}
\end{figure}

Normalizing flows generalize this from linear to nonlinear, learnable maps (Figure~\ref{fig_normalizing_flows}). Let $X \in \mathbb{R}^D$ have unknown density $p_X$, modeled as $X = G(Z)$ for a latent $Z \sim p_Z$ and an invertible $G$ with tractable Jacobian, so that the change-of-variables formula gives exact densities,
\begin{equation}
p_X(x) = p_Z(z)\, \bigl| \det J_G(z) \bigr|^{-1}, \qquad z = G^{-1}(x),
\end{equation}
with $J_G(z) = \partial G(z)/\partial z$. To make $G$ both flexible and tractable it is composed of $L$ simpler bijections, $G = T_L \circ \cdots \circ T_1$, with forward pass $z_k = T_k(z_{k-1})$ and inverse $z_{k-1} = T_k^{-1}(z_k)$; the Jacobian determinant factorizes by the chain rule, $\det J_G(z) = \prod_{k=1}^L \det\!\bigl(\partial T_k(z_{k-1})/\partial z_{k-1}\bigr)$, so the log-likelihood of observed data is
\begin{equation}
\log p_X(x) = \log p_Z(G^{-1}(x)) - \log \bigl| \det J_G(G^{-1}(x)) \bigr|,
\end{equation}
maximized to fit $G$. Common building blocks are affine couplings and invertible linear maps; the framework originates with \citet{dinh2015nice} and the Real NVP model \citep{dinh2017density} and has been applied to conditional density estimation \citep{trippe2018conditional} and posterior approximation \citep{rezende2015variational, muller2019neural}. An invertible neural network \citep{dinh2017density} is such a bijection with tractable Jacobian: triangular constructions give efficient determinants \citep{song2019mintnet}, common architectures can have unstable inverses \citep{behrmann2021understanding}, and HINT networks \citep{kruse2021hint} transport a complex posterior toward a simple base measure for direct posterior sampling.

\begin{figure}[H]
\begin{center}
\resizebox{\textwidth}{!}{%
\begin{tikzpicture}[
    node distance=1.8cm,
    transform/.style={rectangle, draw, rounded corners, fill=gray!10, minimum width=1.3cm, minimum height=0.9cm},
    latent/.style={circle, draw, fill=blue!10, minimum size=1cm},
    obs/.style={circle, draw, fill=red!10, minimum size=1cm},
    arrow/.style={-Latex, thick, color=black!70},
    invarrow/.style={dotted, -Latex, thick, color=blue!50},
    jacarrow/.style={dashed, -Latex, thick, color=purple!40},
    font=\small
]

\node[latent] (z) {$Z$};
\node[transform, right=of z] (t1) {$T_1$};
\node[transform, right=of t1] (t2) {$T_2$};
\node[right=of t2, font=\normalsize] (dots) {$\cdots$};
\node[transform, right=of dots] (tk) {$T_L$};
\node[obs, right=of tk] (x) {$X$};

\draw[arrow] (z) -- node[above, yshift=2pt] {$z_0$} (t1);
\draw[arrow] (t1) -- node[above, yshift=2pt] {$z_1$} (t2);
\draw[arrow] (t2) -- node[above, yshift=2pt] {$z_2$} (dots);
\draw[arrow] (dots) -- node[above, yshift=2pt] {$z_{L-1}$} (tk);
\draw[arrow] (tk) -- node[above, yshift=2pt] {$z_L$} (x);

\draw[jacarrow] (t1.north) to[out=80,in=100] node[above, font=\footnotesize, yshift=2pt] {$\det J_{T_1}$} (z.north);
\draw[jacarrow] (t2.north) to[out=80,in=100] node[above, font=\footnotesize, yshift=2pt] {$\det J_{T_2}$} (t1.north);
\draw[jacarrow] (tk.north) to[out=80,in=100] node[above, font=\footnotesize, yshift=2pt] {$\det J_{T_L}$} (dots.north);

\draw[invarrow] (x) -- node[below, yshift=-2pt] {$T_L^{-1}$} (tk);
\draw[invarrow] (tk) -- node[below, yshift=-2pt] {$T_{L-1}^{-1}$} (dots);
\draw[invarrow] (dots) -- node[below, yshift=-2pt] {$T_2^{-1}$} (t2);
\draw[invarrow] (t2) -- node[below, yshift=-2pt] {$T_1^{-1}$} (t1);
\draw[invarrow] (t1) -- node[below, yshift=-2pt] {$T_1^{-1}$} (z);

\node[above=0.3cm of z, font=\footnotesize, align=center] {$p_Z(z)$};
\node[above=0.3cm of x, font=\footnotesize, align=center] {$p_X(x) = p_Z(z)\,|\det J_G(z)|^{-1}$};

\end{tikzpicture}
}

\end{center}
\caption{Normalizing flow. A composition of invertible maps $T_1, \ldots, T_L$ transports a simple base density $p_Z$ to the data density $p_X$; the inverse pass and the factorized Jacobian determinant give exact likelihoods.}
\label{fig_normalizing_flows}
\end{figure}

In the parameter--outcome class a conditional flow $f_\phi(\cdot \mid Y)$ (with \(\phi\) denoting network parameters) learns the bijection $\theta = f_\phi^{-1}(Z \mid Y)$ from a base law $Z \sim p_Z$ to the posterior, with the explicit objective
\[
\max_\phi \frac{1}{N}\sum_{i=1}^N \Bigl[ \log p_Z\bigl(f_\phi(\theta_i \mid Y_i)\bigr) + \log \bigl|\det J_{f_\phi}(\theta_i \mid Y_i)\bigr| \Bigr]
\]
over simulated pairs $\{(\theta_i, Y_i)\}_{i=1}^N$, giving the exact approximate-posterior density $\hat{p}(\theta \mid Y) = p_Z(f_\phi(\theta \mid Y))\,|\det J_{f_\phi}(\theta \mid Y)|$, an advantage over the quantile parameterization of generative Bayesian computation, which returns quantiles but no closed-form density \citep{papamakarios2019sequential, greenberg2019automatic}. The invertibility requirement forces $\dim(\theta) = \dim(Z)$ (or padding) and, for posteriors with well-separated modes, a continuous path between modes through the base law that can distort their shapes, a constraint that quantile and diffusion estimators avoid. The same construction serves the other two classes by changing the conditioning: an unconditional flow $X = G(Z)$ is an outcome generator whose exact likelihood scores draws (used, for example, to flag low-likelihood network traffic as anomalous), and a covariate-conditioned flow $Y = G(X, Z)$ is a predictive generator with exact conditional density.

\subsection{Generators for Predictive Distributions}

The third generator class targets prediction with uncertainty quantification. The data object is an input--output sample or simulator table
\[
\{(X_i,Y_i)\}_{i=1}^N,
\]
and the target is the conditional outcome distribution
\[
Y\mid X=x \stackrel{d}{=} G(x,Z), \qquad Z \sim \mathrm{Unif}(0,1).
\]
This is a forward generator rather than an inverse generator: the input is a covariate vector, design point, parameter setting, or simulator configuration, and the output is a predictive distribution for future outcomes.

Classical regression, Gaussian process emulation \citep{kennedy2001bayesian}, conditional flows, conditional variational autoencoders, conditional adversarial generators, and conditional diffusion models all fit within this class when their output is a conditional distribution for \(Y\mid X\). Their differences are operational. Gaussian process emulators give smooth interpolation and uncertainty through covariance modeling, but require specifying covariance structure and can scale poorly. Conditional flows provide exact density evaluation if an invertible map is appropriate. Conditional variational autoencoders and adversarial generators produce fast conditional samples but require empirical calibration checks. Conditional diffusion models are strong for complex multimodal predictive distributions, but sampling remains sequential and expensive.

Quantile neural networks are direct predictive generators when the target is a set of conditional quantiles or prediction intervals. Exchanging the roles of \(\theta\) and \(Y\) in generative Bayesian computation detailed in the next section gives the forward map \(Y=G(X, Z)\). This role reversal is important: the same algorithm can estimate a posterior generator from \((\theta,Y)\) pairs or a predictive generator from \((X,Y)\) pairs, but the statistical interpretation changes from inverse inference to forward prediction. Conformalized quantile regression \citep{romano2019conformalized} augments such a quantile generator with a distribution-free calibration step, delivering finite-sample marginal coverage \citep{vovk2005algorithmic, lei2018distribution}. Engression \citep{shen2024engression} is a closely related distributional-regression generator that targets the conditional law through a different training objective.

Another example is conditional normalizing flow, detailed in the parameter--outcome subsection above, which gives full predictive distributions for time-to-event outcomes. With survival time $Y \in \mathbb{R}_+$ and covariates $X \in \mathbb{R}^d$ (age, treatment indicator, biomarkers), the flow parameterizes $Z = f_\phi(Y \mid X)$, mapping survival times to a base law $Z \sim \mathcal{N}(0,1)$ conditional on covariates, with conditional density
\[
p(y \mid x) = p_Z\bigl(f_\phi(y \mid x)\bigr)\, \bigl|\partial f_\phi(y \mid x)/\partial y\bigr|,
\]
maximized over observed pairs, with censoring handled by integrating the density over the censored region. The conditional quantile function follows directly as $F^{-1}_{Y \mid X}(\tau) = f_\phi^{-1}(\Phi^{-1}(\tau) \mid X)$, giving median survival, interquartile intervals, and any quantile from a single fit. This is the predictive construction $Y = G(X, Z)$ of this class with exact density. Compared with the Cox proportional-hazards model \citep{cox1972regression}, which models the hazard parametrically and provides no direct density, the conditional flow is nonparametric in the covariate--outcome relationship and supplies the full conditional density and all quantiles, at the cost of the invertibility constraint and of training the flow rather than fitting a partial likelihood.

\section{Generative Bayesian Computation}\label{sec:gbc}

This section presents generative Bayesian computation, a framework for the second generator class: posterior generation from simulated parameter--outcome pairs. The method learns the inverse posterior map directly using quantile-based deep neural networks. The same quantile-network architecture, applied with the roles of $\theta$ and $Y$ exchanged, yields a forward emulator $Y = G(X, \tau)$ for predictive inference; this dual construction is used in Section~\ref{sec:ebola}.

The framework is applicable whether or not a tractable likelihood is available. When a forward model $Y = f(\theta)$ is available but the likelihood $p(Y \mid \theta)$ is intractable, or when Markov chain Monte Carlo posterior sampling is too costly, generative Bayesian computation bypasses density evaluation and iterative sampling entirely by relying on forward simulation alone. The central object is the inverse posterior map in conditional form
\[
\theta = F^{-1}_{\theta \mid Y = y}(\tau), \qquad \tau \sim \mathrm{Unif}(0,1),
\]
which we will write equivalently as $\theta = F^{-1}_{\theta \mid y}(\tau)$ when the conditioning variable is unambiguous. For scalar $\theta$, this is the conditional quantile function introduced in Section~\ref{sec:quantile_estimation}.

\subsubsection*{An optimal-transport objective for vector-valued parameters} 
For vector-valued $\theta$ the scalar inverse cumulative distribution function is no longer unique. \citet{carlier2016vector} give a coordinate-free generalization based on the optimal-transport map of \citet{brenier1991polar}, which characterizes a vector quantile function through the gradient of a convex potential. We instead adopt an autoregressive factorization that fixes a coordinate ordering and uses scalar conditional quantiles at each stage:
\[
\theta = \left( F^{-1}_{\theta_1 \mid Y=y}(\tau_{1}),\ F^{-1}_{\theta_2 \mid \theta_1, Y = y}(\tau_{2}),\ \ldots,\ F^{-1}_{\theta_d \mid \theta_1, \ldots, \theta_{d-1}, Y = y}(\tau_d) \right),
\]
where each conditional quantile is evaluated at the observed data value $Y=y$ and, after the first coordinate, at the previously generated parameter coordinates. Two properties of this choice should be stated plainly. First, it is \emph{order-dependent}: the factorization, and hence the resulting joint generator, depends on the coordinate ordering. One-dimensional marginal quantiles of each coordinate are well defined, but joint credible regions and the shape of marginal intervals are not invariant to the ordering, and a practitioner reporting such summaries should fix and disclose the ordering. The construction of \citet{brenier1991polar} avoids this at the cost of solving an optimal-transport problem. Second, \emph{approximation error propagates}: each stage conditions on previously generated (hence imperfect) coordinates, so errors compound along the chain. We are not aware of a sharp rate analysis for the $d$-dimensional autoregressive estimator (the rate stated below is for the scalar map), and we emphasize both the ordering dependence and the error propagation as open issues. \footnote{Each conditional quantile is parameterized by a neural network trained on simulated pairs $\{(\theta_i, Y_i)\}_{i=1}^N$. The training objective is the pinball loss defined in Section~\ref{sec:quantile_estimation}; Appendix~\ref{app_bayes_rule} describes the quantile-theoretic foundations. An open-source implementation of the generative Bayesian computation framework is available in the \texttt{gbc} Python package.\url{https://github.com/VadimSokolov/gbc}. }

The Brenier route can be turned into a trainable objective that keeps the deterministic-transport structure of the scalar quantile map. By Brenier's theorem, for an absolutely continuous baseline $\mu$ the optimal transport map carrying $\mu$ onto the posterior $\pi(\cdot \mid y)$ under quadratic cost is the gradient of a convex potential \citep{brenier1991polar, mccann1995existence, villani2009optimal},
\[
\theta = \nabla \psi(U, y), \qquad \psi(\cdot, y)\ \text{convex}, \qquad (\nabla \psi)_{\#}\,\mu = \pi(\cdot \mid y), \quad U \sim \mu,
\]
and the map is $\mu$-almost-everywhere unique, so the latent $U$ is identified exactly as the probability level $\tau$ is in one dimension. The scalar check loss no longer applies, but a strictly proper scoring rule takes its place: the energy score \citep{szekely2013energy, gneiting2007strictly}
\[
S(P, \theta) = \mathbb{E}_{Z \sim P}\lVert Z - \theta \rVert - \tfrac{1}{2}\,\mathbb{E}_{Z, Z' \sim P}\lVert Z - Z' \rVert,
\]
with $Z, Z'$ independent draws from $P$, has expected value uniquely minimized at $P = \pi(\cdot \mid y)$, so fitting $\nabla \psi$ against it is consistent for the posterior. This preserves the three defining features of generative Bayesian computation, deterministic transport, a fixed proper objective, and the absence of a discriminator, in dimension greater than one, at the cost of solving an optimal-transport problem in place of a sequence of scalar quantile regressions.

Neural-network estimation of full quantile functions was popularized in distributional reinforcement learning, where \citet{dabney2017distributional} represent the distribution of future rewards rather than only its mean. We use that literature only for its numerical mechanism: embedding a quantile index and training the network with the pinball loss. In the present paper the conditioning variable is the observed data $Y$, the target is the Bayesian parameter $\theta$, and no reinforcement-learning objective or utility function is used.

\subsubsection*{Quantile neural networks} The posterior distribution can be represented through its inverse cumulative distribution function, following the classical probability integral transform and the quantile-function viewpoint developed by \citet{parzen2004quantile}, who frames quantile methods as direct alternatives to density estimation:
\[
\theta = F^{-1}_{\theta \mid Y}(\tau), \quad \tau \sim \mathrm{Unif}(0,1).
\]
More generally, when a sufficient statistic $S(Y)$ is used, this fits within the inverse map framework $\theta = G(S(Y), \tau)$, where the posterior is characterized by
\[
\theta \stackrel{d}{=} G(S(Y), \tau), \quad \tau \sim \mathrm{Unif}(0,1).
\]
A deep neural network parameterizes $G$, taking both the data summary $S(Y)$ and the quantile level $\tau$ as inputs and producing the corresponding posterior quantile as output. The network is trained by minimizing the pinball loss (Section~\ref{sec:quantile_estimation}) over simulated pairs $\{(\theta_i, Y_i)\}_{i=1}^N$. The Kolmogorov--Arnold representation theorem \citep{kolmogorov1957representation} guarantees that any multivariate continuous function can be expressed as a superposition of univariate functions; \cite{schmidt-hieber2020nonparametric} establish convergence rates for deep rectified-linear-unit networks approximating such functions. By exchanging the roles of $\theta$ and $Y$, the same construction gives a forward emulator $Y = G(X, \tau)$ for predictive inference (Section~\ref{sec:ebola}), with identical network architecture and training objective.

Quantile neural networks differ from normalizing flows. Flows require an invertible architecture and produce exact densities via the change-of-variables formula \citep{Rezende15, dinh2015nice, dinh2017density, bond-taylor2022deep}, but this constraint limits architectural flexibility. Quantile networks impose no invertibility requirement: they directly approximate the inverse cumulative distribution function without modeling the forward density. This makes them applicable when the parameter space is lower-dimensional than the data space (a common setting in Bayesian inference) and allows the use of standard feedforward architectures \citep{kingma2022autoencoding}. The trade-off is that quantile networks do not provide a closed-form density for the approximate posterior.

\subsubsection*{Convergence rates} 

The approximation quality of the generative Bayesian computation quantile estimator is governed by the smoothness of the posterior quantile function $F^{-1}_{\theta \mid y}(\tau)$ and the number of simulated training pairs $N$. Suppose the posterior quantile function, viewed as a map $(\tau, y) \mapsto F^{-1}_{\theta \mid y}(\tau)$, is a composite function with $L$ layers, where the $\ell$-th layer acts on $d_\ell$ variables with H\"{o}lder smoothness index $\beta_\ell$. The effective dimension at each layer is $d_\ell$, and the effective smoothness after composition is $\beta_\ell^* = \beta_\ell \prod_{k=\ell+1}^{L} \min(\beta_k, 1)$. Under these conditions, deep rectified-linear-unit network estimators of the posterior quantile function achieve the convergence rate
\[
\mathbb{E}\bigl[\lVert \hat{F}^{-1}_N - F^{-1}_{\theta \mid y} \rVert^2\bigr] = O\!\left( \max_{1 \le \ell \le L} N^{-2\beta_\ell^* / (2\beta_\ell^* + d_\ell)} \right),
\]
where the expectation is over the simulated training pairs. The conditional-quantile-regression form of this rate is due to \citet{padilla2022quantile}; the multi-quantile training case is treated by \citet{moon2021learning}; both rest on the composition theory of \citet{schmidt-hieber2020nonparametric}. See Appendix~\ref{app_bayes_risk} for the Bayes risk framework underlying these bounds. Two caveats are important. First, this is an approximation/statistical rate for the \emph{estimator class}; it excludes optimization error and is not specific to the cosine-embedding architecture used below, so the trained network's error may exceed it. Second, and more important for interpretation, the rate controls the distance between $\hat{F}^{-1}_N$ and the \emph{exact posterior} quantile function, not the distance to the data-generating truth. The total error a practitioner incurs decomposes as
\[
\underbrace{\bigl(\text{exact posterior inferential error}\bigr)}_{\text{governed by data size } n}
\;+\;
\underbrace{\bigl(\hat{F}^{-1}_N - \text{exact posterior}\bigr)}_{\text{governed by simulation budget } N}.
\]
Only the second term is $N$-controlled; the first is the usual posterior concentration, driven by the actual data and unaffected by $N$. The statement that ``$N$ is the binding constraint'' refers to the second term alone. Within that term, the composite structure of the posterior quantile map also shows how generative Bayesian computation can mitigate the curse of dimensionality: if the posterior concentrates on a low-dimensional sufficient statistic $S(Y)$ of intrinsic dimension $d_* \ll \dim(Y)$, the effective dimension in the rate is $d_*$ rather than the ambient data dimension, just as deep networks exploit low-dimensional composition structures \citep{schmidt-hieber2020nonparametric}. This is the formal sense in which learning $S(Y)$ improves statistical properties of generative Bayesian computation. Analogous closed-form rates under the same conditional-quantile risk framework are not directly available for the variational autoencoder evidence-lower-bound objective or for the generative adversarial network objective, since those objectives do not directly control the $L^2$ approximation error of the posterior quantile function; the use of the pinball loss in generative Bayesian computation (which directly targets the conditional quantile function) is what makes this rate analysis tractable \citep{koenker1978regression}.

\subsubsection*{Implicit quantile networks for posterior estimation}

\citet{dabney2018implicit} introduce an implicit quantile network, a deep learning architecture that approximates the full quantile function of a conditional distribution. In our setting, the network estimates $F^{-1}(\tau, Y)$, the $\tau$-quantile of $\theta$ given data $Y$, where $\tau \in (0,1)$ is a quantile index. Sampling from the approximate posterior amounts to drawing $\tau \sim \mathrm{Unif}(0,1)$ and evaluating the network at $(\tau, Y)$.

In the generative Bayesian computation framework, we parameterize the posterior quantile function as a composite network $F^{-1}(\tau, Y) = f(\tau, Y; w)$ (Figure~\ref{fig:gbc_arch}). Following \citet{dabney2018implicit}, this function is decomposed as:
\[
F^{-1}(\tau, Y) = f(\tau, Y; w) = g\left( \psi(Y) \circ \chi(\tau) \right),
\]
where $g$, $\psi$, and $\chi$ are feedforward neural networks, $\circ$ denotes element-wise (Hadamard) multiplication, and $w$ collects all network weights. The quantile embedding $\chi(\tau)$ is constructed using a cosine basis followed by a rectified-linear-unit (ReLU) activation:
\[
\chi_j(\tau) = \mathrm{ReLU} \left( \sum_{k=0}^{p-1} \cos(\pi k \tau) w_{kj} + b_j \right),
\]
providing a smooth encoding of the quantile index that allows the network to share information across nearby quantile levels. The network is estimated by minimizing the pinball loss over a randomly drawn set of quantile levels $\tau \sim \mathrm{Unif}(0,1)$ and simulated pairs $\{(\theta_i, Y_i)\}_{i=1}^N$. This architecture can estimate heteroskedastic and non-Gaussian posteriors without explicit likelihood specification, though it requires a sufficiently large training sample to control approximation error.

\begin{figure}[H]
\centering
\begin{tikzpicture}[
    node distance=0.6cm and 1.2cm,
    box/.style={draw, rounded corners=2pt, minimum width=1.5cm,
                minimum height=0.45cm, font=\scriptsize, align=center},
    io/.style={draw, rounded corners=2pt, minimum width=1.1cm,
               minimum height=0.45cm, font=\scriptsize, fill=gray!12},
    arr/.style={-{Stealth[length=3.5pt]}, thick},
  ]
  \node[io] (x) {$x \in \mathbb{R}^d$};
  \node[io, right=2.4cm of x] (tau) {$\tau \sim \text{Unif}(0,1)$};
  \node[box, below=of x] (fx) {$f_x$: FC+ReLU};
  \node[box, below=of tau] (cos) {$\chi(\tau) = [\cos(j\pi\tau)]_{j=1}^{32}$};
  \node[box, below=of cos] (ftau) {$f_\tau$: FC+ReLU};
  \coordinate (mid) at ($(fx)!0.5!(ftau)$);
  \node[box, below=1.4cm of mid, minimum width=2.1cm,
        fill=blue!8] (merge) {$\odot$ elementwise};
  \node[box, below=of merge, minimum width=2.1cm] (f1) {$f_1$: FC+ReLU};
  \node[box, below left=0.6cm and 0.1cm of f1, fill=green!10] (mu) {$\hat\mu$};
  \node[box, below right=0.6cm and 0.1cm of f1, fill=orange!12] (q) {$\hat q_\tau$};
  \draw[arr] (x) -- (fx);
  \draw[arr] (tau) -- (cos);
  \draw[arr] (cos) -- (ftau);
  \draw[arr] (fx) -- (merge);
  \draw[arr] (ftau) -- (merge);
  \draw[arr] (merge) -- (f1);
  \draw[arr] (f1) -- (mu);
  \draw[arr] (f1) -- (q);
\end{tikzpicture}
\caption{Implicit quantile network used by generative Bayesian computation. The conditioning input $x$ (the data summary $S(Y)$ for posterior inference, or covariates for prediction) and the quantile level $\tau$ are embedded separately, the latter through a cosine feature map $\chi(\tau)$, then combined by an elementwise (Hadamard) product and decoded to the estimated $\tau$-quantile $\hat q_\tau$ and a conditional-mean $\hat\mu$. The branches $f_x$, $f_\tau$, and $f_1$ correspond to $\psi$, $\chi$, and $g$ above; the mean head supports the composite training loss.}
\label{fig:gbc_arch}
\end{figure}

Generative Bayesian computation and the adversarial encoder--decoder both push a simple baseline measure forward onto the target conditional law, but they differ in what map is learned and what objective is optimized, and the differences favor the quantile-transport construction for inference. Appendix~\ref{app_gbc_vs_gan} develops the contrast point by point: a convex risk versus a saddle point, gradient quality, consistency, latent identifiability, density and cumulative-distribution-function access, as well as conditioning and amortization. Table~\ref{tab:gbc_vs_gan} (in that appendix) summarizes the six contrasts side by side, and the closing caveats paragraph there extends the comparison to the multivariate setting of \cite{brenier1991polar}.

An open-source implementation of the generative Bayesian computation framework is available in the \texttt{gbc} Python package (\url{https://github.com/VadimSokolov/gbc}). The Ebola application of Section~\ref{sec:app_ebola_inverse} uses a feedforward implicit-quantile architecture in the style of \citet{dabney2018implicit}: a quantile embedding $\chi(\tau)$ formed from a cosine basis of dimension $p$ followed by a rectified-linear-unit activation, combined with a data branch $\psi(Y)$ via Hadamard product, and decoded by a feedforward head $g$. The composite training objective $\mathcal{L} = \mathcal{L}_{\text{pinball}} + \lambda_1 \mathcal{L}_{\text{mean}} + \lambda_2 \mathcal{L}_{\text{mono}}$ combines the pinball loss with a conditional-mean loss and a monotonicity penalty against quantile crossings; the weights $\lambda_1, \lambda_2 > 0$ are selected by cross-validation. Training uses Adam with a piecewise-constant learning-rate schedule, and the dimensionality reduction step of Section~\ref{sec:app_ebola_inverse} uses partial least squares fit on the simulated $(\theta, Y)$ pairs. The \cite{fadikar2018calibrating} Ebola simulator output ($m = 100$ design points, $100$ stochastic replicates each, $N = 10{,}000$ trajectories of length $56$) and the code used to produce all figures and coverage diagnostics in Section~\ref{sec:ebola} are distributed with the \texttt{gbc} package.

\section{Applications}
\label{sec:ebola}

This section mirrors the three generator classes of Section~\ref{sec:architectures}.  The first application is a parameter--outcome generator for inverse inference in an Ebola agent-based epidemic model. The second is a predictive generator for forward emulation of the same Ebola simulator. The third application is a counterfactual outcome generator with a binary treatment, discussed in Appendix \ref{sec:app_lalonde}. 


\subsection{Parameter--outcome generator: Ebola inverse inference}
\label{sec:app_ebola_inverse}

The first application uses the multi-output agent-based epidemic model documented in \cite{fadikar2018calibrating}. Here the training object is a simulated parameter--outcome table, and the target is the posterior distribution over static simulator inputs given an observed epidemic trajectory. This is the parameter--outcome generator class of Section~\ref{sec:architectures}.

After the 2014-2015 West Africa Ebola outbreak, the Research and Policy for Infectious Disease Dynamics program at the National Institutes of Health convened a workshop to compare forecasting approaches used during the outbreak. A stochastic agent-based model~\citep{ajelli2018rapidd} generated synthetic populations with varying degrees of data accuracy, availability, and intervention measures. Transmission by an infected individual is determined probabilistically based on contact duration and $d=5$ static inputs $(\theta_1,\ldots,\theta_5)$ described in Table~\ref{tab:ebola_params}.

\begin{table}[H]
\centering
\begin{tabular}{c|c|c}
Parameter & Description & Range \\ \hline
$\theta_1$ & Transmission probability & $[3\times10^{-5}, 8\times10^{-5}]$ \\
$\theta_2$ & Initial infected count & $[1, 20]$ \\
$\theta_3$ & Hospital intervention delay  & $[2,10]$ \\
$\theta_4$ & Hospital intervention efficacy  & $[0.1,0.8]$ \\
$\theta_5$ & Travel reduction from intervention & $[0, 2]$ \\
\end{tabular}
\caption{Static inputs for the Ebola agent-based model.}
\label{tab:ebola_params}
\end{table}

\begin{figure}[H]
\centering
\includegraphics[width=.7\linewidth]{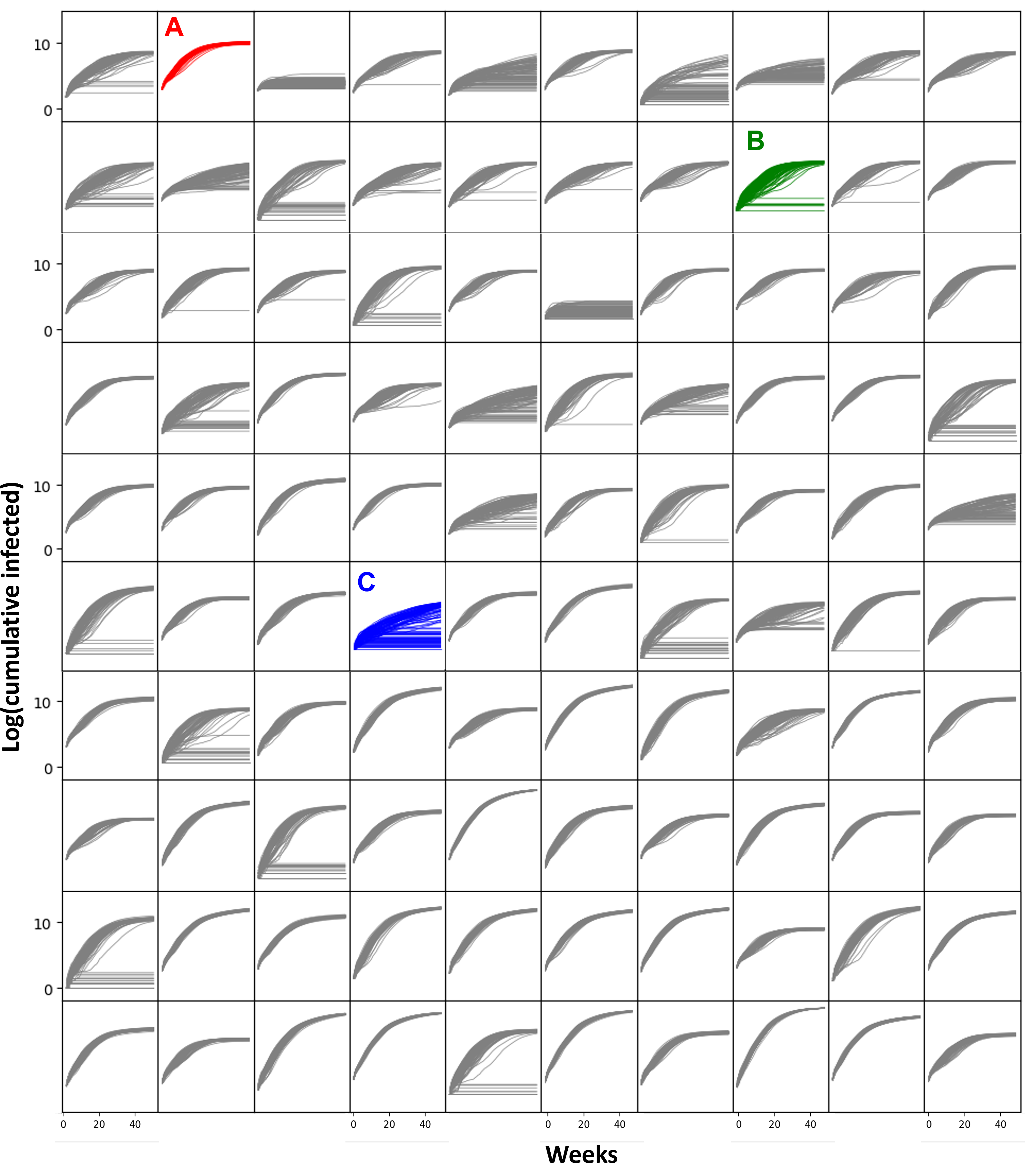}
\caption{Simulated trajectories of cumulative disease incidence across $56$ weeks for all $m=100$ parameter settings, with $100$ stochastic replicates per setting (gray lines). The three parameter settings used as holdouts in the forward-emulation application of Section~\ref{sec:app_ebola_prediction}, labeled A, B, and C, are highlighted in red, green, and blue respectively.}
\label{fig:all_samples}
\end{figure}

A single simulator run outputs cumulative infections over $56$ weeks. We use the data set of \cite{fadikar2018calibrating}: $m=100$ parameter settings generated through a space-filling symmetric Latin hypercube design, with $100$ stochastic replicates per setting, for a total of $N=10{,}000$ simulated epidemic trajectories. The log trajectories are shown in Figure~\ref{fig:all_samples}.

Given an observed trajectory $Y$, the inverse-inference target is $p(\theta \mid Y)$. We train a quantile neural network on simulated pairs $\{(\theta_i, Y_i)\}_{i=1}^N$ using partial least squares for dimensionality reduction of the 56-dimensional output. The network is trained with a composite loss
\[
\mathcal{L} = \mathcal{L}_{\text{pinball}} + \lambda_1 \mathcal{L}_{\text{mean}} + \lambda_2 \mathcal{L}_{\text{mono}},
\]
where $\mathcal{L}_{\text{pinball}}$ is the pinball loss of \cite{koenker1978regression} targeting the conditional quantile function, $\mathcal{L}_{\text{mean}}$ is an $L_1$ loss on the conditional mean (stabilizing training), and $\mathcal{L}_{\text{mono}}$ is a monotonicity regularizer penalizing quantile crossings; the regularization weights $\lambda_1, \lambda_2 > 0$ are selected by cross-validation. The network architecture follows \cite{dabney2018implicit}.

\begin{figure}[H]
\centering
\includegraphics[width=\linewidth]{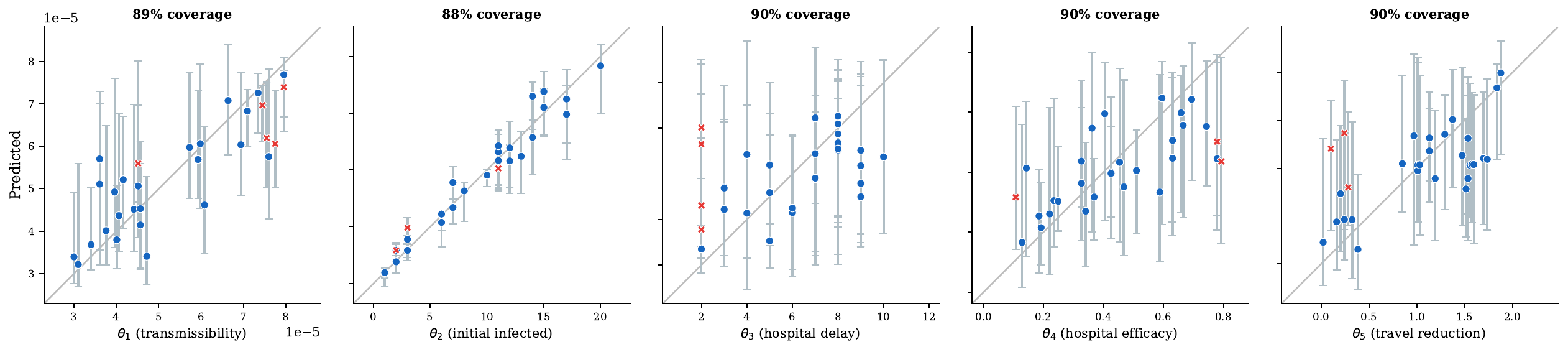}
\caption{Predicted versus true parameter values for 30 randomly selected observations from the 2,000-observation held-out test set. Each panel shows one model parameter $\theta_j$: the posterior median (dots) against the true value, with 90\% credible intervals (vertical bars) from the network's 0.05 and 0.95 quantile outputs. Blue dots indicate the true value falls within the interval; red dots indicate miscoverage. Coverage rates shown are empirical marginal per-coordinate rates computed over the full 2{,}000-observation test set; the plotted points are a 30-observation display sample.}
\label{fig:thetahist1}
\end{figure}

The generative Bayesian computation model achieves empirical 90\% marginal per-coordinate credible interval coverage of 89\% ($\theta_1$), 88\% ($\theta_2$), 90\% ($\theta_3$), 90\% ($\theta_4$), and 90\% ($\theta_5$) on 2,000 held-out test observations (an 80/20 split of the $N = 10{,}000$ simulated trajectories). All five marginal coverages are within $2$ percentage points of the nominal $90\%$ level (standard error $\approx 0.7\%$ at $n = 2{,}000$). The intervals are formed independently for each coordinate from the network's 0.05 and 0.95 quantile outputs and do not certify joint posterior coverage. Figure~\ref{fig:thetahist1} shows posterior medians and intervals for 30 randomly selected held-out observations. Parameter $\theta_5$ is recovered accurately, while $\theta_3$ and $\theta_4$ exhibit greater posterior uncertainty consistent with weaker identifiability in the simulator output; the marginal coverage is nevertheless close to nominal. The \texttt{gbc} package provides post-hoc conformal calibration when stricter coverage guarantees are required, and Section~\ref{sec:app_ebola_inverse_calibration} below probes the calibration further with simulation-based-calibration rank histograms that reveal structural mismatches the marginal coverage statistic alone does not detect.

The coverage statistics reported above measure frequentist coverage averaged over the simulated joint $(\theta, Y)$ in the held-out split. In the language of \citet{talts2018validating}, this is simulation-based calibration of the \emph{average} posterior, and it is the standard diagnostic for amortized neural inference. It is not the same as a check that the approximation $\hat{p}(\theta \mid Y = y)$ agrees with the exact posterior $p(\theta \mid Y = y)$ at any particular observed $y$; calibration averages can mask systematic errors at specific $y$, including under- or over-dispersion that cancels across the test set. Moreover, the pinball-loss training criterion controls the conditional quantile of each coordinate of $\theta$ marginally, but does not by itself certify a coherent joint posterior over $\theta$ once the autoregressive factorization of Section~\ref{sec:gbc} is invoked. Two complementary checks address this. First, rank histograms in the sense of \citet{talts2018validating} test uniformity of the simulation-based calibration ranks per coordinate. Second, where a closed-form posterior is available (a conjugate model is a natural choice) one can overlay generative Bayesian computation samples directly against the exact posterior to test agreement at specific $y$. These two checks are complementary to the coverage statistics reported above and should accompany them in routine practice.

\begin{figure}[H]
\centering
\begingroup
\setlength{\tabcolsep}{1pt}
\begin{tabular}{@{}ccccc@{}}
\includegraphics[width=0.18\linewidth]{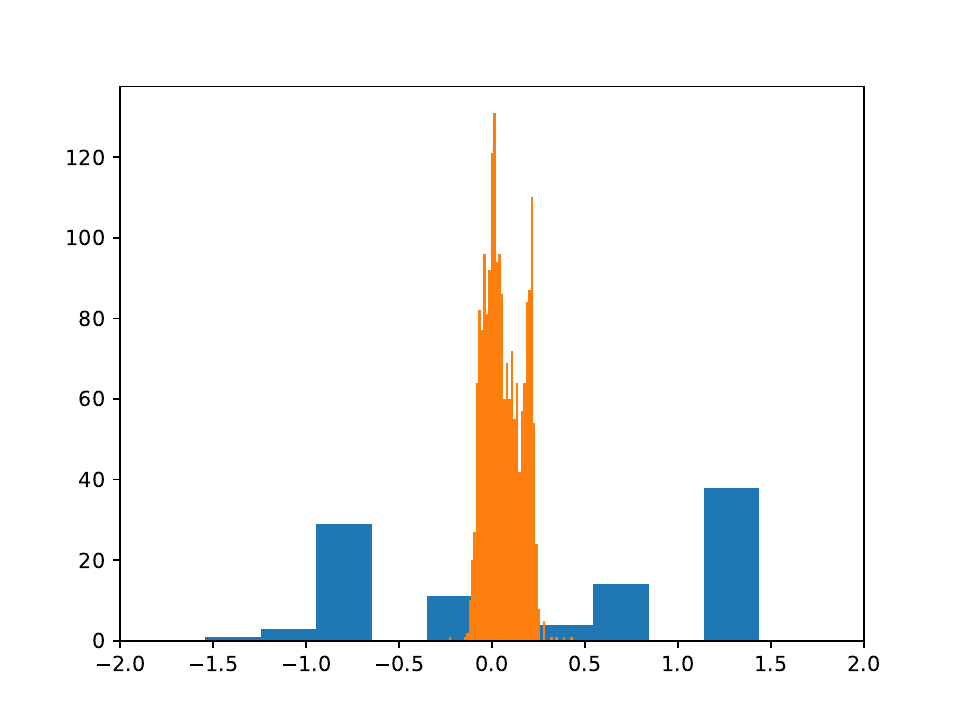} & \includegraphics[width=0.18\linewidth]{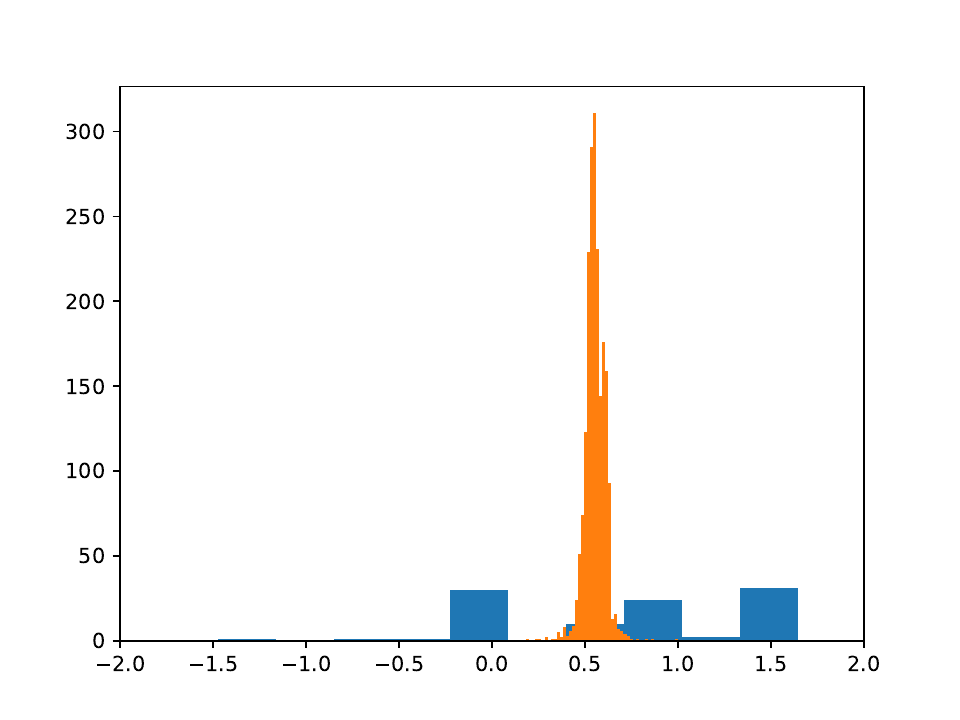} &
\includegraphics[width=0.18\linewidth]{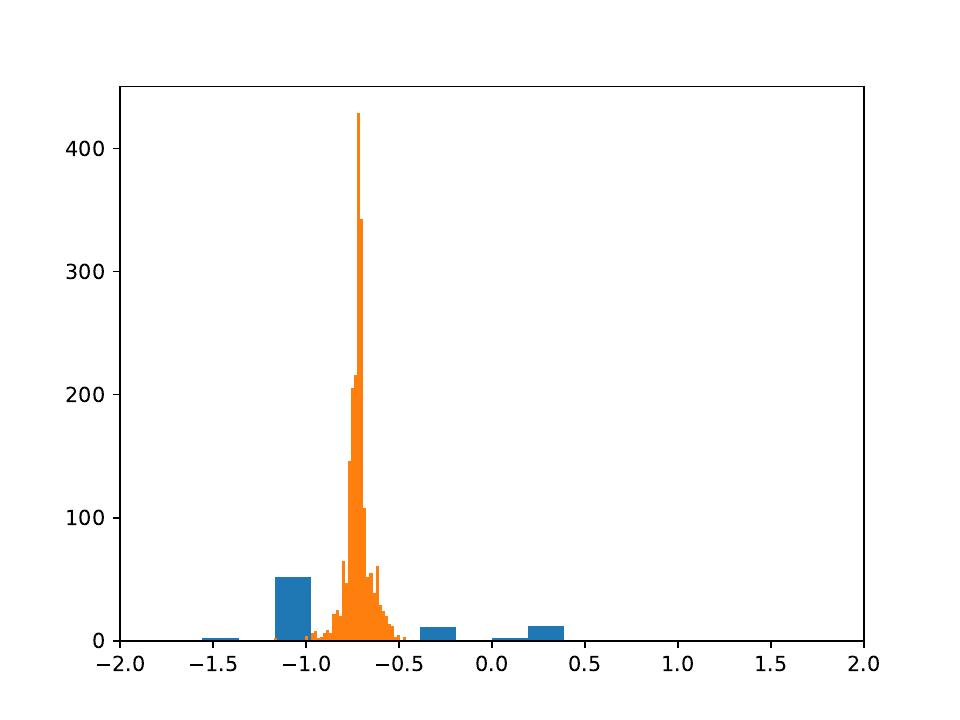} & \includegraphics[width=0.18\linewidth]{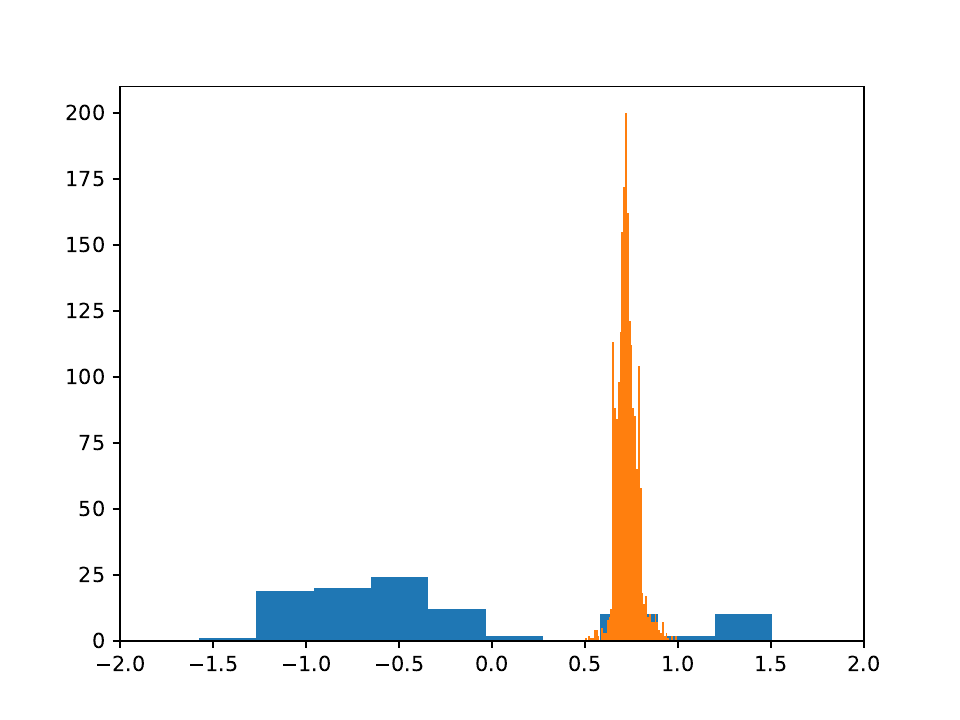} &
\includegraphics[width=0.18\linewidth]{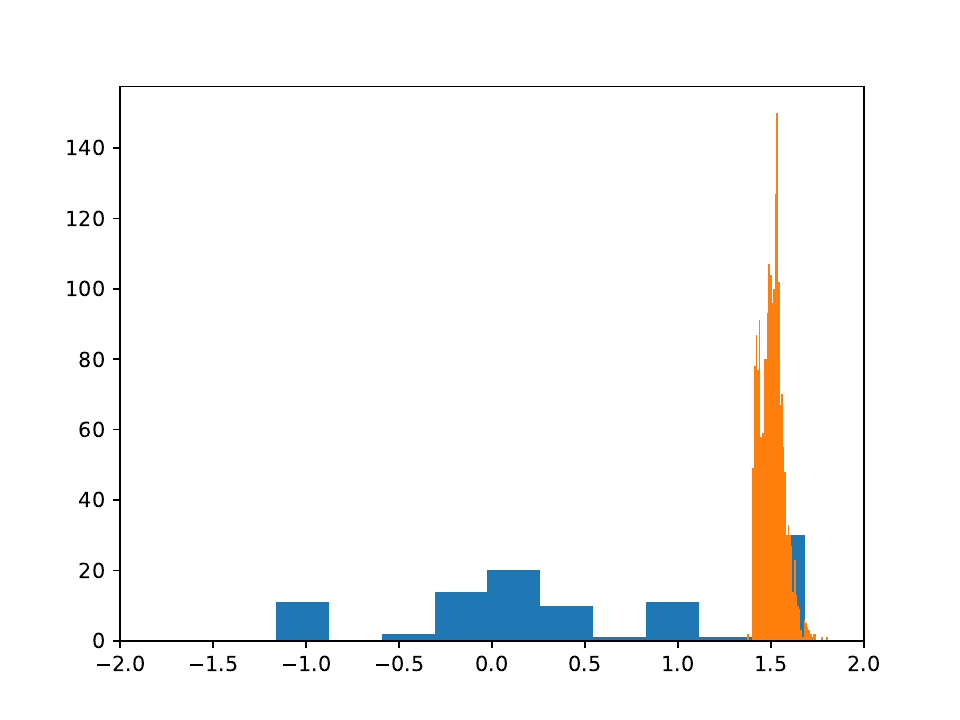}\\
$\theta_1$ & $\theta_2$ & $\theta_3$ & $\theta_4$ & $\theta_5$
\end{tabular}
\endgroup
\caption{Approximate posterior histograms for observation $y^{(7080)}$. Each panel overlays the generative Bayesian computation posterior samples (orange; generated by drawing $\tau \sim \mathrm{Unif}(0,1)$ and evaluating the trained quantile network) with a nearest-neighbor reference distribution (blue; the parameter values from the 100 agent-based model simulations whose output trajectories are closest to $y^{(7080)}$). The reference distribution is an approximate posterior proxy and may be overdispersed due to the limited number of neighbors and imperfect matching in output space.}
\label{fig:thetahist2}
\end{figure}

For a single held-out observation, Figure~\ref{fig:thetahist2} overlays the generative Bayesian computation posterior samples on a nearest-neighbor reference distribution drawn from the closest training trajectories, providing a second view of the per-coordinate fit on one fixed $y$ rather than the cross-observation summary of Figure~\ref{fig:thetahist1}.

The relevant baseline is the quantile kriging approach of \cite{fadikar2018calibrating}. It fits a Gaussian process emulator on the augmented input $(\theta,\tau)$ using the model-assisted calibration framework of \cite{higdon2008computer}. Their approach requires specifying covariance structure, basis functions, and nugget parameters, with calibration performed by Markov chain Monte Carlo. Both approaches use the same 10,000 simulator runs. After training, the quantile network produces each posterior draw by one network evaluation, whereas Gaussian-process calibration requires a separate Markov chain Monte Carlo calibration step. Approximate Bayesian computation would require repeated nearest-neighbor or rejection calculations for each observation, and MCMC with simulated likelihood would call the simulator $O(N_\text{chain})$ times per observation. Appendix~\ref{app_cost} tabulates these costs and their trade-offs.

Appendix~\ref{app_calibration} reports three further calibration checks of the generative Bayesian computation posterior on the Ebola inverse problem: a closed-form conjugate benchmark where the exact posterior is available (Kolmogorov--Smirnov statistic between $0.032$ and $0.040$ against the exact Beta), per-parameter simulation-based calibration rank histograms (the rank distribution rejects uniformity for $\theta_2$--$\theta_5$ even though marginal coverage is near-nominal, so the two diagnostics are complementary), and a detailed comparison against a flow-based neural posterior estimator on the same simulated table (the quantile network has $4$--$19$ percentage points higher marginal coverage than a masked autoregressive flow at default settings).

\subsection{Predictive generator: Ebola forward emulation}
\label{sec:app_ebola_prediction}

The second application uses the same Ebola simulator but changes the generator class. The input is now a parameter setting, and the target is the predictive distribution of the 56-week epidemic trajectory. This is a forward prediction problem, not inverse inference. The relevant training object is an input--output table indexed by the simulator input and a quantile level.

\begin{figure}[H]
\centering
\includegraphics[width=.5\linewidth]{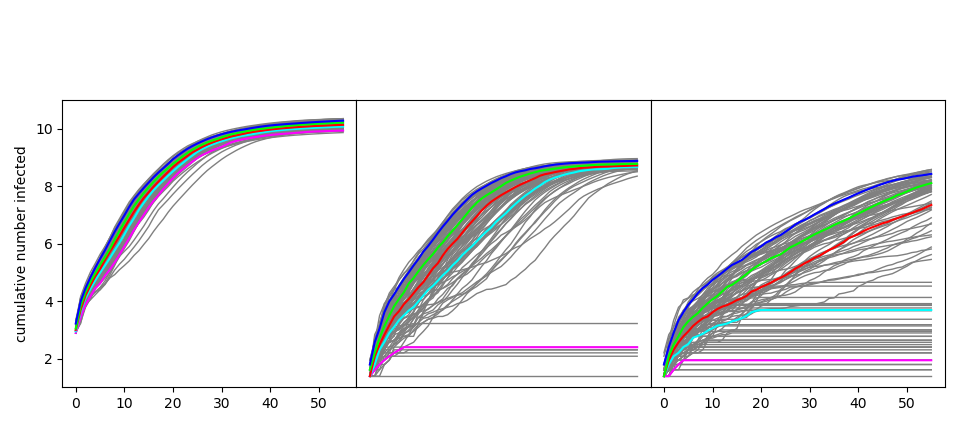}
\caption{For each holdout scenario, labeled A, B, and C, the $100$ simulated trajectories of cumulative disease incidence over $56$ weeks are shown as gray lines. The five colored lines represent the $[0.05, 0.275, 0.5, 0.725, 0.95]$ quantiles, indexed by the additional input $\tau$.}
\label{fig:holdout_q}
\end{figure}

Different Ebola parameter settings produce strongly divergent behaviors: some trajectories follow a mean trend while others exhibit bimodality, heteroskedasticity, or remain flat. To handle this stochastic variability, \cite{fadikar2018calibrating} replace the 100 replicates at each parameter setting with $n_\tau = 5$ quantile-based summary trajectories at levels $\tau \in \{0.05, 0.275, 0.5, 0.725, 0.95\}$. This reduces the stochastic output to a deterministic function of the augmented input
\[
\tilde{\theta} = \left[\theta_1, \theta_2, \theta_3, \theta_4, \theta_5, \tau \right].
\]
For testing, three parameter settings, labeled A, B, and C, and their respective $n=100$ simulated outputs were excluded from the training set. Figure~\ref{fig:holdout_q} shows the five empirical quantiles of these holdout scenarios.

\begin{figure}[H]
\centering
\includegraphics[width=\linewidth]{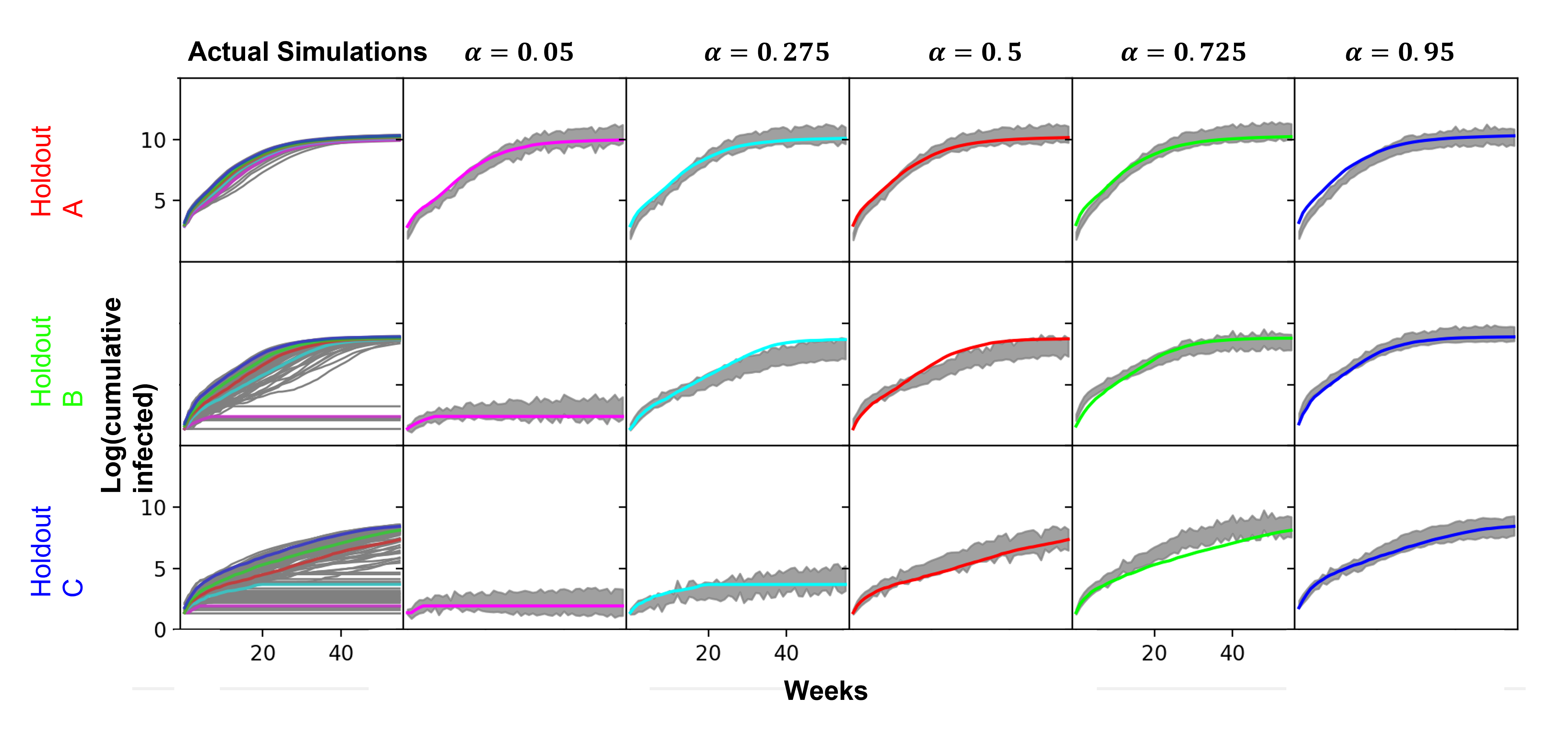}
\caption{Forward emulation with a deep neural network and Gaussian process model. The first column shows the 100 agent-based model replicates and their empirical quantiles for each holdout scenario. The remaining five columns show the model's predicted median (colored line) and 90\% prediction interval (gray band) for each quantile level $\tau$.}
\label{fig:ebola_res}
\end{figure}

Figure~\ref{fig:ebola_res} shows results from a deep neural network combined with a Gaussian process emulator. The neural network learns a nonlinear reduction of the 56-dimensional output into a 3-dimensional latent score, and independent Gaussian process regressions are then fitted to the latent scores. This is a predictive generator: the output is a conditional distribution over future trajectories given a parameter setting and quantile index. It is closest to the prediction class in Section~\ref{sec:architectures}, not to a normalizing flow, because the reduction is not constrained to be invertible and no exact density is evaluated.

A visual inspection of the lower quantiles ($\tau = 0.05, 0.275$) suggests potential improvement in lower-quantile calibration relative to the quantile kriging approach of \cite{fadikar2018calibrating}, with the true quantile trajectory remaining inside the 90\% predictive band across the full 56-week timeline for several cases. This is a qualitative observation on a single application, not a formal claim: a quantitative comparison via root mean squared error or coverage on a held-out evaluation set would be needed to confirm or refute the impression, and we have not conducted such a comparison. The key conceptual distinction from the previous subsection is that this generator maps inputs forward to outcomes, whereas the inverse-inference generator maps observed outcomes back to plausible parameters.

\section{Discussion}
\label{sec:discussion}

We have framed generative Bayesian computation within three generator classes. These are counterfactual outcome distributions, posterior distributions learned from simulated parameter and outcome pairs, and predictive distributions. The unifying idea, formalized through the Kallenberg noise outsourcing theorem, is that all three classes can be recast as learning a deterministic function of inputs and independent reference noise. In compact form, $Y = G(c,Z)$ for outcome generation, $\theta = G(Y,Z)$ for parameter generation, and $Y = G(X,Z)$ for prediction.

The main organizational point is that methods should be evaluated by the generator they are asked to estimate. For counterfactual outcome distributions, the central issue is identification of $Y(a)$ and calibration of the conditional outcome distribution; adversarial, diffusion, variational, flow, and quantile generators are numerical choices after the causal assumptions have been stated. For simulated parameter--outcome pairs, the central issue is inverse inference: the generator must transform an observed outcome into posterior draws, quantiles, or densities for $\theta$. Approximate Bayesian computation, fiducial inference, conditional flows, conditional diffusion, and generative Bayesian computation all belong here when trained on $(\theta,Y)$ pairs. For prediction problems, the generator is forward-looking: it maps covariates, design points, or simulator inputs into a predictive distribution for $Y$.

We present Generative Bayesian computation which is primarily a parameter--outcome generator. It estimates posterior quantiles directly from simulated pairs, avoiding both likelihood evaluation and invertibility constraints. Its role in prediction is obtained by exchanging the roles of parameter and outcome, as in the Ebola forward-emulation application. Its role in counterfactual inference is obtained by changing the training object to causal triples $(\tilde{S}(X),A,Y)$ and interpreting the fitted quantile map as $G(q,\tilde{s},a)$.  The main trade-off is not the split between neural and non-neural methods. Rejection methods are transparent and simulation-based, but they scale poorly with the dimension of the summary statistic. Flows provide density evaluation, at the cost of invertible maps and tractable Jacobians. Diffusion methods handle multimodality but require sequential sampling. Variational autoencoders and adversarial generators produce fast samples, yet they require calibration checks and may miss tail behavior or modes. Quantile generators directly target conditional quantiles and intervals, but they do not provide a closed-form density. The appropriate choice depends on whether the application requires counterfactual outcome distributions, parameter posteriors, or predictive distributions.

A separate axis of comparison concerns whether the generator is trained once on a fixed simulation budget (amortized inference) or refined adaptively for the specific observation of interest (sequential inference). Amortized methods, including generative Bayesian computation, conditional normalizing flows, and conditional diffusion, fit the generator on a single set of $N$ simulated pairs and produce posteriors for any new observation by direct evaluation; the simulation budget is shared across observations and the per-observation inference cost is one function evaluation. Sequential methods, including sequential neural posterior estimation \citep{papamakarios2016fast, lueckmann2017flexible, greenberg2019automatic}, sequential neural likelihood, and sequential neural ratio estimation, focus the simulation budget around the observed datum by drawing new simulations from a proposal that adapts to the current posterior approximation; this can be statistically more efficient per simulator call but discards the amortized property. The choice between the two depends on whether multiple observations are expected (amortized methods are preferred) or only a single observation is of interest with a tight simulator budget (sequential methods may pay off). Appendix~\ref{app_cost} compares these costs explicitly for the Ebola application.

The architectural matrix of Appendix ~\ref{sec:arch_cross} (Table~\ref{tab:method_class}) can be reduced to a short guide to the practical question facing the analyst. If a tractable density of the approximate posterior is required (for likelihood-ratio tests, model selection, or hierarchical embedding), a normalizing flow trained as a neural posterior or likelihood estimator is the natural choice. If only credible intervals and posterior quantiles are needed, a quantile-based generator such as generative Bayesian computation avoids the invertibility constraint and the change-of-variables Jacobian, and is straightforward to calibrate post-hoc by conformal procedures. If the target distribution is multimodal, a diffusion or score-based generator handles modes more gracefully than a single-mode-friendly variational autoencoder. If the parameter or outcome is high-dimensional and a coordinate ordering is unattractive, a Brenier optimal-transport map or a multivariate flow is preferable to the autoregressive factorization of Section~\ref{sec:gbc}. For counterfactual outcome questions the choice of architecture is secondary to the causal assumptions; once those are stated, any conditional generator can serve as the estimator of $G$. The single highest-leverage point in all three classes is to learn a low-dimensional sufficient summary of the data before fitting the generator: this reduces the effective dimension in the rate of Section~\ref{sec:gbc} and improves both calibration and stability.

Each generator class has distinct unresolved issues. Counterfactual generators inherit the identification challenges of causal inference: unconfoundedness, positivity, and support overlap. Parameter--outcome generators require posterior calibration diagnostics, better handling of high-dimensional parameters, and robustness to simulator misspecification. Predictive generators require conditional calibration under covariate shift and scalable uncertainty quantification for high-dimensional outputs. Across all three classes, the central statistical problem is to connect approximation error of the learned generator to the inferential object the researcher reports.

\bibliographystyle{apalike}
\bibliography{GenBayes,polsok}

@article{cox1972regression,
  title={Regression models and life-tables},
  author={Cox, David R},
  journal={Journal of the Royal Statistical Society: Series B (Methodological)},
  volume={34},
  number={2},
  pages={187--202},
  year={1972},
  doi = {10.1007/978-1-4612-4380-9_37}
}

@article{hyvarinen2005estimation,
  title={Estimation of non-normalized statistical models by score matching},
  author={Hyv\"{a}rinen, Aapo},
  journal={Journal of Machine Learning Research},
  volume={6},
  pages={695--709},
  year={2005}
}

@article{salimans2016improved,
  title={Improved techniques for training {GANs}},
  author={Salimans, Tim and Goodfellow, Ian and Zaremba, Wojciech and Cheung, Vicki and Radford, Alec and Chen, Xi},
  journal={Advances in Neural Information Processing Systems},
  volume={29},
  year={2016},
  eprint = {1606.03498}
}

@article{ho2022classifier,
  title={Classifier-free diffusion guidance},
  author={Ho, Jonathan and Salimans, Tim},
  journal={arXiv preprint arXiv:2207.12598},
  year={2022},
  eprint = {2207.12598v1}
}

@article{kolmogorov1957representation,
  title={On the representation of continuous functions of many variables by superposition of continuous functions of one variable and addition},
  author={Kolmogorov, Andrey N},
  journal={Doklady Akademii Nauk SSSR},
  volume={114},
  pages={953--956},
  year={1957},
  doi = {10.1090/trans2/028/04}
}

@article{cranmer2020frontier,
  title={The frontier of simulation-based inference},
  author={Cranmer, Kyle and Brehmer, Johann and Louppe, Gilles},
  journal={Proceedings of the National Academy of Sciences},
  volume={117},
  number={48},
  pages={30055--30062},
  year={2020},
  doi = {10.1073/pnas.1912789117}
}

@article{radev2020bayesflow,
  title={{BayesFlow}: Learning complex stochastic models with invertible neural networks},
  author={Radev, Stefan T and Mertens, Ulf K and Voss, Andreas and Ardizzone, Lynton and K{\"o}the, Ullrich},
  journal={IEEE Transactions on Neural Networks and Learning Systems},
  volume={33},
  number={4},
  pages={1452--1466},
  year={2020},
  eprint = {2003.06281v4}
}

@article{higdon2008computer,
  title={Computer model calibration using high-dimensional output},
  author={Higdon, Dave and Gattiker, James and Williams, Brian and Rightley, Maria},
  journal={Journal of the American Statistical Association},
  volume={103},
  number={482},
  pages={570--583},
  year={2008},
  doi = {10.1198/016214507000000888}
}

@article{ho2020denoising,
  title={Denoising diffusion probabilistic models},
  author={Ho, Jonathan and Jain, Ajay and Abbeel, Pieter},
  journal={Advances in Neural Information Processing Systems},
  volume={33},
  pages={6840--6851},
  year={2020},
  eprint = {2006.11239v2}
}

@inproceedings{song2021scorebased,
  title={Score-based generative modeling through stochastic differential equations},
  author={Song, Yang and Sohl-Dickstein, Jascha and Kingma, Diederik P and Kumar, Abhishek and Ermon, Stefano and Poole, Ben},
  booktitle={International Conference on Learning Representations},
  year={2021},
  eprint = {2011.13456v2}
}

@inproceedings{song2023consistency,
  title={Consistency models},
  author={Song, Yang and Dhariwal, Prafulla and Chen, Mark and Sutskever, Ilya},
  booktitle={International Conference on Machine Learning},
  pages={32211--32252},
  year={2023},
  eprint = {2303.01469v2}
}

@article{geffner2023compositional,
  title={Compositional score modeling for simulation-based inference},
  author={Geffner, Tomas and Papamakarios, George and Mnih, Andriy},
  journal={International Conference on Machine Learning},
  pages={11098--11116},
  year={2023},
  eprint = {2209.14249}
}

@article{sharrock2024sequential,
  title={Sequential neural score estimation: Likelihood-free inference with conditional score based diffusion models},
  author={Sharrock, Louis and Simons, Jack and Liu, Song and Sherlock, Chris},
  journal={International Conference on Machine Learning},
  year={2024},
  eprint = {2210.04872}
}

@inproceedings{greenberg2019automatic,
  title={Automatic posterior transformation for likelihood-free inference},
  author={Greenberg, David and Nonnenmacher, Marcel and Macke, Jakob},
  booktitle={International Conference on Machine Learning},
  pages={2404--2414},
  year={2019},
  eprint = {1905.07488}
}

@article{lueckmann2017flexible,
  title={Flexible statistical inference for mechanistic models of neural dynamics},
  author={Lueckmann, Jan-Matthis and Goncalves, Pedro J and Bassetto, Giacomo and {\"O}cal, Kaan and Nonnenmacher, Marcel and Macke, Jakob H},
  journal={Advances in Neural Information Processing Systems},
  volume={30},
  year={2017},
  eprint = {1711.01861}
}

@article{koenker1978regression,
  title={Regression quantiles},
  author={Koenker, Roger and Bassett Jr, Gilbert},
  journal={Econometrica},
  volume={46},
  number={1},
  pages={33--50},
  year={1978}
}

@article{moon2021learning,
  title={Learning multiple quantiles with neural networks},
  author={Moon, Sang Jun and Jeon, Jong-June and Lee, Jason Sang Hun and Kim, Yongdai},
  journal={Journal of Computational and Graphical Statistics},
  volume={30},
  number={4},
  pages={1238--1248},
  year={2021},
  doi = {10.1080/10618600.2021.1909601}
}

@article{ajelli2018rapidd,
  title = {The {{RAPIDD Ebola}} Forecasting Challenge: {{Model}} Description and Synthetic Data Generation},
  shorttitle = {The {{RAPIDD Ebola}} Forecasting Challenge},
  author = {Ajelli, Marco and Zhang, Qian and Sun, Kaiyuan and Merler, Stefano and Fumanelli, Laura and Chowell, Gerardo and Simonsen, Lone and Viboud, Cecile and Vespignani, Alessandro},
  year = {2018},
  month = mar,
  journal = {Epidemics},
  series = {The {{RAPIDD Ebola Forecasting Challenge}}},
  volume = {22},
  pages = {3--12},
  doi = {10.1016/j.epidem.2017.09.001}
}

@article{akesson2021convolutional,
  title = {Convolutional Neural Networks as Summary Statistics for Approximate Bayesian Computation},
  author = {Akesson, Mattias and Singh, Prashant and Wrede, Fredrik and Hellander, Andreas},
  year = {2021},
  journal = {IEEE/ACM Transactions on Computational Biology and Bioinformatics},
  publisher = {IEEE},
  doi = {10.1109/tcbb.2021.3108695}
}

@article{albert2022learning,
  title = {Learning Summary Statistics for Bayesian Inference with Autoencoders},
  author = {Albert, Carlo and Ulzega, Simone and Ozdemir, Firat and {Perez-Cruz}, Fernando and Mira, Antonietta},
  year = {2022},
  journal = {SciPost Physics Core},
  volume = {5},
  number = {3},
  pages = {043},
  doi = {10.21468/scipostphyscore.5.3.043}
}

@article{bach2024highdimensional,
  title = {High-Dimensional Analysis of Double Descent for Linear Regression with Random Projections},
  author = {Bach, Francis},
  year = {2024},
  journal = {SIAM Journal on Mathematics of Data Science},
  volume = {6},
  number = {1},
  pages = {26--50},
  publisher = {SIAM},
  doi = {10.1137/23m1558781}
}

@article{baker2022analyzing,
  title = {Analyzing Stochastic Computer Models: A Review with Opportunities},
  author = {Baker, Evan and Barbillon, Pierre and Fadikar, Arindam and Gramacy, Robert B and Herbei, Radu and Higdon, David and Huang, Jiangeng and Johnson, Leah R and Ma, Pulong and Mondal, Anirban and others},
  year = {2022},
  journal = {Statistical Science},
  volume = {37},
  number = {1},
  pages = {64--89},
  publisher = {Institute of Mathematical Statistics},
  doi = {10.1214/21-sts822}
}

@article{barron1993universal,
  title = {Universal Approximation Bounds for Superpositions of a Sigmoidal Function},
  author = {Barron, Andrew R},
  year = {1993},
  journal = {IEEE Transactions on Information theory},
  volume = {39},
  number = {3},
  pages = {930--945},
  publisher = {IEEE},
  doi = {10.1109/18.256500}
}

@article{beaumont2002approximate,
  title = {Approximate {{Bayesian}} Computation in Population Genetics},
  author = {Beaumont, Mark A and Zhang, Wenyang and Balding, David J},
  year = {2002},
  journal = {Genetics},
  volume = {162},
  number = {4},
  pages = {2025--2035},
  publisher = {Oxford University Press}
}

@inproceedings{behrmann2021understanding,
  title = {Understanding and {{Mitigating Exploding Inverses}} in {{Invertible Neural Networks}}},
  booktitle = {Proceedings of {{The}} 24th {{International Conference}} on {{Artificial Intelligence}} and {{Statistics}}},
  author = {Behrmann, Jens and Vicol, Paul and Wang, Kuan-Chieh and Grosse, Roger and Jacobsen, Joern-Henrik},
  year = {2021},
  month = mar,
  pages = {1792--1800},
  publisher = {PMLR},
  eprint = {2006.09347}
}

@inproceedings{belkin2019does,
  title = {Does Data Interpolation Contradict Statistical Optimality?},
  booktitle = {Proceedings of the {{Twenty-Second International Conference}} on {{Artificial Intelligence}} and {{Statistics}}},
  author = {Belkin, Mikhail and Rakhlin, Alexander and Tsybakov, Alexandre B.},
  year = {2019},
  month = apr,
  pages = {1611--1619},
  publisher = {PMLR},
  eprint = {1806.09471}
}

@article{bernton2019approximate,
  title = {Approximate {{Bayesian}} Computation with the {{Wasserstein}} Distance},
  author = {Bernton, Espen and Jacob, Pierre E. and Gerber, Mathieu and Robert, Christian P.},
  year = {2019},
  month = apr,
  journal = {Journal of the Royal Statistical Society: Series B},
  volume = {81},
  number = {2},
  pages = {235--269},
  doi = {10.1111/rssb.12312}
}

@article{bhadra2021merging,
  title = {Merging Two Cultures: Deep and Statistical Learning},
  author = {Bhadra, Anindya and Datta, Jyotishka and Polson, Nick and Sokolov, Vadim and Xu, Jianeng},
  year = {2021},
  journal = {arXiv preprint arXiv:2110.11561},
  eprint = {2110.11561},
  archiveprefix = {arXiv},
  doi = {10.1002/wics.1647}
}

@article{bond-taylor2022deep,
  title = {Deep {{Generative Modelling}}: {{A Comparative Review}} of {{VAEs}}, {{GANs}}, {{Normalizing Flows}}, {{Energy-Based}} and {{Autoregressive Models}}},
  shorttitle = {Deep {{Generative Modelling}}},
  author = {{Bond-Taylor}, Sam and Leach, Adam and Long, Yang and Willcocks, Chris G.},
  year = {2022},
  month = nov,
  journal = {IEEE Transactions on Pattern Analysis and Machine Intelligence},
  volume = {44},
  number = {11},
  pages = {7327--7347},
  doi = {10.1109/tpami.2021.3116668}
}

@incollection{brillinger2012generalized,
  title = {A {{Generalized Linear Model With}} ``{{Gaussian}}'' {{Regressor Variables}}},
  booktitle = {Selected {{Works}} of {{David Brillinger}}},
  author = {Brillinger, David R.},
  editor = {Guttorp, Peter and Brillinger, David},
  year = {2012},
  series = {Selected {{Works}} in {{Probability}} and {{Statistics}}},
  pages = {589--606},
  publisher = {Springer},
  address = {New York, NY},
  isbn = {978-1-4614-1344-8},
  doi = {10.1007/978-1-4614-1344-8_34}
}

@article{carlier2016vector,
  title = {Vector Quantile Regression: {{An}} Optimal Transport Approach},
  shorttitle = {Vector Quantile Regression},
  author = {Carlier, Guillaume and Chernozhukov, Victor and Galichon, Alfred},
  year = {2016},
  month = jun,
  journal = {The Annals of Statistics},
  volume = {44},
  number = {3},
  pages = {1165--1192},
  publisher = {Institute of Mathematical Statistics}
}

@article{coppejans2004kolmogorovs,
  title = {On {{Kolmogorov}}'s Representation of Functions of Several Variables by Functions of One Variable},
  author = {Coppejans, Mark},
  year = {2004},
  month = nov,
  journal = {Journal of Econometrics},
  volume = {123},
  number = {1},
  pages = {1--31},
  doi = {10.1016/j.jeconom.2003.10.026}
}

@misc{dabney2017distributional,
  title = {Distributional {{Reinforcement Learning}} with {{Quantile Regression}}},
  author = {Dabney, Will and Rowland, Mark and Bellemare, Marc G. and Munos, R{\'e}mi},
  year = {2017},
  month = oct,
  number = {arXiv:1710.10044},
  eprint = {1710.10044},
  primaryclass = {cs, stat},
  institution = {arXiv},
  archiveprefix = {arXiv},
  doi = {10.7551/mitpress/14207.001.0001}
}

@misc{dabney2018implicit,
  title = {Implicit {{Quantile Networks}} for {{Distributional Reinforcement Learning}}},
  author = {Dabney, Will and Ostrovski, Georg and Silver, David and Munos, R{\'e}mi},
  year = {2018},
  month = jun,
  number = {arXiv:1806.06923},
  eprint = {1806.06923},
  primaryclass = {cs, stat},
  institution = {arXiv},
  archiveprefix = {arXiv}
}

@article{diggle1984monte,
  title = {Monte {{Carlo Methods}} of {{Inference}} for {{Implicit Statistical Models}}},
  author = {Diggle, Peter J. and Gratton, Richard J.},
  year = {1984},
  journal = {Journal of the Royal Statistical Society. Series B (Methodological)},
  volume = {46},
  number = {2},
  eprint = {2345504},
  eprinttype = {jstor},
  pages = {193--227},
  publisher = {[Royal Statistical Society, Wiley]}
}

@misc{dinh2015nice,
  title = {{{NICE}}: {{Non-linear Independent Components Estimation}}},
  shorttitle = {{{NICE}}},
  author = {Dinh, Laurent and Krueger, David and Bengio, Yoshua},
  year = {2015},
  month = apr,
  number = {arXiv:1410.8516},
  eprint = {1410.8516},
  primaryclass = {cs},
  publisher = {arXiv},
  archiveprefix = {arXiv}
}

@misc{dinh2017density,
  title = {Density Estimation Using {{Real NVP}}},
  author = {Dinh, Laurent and {Sohl-Dickstein}, Jascha and Bengio, Samy},
  year = {2017},
  month = feb,
  number = {arXiv:1605.08803},
  eprint = {1605.08803},
  primaryclass = {cs},
  publisher = {arXiv},
  archiveprefix = {arXiv}
}

@article{drovandi2011approximate,
  title = {Approximate {{Bayesian}} Computation Using Indirect Inference},
  author = {Drovandi, Christopher C. and Pettitt, Anthony N. and Faddy, Malcolm J.},
  year = {2011},
  journal = {Journal of the Royal Statistical Society: Series C (Applied Statistics)},
  volume = {60},
  number = {3},
  pages = {317--337}
}

@article{drovandi2015bayesian,
  title = {Bayesian {{Indirect Inference Using}} a {{Parametric Auxiliary Model}}},
  author = {Drovandi, Christopher C. and Pettitt, Anthony N. and Lee, Anthony},
  year = {2015},
  month = feb,
  journal = {Statistical Science},
  volume = {30},
  number = {1},
  pages = {72--95},
  publisher = {Institute of Mathematical Statistics},
  doi = {10.1214/14-sts498}
}

@article{fadikar2018calibrating,
  title = {Calibrating a {{Stochastic}}, {{Agent-Based Model Using Quantile-Based Emulation}}},
  author = {Fadikar, {\relax Arindam}. and Higdon, {\relax Dave}. and Chen, {\relax Jiangzhuo}. and Lewis, {\relax Bryan}. and Venkatramanan, {\relax Srinivasan}. and Marathe, {\relax Madhav}.},
  year = {2018},
  month = jan,
  journal = {SIAM/ASA Journal on Uncertainty Quantification},
  volume = {6},
  number = {4},
  pages = {1685--1706},
  doi = {10.1137/17m1161233}
}

@article{hahn2020bayesian,
  title = {Bayesian {{Regression Tree Models}} for {{Causal Inference}}: {{Regularization}}, {{Confounding}}, and {{Heterogeneous Effects}} (with {{Discussion}})},
  shorttitle = {Bayesian {{Regression Tree Models}} for {{Causal Inference}}},
  author = {Hahn, P. Richard and Murray, Jared S. and Carvalho, Carlos M.},
  year = {2020},
  month = sep,
  journal = {Bayesian Analysis},
  volume = {15},
  number = {3},
  pages = {965--1056},
  publisher = {International Society for Bayesian Analysis}
}

@article{igelnik2003kolmogorovs,
  title = {Kolmogorov's Spline Network},
  author = {Igelnik, B. and Parikh, N.},
  year = {2003},
  month = jul,
  journal = {IEEE Transactions on Neural Networks},
  volume = {14},
  number = {4},
  pages = {725--733},
  doi = {10.1109/tnn.2003.813830}
}

@article{jiang2017learning,
  title = {Learning {{Summary Statistic For Approximate Bayesian Computation Via Deep Neural Network}}},
  author = {Jiang, Bai and Wu, Tung-Yu and Zheng, Charles and Wong, Wing H.},
  year = {2017},
  journal = {Statistica Sinica},
  volume = {27},
  number = {4},
  eprint = {26384090},
  eprinttype = {jstor},
  pages = {1595--1618},
  publisher = {Institute of Statistical Science, Academia Sinica},
  doi = {10.5705/ss.202015.0340}
}

@inproceedings{jiang2018approximate,
  title = {Approximate {{Bayesian Computation}} with {{Kullback-Leibler Divergence}} as {{Data Discrepancy}}},
  booktitle = {Proceedings of the {{Twenty-First International Conference}} on {{Artificial Intelligence}} and {{Statistics}}},
  author = {Jiang, Bai and {Wu, Tung-Yu} and {Wing Hung Wong}},
  year = {2018},
  month = mar,
  pages = {1711--1721},
  publisher = {PMLR},
  eprint = {1510.02175v3}
}

@book{kallenberg1997foundations,
  title = {Foundations of {{Modern Probability}}},
  author = {Kallenberg, Olav},
  year = {1997},
  month = jan,
  edition = {2nd ed. edition},
  publisher = {Springer},
  isbn = {978-0-387-94957-4},
  doi = {10.1007/978-3-030-61871-1}
}

@misc{kingma2022autoencoding,
  title = {Auto-{{Encoding Variational Bayes}}},
  author = {Kingma, Diederik P. and Welling, Max},
  year = {2022},
  month = dec,
  number = {arXiv:1312.6114},
  eprint = {1312.6114},
  primaryclass = {cs, stat},
  publisher = {arXiv},
  archiveprefix = {arXiv}
}

@article{kolmogorov1942definition,
  title = {Definition of Center of Dispersion and Measure of Accuracy from a Finite Number of Observations (in {{Russian}})},
  author = {Kolmogorov, {\relax AN}},
  year = {1942},
  journal = {Izv. Akad. Nauk SSSR Ser. Mat.},
  volume = {6},
  pages = {3--32}
}

@article{kruse2021hint,
  title = {{{HINT}}: {{Hierarchical Invertible Neural Transport}} for {{Density Estimation}} and {{Bayesian Inference}}},
  shorttitle = {{{HINT}}},
  author = {Kruse, Jakob and Detommaso, Gianluca and K{\"o}the, Ullrich and Scheichl, Robert},
  year = {2021},
  month = may,
  journal = {Proceedings of the AAAI Conference on Artificial Intelligence},
  volume = {35},
  number = {9},
  pages = {8191--8199},
  eprint = {1905.10687v4}
}

@misc{mackay1999maximum,
  title = {Maximum Likelihood and Covariant Algorithms for Independent Component Analysis},
  author = {MacKay, David JC},
  year = {1999},
  publisher = {Citeseer},
  archiveprefix = {Citeseer},
  url = {https://www.inference.org.uk/mackay/ica.pdf}
}

@misc{montanelli2020error,
  title = {Error Bounds for Deep {{ReLU}} Networks Using the {{Kolmogorov--Arnold}} Superposition Theorem},
  author = {Montanelli, Hadrien and Yang, Haizhao},
  year = {2020},
  month = may,
  number = {arXiv:1906.11945},
  eprint = {1906.11945},
  primaryclass = {cs, math},
  publisher = {arXiv},
  archiveprefix = {arXiv}
}

@article{muller2019neural,
  title = {Neural {{Importance Sampling}}},
  author = {M{\"u}ller, Thomas and Mcwilliams, Brian and Rousselle, Fabrice and Gross, Markus and Nov{\'a}k, Jan},
  year = {2019},
  month = oct,
  journal = {ACM Trans. Graph.},
  volume = {38},
  number = {5},
  pages = {145:1--145:19},
  doi = {10.1145/3341156}
}

@article{nunes2010optimal,
  title = {On {{Optimal Selection}} of {{Summary Statistics}} for {{Approximate Bayesian Computation}}},
  author = {Nunes, Matthew A. and Balding, David J.},
  year = {2010},
  month = sep,
  journal = {Statistical Applications in Genetics and Molecular Biology},
  volume = {9},
  number = {1},
  publisher = {De Gruyter},
  doi = {10.2202/1544-6115.1576}
}

@misc{ostrovski2018autoregressive,
  title = {Autoregressive {{Quantile Networks}} for {{Generative Modeling}}},
  author = {Ostrovski, Georg and Dabney, Will and Munos, R{\'e}mi},
  year = {2018},
  month = jun,
  number = {arXiv:1806.05575},
  eprint = {1806.05575},
  primaryclass = {cs, stat},
  institution = {arXiv},
  archiveprefix = {arXiv}
}

@article{padilla2022quantile,
  title = {Quantile Regression with {{ReLU}} Networks: Estimators and Minimax Rates},
  shorttitle = {Quantile Regression with {{ReLU}} Networks},
  author = {Padilla, Oscar Hernan Madrid and Tansey, Wesley and Chen, Yanzhen},
  year = {2022},
  month = jan,
  journal = {The Journal of Machine Learning Research},
  volume = {23},
  number = {1},
  pages = {247:11251--247:11292},
  eprint = {2010.08236}
}

@inproceedings{papamakarios2016fast,
  title = {Fast {\textbackslash}epsilon -Free {{Inference}} of {{Simulation Models}} with {{Bayesian Conditional Density Estimation}}},
  booktitle = {Advances in {{Neural Information Processing Systems}}},
  author = {Papamakarios, George and Murray, Iain},
  year = {2016},
  volume = {29},
  publisher = {Curran Associates, Inc.}
}

@inproceedings{papamakarios2019sequential,
  title = {Sequential Neural Likelihood: {{Fast}} Likelihood-Free Inference with Autoregressive Flows},
  booktitle = {The 22nd International Conference on Artificial Intelligence and Statistics},
  author = {Papamakarios, George and Sterratt, David and Murray, Iain},
  year = {2019},
  pages = {837--848},
  organization = {PMLR},
  eprint = {1805.07226}
}

@inproceedings{park2016k2abc,
  title = {K2-{{ABC}}: {{Approximate Bayesian Computation}} with {{Kernel Embeddings}}},
  shorttitle = {K2-{{ABC}}},
  booktitle = {Proceedings of the 19th {{International Conference}} on {{Artificial Intelligence}} and {{Statistics}}},
  author = {Park, Mijung and Jitkrittum, Wittawat and Sejdinovic, Dino},
  year = {2016},
  month = may,
  pages = {398--407},
  publisher = {PMLR},
  eprint = {1502.02558}
}

@article{parzen2004quantile,
  title = {Quantile {{Probability}} and {{Statistical Data Modeling}}},
  author = {Parzen, Emanuel},
  year = {2004},
  journal = {Statistical Science},
  volume = {19},
  number = {4},
  eprint = {4144436},
  eprinttype = {jstor},
  pages = {652--662},
  publisher = {Institute of Mathematical Statistics}
}

@article{pastorello2003iterative,
  title = {Iterative and Recursive Estimation in Structural Nonadaptive Models},
  author = {Pastorello, Sergio and Patilea, Valentin and Renault, Eric},
  year = {2003},
  journal = {Journal of Business \& Economic Statistics},
  volume = {21},
  number = {4},
  pages = {449--509},
  publisher = {Taylor \& Francis},
  doi = {10.1198/073500103288619124}
}

@inproceedings{polson2018posterior,
  title = {Posterior {{Concentration}} for {{Sparse Deep Learning}}},
  booktitle = {Advances in {{Neural Information Processing Systems}}},
  author = {Polson, Nicholas G and Ro{\v c}kov{\'a}, Veronika},
  year = {2018},
  volume = {31},
  publisher = {Curran Associates, Inc.},
  eprint = {1803.09138}
}

@article{polson2021deep,
  title = {Deep {{Learning Partial Least Squares}}},
  author = {Polson, Nicholas and Sokolov, Vadim and Xu, Jianeng},
  year = {2021},
  journal = {arXiv preprint arXiv:2106.14085},
  eprint = {2106.14085},
  archiveprefix = {arXiv}
}

@inproceedings{rezende2015variational,
  title = {Variational {{Inference}} with {{Normalizing Flows}}},
  booktitle = {Proceedings of the 32nd {{International Conference}} on {{Machine Learning}}},
  author = {Rezende, Danilo and Mohamed, Shakir},
  year = {2015},
  month = jun,
  pages = {1530--1538},
  publisher = {PMLR}
}

@article{schmidt-hieber2020nonparametric,
  title = {Nonparametric Regression Using Deep Neural Networks with {{ReLU}} Activation Function},
  author = {{Schmidt-Hieber}, Johannes},
  year = {2020},
  month = aug,
  journal = {The Annals of Statistics},
  volume = {48},
  number = {4},
  pages = {1875--1897},
  publisher = {Institute of Mathematical Statistics}
}

@article{schultz2022bayesian,
  title = {Bayesian {{Calibration}} for {{Activity Based Models}}},
  author = {Schultz, Laura and Auld, Joshua and Sokolov, Vadim},
  year = {2022},
  journal = {arXiv preprint arXiv:2203.04414},
  eprint = {2203.04414},
  archiveprefix = {arXiv}
}

@misc{shen2021deep,
  title = {Deep {{Quantile Regression}}: {{Mitigating}} the {{Curse}} of {{Dimensionality Through Composition}}},
  shorttitle = {Deep {{Quantile Regression}}},
  author = {Shen, Guohao and Jiao, Yuling and Lin, Yuanyuan and Horowitz, Joel L. and Huang, Jian},
  year = {2021},
  month = jul,
  eprint = {2107.04907}
}

@misc{sohl-dickstein2015deep,
  title = {Deep {{Unsupervised Learning}} Using {{Nonequilibrium Thermodynamics}}},
  author = {{Sohl-Dickstein}, Jascha and Weiss, Eric A. and Maheswaranathan, Niru and Ganguli, Surya},
  year = {2015},
  month = nov,
  number = {arXiv:1503.03585},
  eprint = {1503.03585},
  primaryclass = {cond-mat, q-bio, stat},
  publisher = {arXiv},
  archiveprefix = {arXiv}
}

@article{song2019mintnet,
  title = {{{MintNet}}: {{Building Invertible Neural Networks}} with {{Masked Convolutions}}},
  shorttitle = {{{MintNet}}},
  author = {Song, Yang and Meng, Chenlin and Ermon, Stefano},
  year = {2019},
  month = oct,
  journal = {arXiv:1907.07945 [cs, stat]},
  eprint = {1907.07945},
  primaryclass = {cs, stat},
  archiveprefix = {arXiv}
}

@article{trippe2018conditional,
  title = {Conditional {{Density Estimation}} with {{Bayesian Normalising Flows}}},
  author = {Trippe, Brian L. and Turner, Richard E.},
  year = {2018},
  month = feb,
  journal = {arXiv:1802.04908 [stat]},
  eprint = {1802.04908},
  primaryclass = {stat},
  archiveprefix = {arXiv}
}

@article{marin2012approximate,
  title = {Approximate {{Bayesian}} Computational Methods},
  author = {Marin, Jean-Michel and Pudlo, Pierre and Robert, Christian P. and Ryder, Robin J.},
  year = {2012},
  journal = {Statistics and Computing},
  volume = {22},
  number = {6},
  pages = {1167--1180},
  publisher = {Springer},
  doi = {10.1007/s11222-011-9288-2}
}

@article{brenier1991polar,
  title = {Polar Factorization and Monotone Rearrangement of Vector-Valued Functions},
  author = {Brenier, Yann},
  year = {1991},
  journal = {Communications on Pure and Applied Mathematics},
  volume = {44},
  number = {4},
  pages = {375--417},
  doi = {10.1002/cpa.3160440402}
}

@inproceedings{yoon2018ganite,
  title={{GANITE}: Estimation of individualized treatment effects using generative adversarial nets},
  author={Yoon, Jinsung and Jordon, James and van der Schaar, Mihaela},
  booktitle={International Conference on Learning Representations},
  year={2018}
}

@article{chernozhukov2018double,
  title={Double/debiased machine learning for treatment and structural parameters},
  author={Chernozhukov, Victor and Chetverikov, Denis and Demirer, Mert and Duflo, Esther and Hansen, Christian and Newey, Whitney and Robins, James},
  journal={The Econometrics Journal},
  volume={21},
  number={1},
  pages={C1--C68},
  year={2018},
  doi = {10.3386/w23564}
}

@article{chernozhukov2005iv,
  title={An {IV} model of quantile treatment effects},
  author={Chernozhukov, Victor and Hansen, Christian},
  journal={Econometrica},
  volume={73},
  number={1},
  pages={245--261},
  year={2005},
  doi = {10.1111/j.1468-0262.2005.00570.x}
}

@inproceedings{hermans2020likelihood,
  title={Likelihood-free {MCMC} with amortized approximate ratio estimators},
  author={Hermans, Joeri and Begy, Volodimir and Louppe, Gilles},
  booktitle={International Conference on Machine Learning},
  pages={4239--4248},
  year={2020}
}

@inproceedings{durkan2020contrastive,
  title={On contrastive learning for likelihood-free inference},
  author={Durkan, Conor and Murray, Iain and Papamakarios, George},
  booktitle={International Conference on Machine Learning},
  pages={2771--2781},
  year={2020},
  eprint = {2002.03712}
}

@article{tejerocantero2020sbi,
  title={sbi: A toolkit for simulation-based inference},
  author={Tejero-Cantero, Alvaro and Boelts, Jan and Deistler, Michael and Lueckmann, Jan-Matthis and Durkan, Conor and Gon{\c{c}}alves, Pedro J and Greenberg, David S and Macke, Jakob H},
  journal={Journal of Open Source Software},
  volume={5},
  number={52},
  pages={2505},
  year={2020},
  doi = {10.21105/joss.02505}
}

@inproceedings{lueckmann2021benchmarking,
  title={Benchmarking simulation-based inference},
  author={Lueckmann, Jan-Matthis and Boelts, Jan and Greenberg, David and Goncalves, Pedro and Macke, Jakob},
  booktitle={Proceedings of the 24th International Conference on Artificial Intelligence and Statistics},
  pages={343--351},
  year={2021},
  eprint = {2101.04653}
}

@inproceedings{romano2019conformalized,
  title={Conformalized quantile regression},
  author={Romano, Yaniv and Patterson, Evan and Cand{\`e}s, Emmanuel},
  booktitle={Advances in Neural Information Processing Systems},
  volume={32},
  year={2019}
}

@book{vovk2005algorithmic,
  title={Algorithmic Learning in a Random World},
  author={Vovk, Vladimir and Gammerman, Alexander and Shafer, Glenn},
  year={2005},
  publisher={Springer},
  doi = {10.1007/978-3-031-06649-8}
}

@article{lei2018distribution,
  title={Distribution-free predictive inference for regression},
  author={Lei, Jing and G'Sell, Max and Rinaldo, Alessandro and Tibshirani, Ryan J and Wasserman, Larry},
  journal={Journal of the American Statistical Association},
  volume={113},
  number={523},
  pages={1094--1111},
  year={2018},
  doi = {10.1080/01621459.2017.1307116}
}

@article{shen2024engression,
  title={Engression: Extrapolation through the lens of distributional regression},
  author={Shen, Xinwei and Meinshausen, Nicolai},
  journal={Journal of the Royal Statistical Society: Series B},
  year={2024},
  doi = {10.1093/jrsssb/qkae108}
}

@article{talts2018validating,
  title={Validating {B}ayesian inference algorithms with simulation-based calibration},
  author={Talts, Sean and Betancourt, Michael and Simpson, Daniel and Vehtari, Aki and Gelman, Andrew},
  journal={arXiv preprint arXiv:1804.06788},
  year={2018},
  eprint = {1804.06788}
}

@article{xie2013confidence,
  title={Confidence distribution, the frequentist distribution estimator of a parameter: A review},
  author={Xie, Min-ge and Singh, Kesar},
  journal={International Statistical Review},
  volume={81},
  number={1},
  pages={3--39},
  year={2013},
  doi = {10.1111/insr.12000}
}

@article{kennedy2001bayesian,
  title={Bayesian calibration of computer models},
  author={Kennedy, Marc C and O'Hagan, Anthony},
  journal={Journal of the Royal Statistical Society: Series B},
  volume={63},
  number={3},
  pages={425--464},
  year={2001},
  doi = {10.1111/1467-9868.00294}
}

@inproceedings{arjovsky2017wasserstein,
  title={Wasserstein generative adversarial networks},
  author={Arjovsky, Martin and Chintala, Soumith and Bottou, L\'{e}on},
  booktitle={International Conference on Machine Learning},
  pages={214--223},
  year={2017},
  organization={PMLR}
}

@inproceedings{donahue2017adversarial,
  title={Adversarial feature learning},
  author={Donahue, Jeff and Kr\"{a}henb\"{u}hl, Philipp and Darrell, Trevor},
  booktitle={International Conference on Learning Representations (ICLR)},
  year={2017},
  eprint = {1605.09782}
}

@inproceedings{dumoulin2017adversarially,
  title={Adversarially learned inference},
  author={Dumoulin, Vincent and Belghazi, Ishmael and Poole, Ben and Mastropietro, Olivier and Lamb, Alex and Arjovsky, Martin and Courville, Aaron},
  booktitle={International Conference on Learning Representations (ICLR)},
  year={2017},
  eprint = {1606.00704}
}

@inproceedings{larsen2016autoencoding,
  title={Autoencoding beyond pixels using a learned similarity metric},
  author={Larsen, Anders Boesen Lindbo and S\o{}nderby, S\o{}ren Kaae and Larochelle, Hugo and Winther, Ole},
  booktitle={International Conference on Machine Learning},
  pages={1558--1566},
  year={2016},
  organization={PMLR}
}

@article{makhzani2015adversarial,
  title={Adversarial autoencoders},
  author={Makhzani, Alireza and Shlens, Jonathon and Jaitly, Navdeep and Goodfellow, Ian and Frey, Brendan},
  journal={arXiv preprint arXiv:1511.05644},
  year={2015},
  eprint = {1511.05644}
}

@article{szekely2013energy,
  title={Energy statistics: A class of statistics based on distances},
  author={Sz\'{e}kely, G\'{a}bor J and Rizzo, Maria L},
  journal={Journal of Statistical Planning and Inference},
  volume={143},
  number={8},
  pages={1249--1272},
  year={2013},
  publisher={Elsevier},
  doi = {10.1016/j.jspi.2013.03.018}
}

@book{villani2009optimal,
  title={Optimal Transport: Old and New},
  author={Villani, C\'{e}dric},
  volume={338},
  year={2009},
  publisher={Springer},
  doi = {10.1007/978-3-540-71050-9}
}

@article{mccann1995existence,
  title={Existence and uniqueness of monotone measure-preserving maps},
  author={McCann, Robert J},
  journal={Duke Mathematical Journal},
  volume={80},
  number={2},
  pages={309--323},
  year={1995},
  publisher={Duke University Press},
  doi = {10.1215/s0012-7094-95-08013-2}
}

@article{gneiting2007strictly,
  title={Strictly proper scoring rules, prediction, and estimation},
  author={Gneiting, Tilmann and Raftery, Adrian E},
  journal={Journal of the American statistical Association},
  volume={102},
  number={477},
  pages={359--378},
  year={2007},
  publisher={Taylor \& Francis},
  doi = {10.21236/ada454828}
}

@article{goodfellow2020generative,
  title={Generative adversarial networks},
  author={Goodfellow, Ian and Pouget-Abadie, Jean and Mirza, Mehdi and Xu, Bing and Warde-Farley, David and Ozair, Sherjil and Courville, Aaron and Bengio, Yoshua},
  journal={Communications of the ACM},
  volume={63},
  number={11},
  pages={139--144},
  year={2020},
  publisher={ACM New York, NY, USA},
  doi = {10.1145/3422622}
}

@article{stroud2003nonlinear,
  title={Nonlinear state-space models with state-dependent variances},
  author={Stroud, Jonathan R and M{\"u}ller, Peter and Polson, Nicholas G},
  journal={Journal of the American Statistical Association},
  volume={98},
  number={462},
  pages={377--386},
  year={2003},
  publisher={Taylor \& Francis},
  doi = {10.1198/016214503000161}
}

@article{Rezende15,
title={Variational inference with normalizing flows},
author={Rezende, Danilo Jimenez and Mohamed, Shakir},
journal={arXiv preprint arXiv:1505.05770},
year={2015},
  eprint = {1505.05770}
}

@incollection{sisson2018overview,
  title={Overview of ABC},
  author={Sisson, Scott A and Fan, Yanan and Beaumont, Mark A},
  booktitle={Handbook of approximate Bayesian computation},
  pages={3--54},
  year={2018},
  publisher={Chapman and Hall/CRC},
  doi = {10.1201/9781315117195-1}
}

@inproceedings{kingma2014stochastic,
  title={Stochastic gradient VB and the variational auto-encoder},
  author={Kingma, Diederik P and Welling, Max},
  booktitle={Second international conference on learning representations, ICLR},
  volume={19},
  pages={121},
  year={2014}
}

@article{goodfellow2014generative,
  title={Generative adversarial nets},
  author={Goodfellow, Ian J and Pouget-Abadie, Jean and Mirza, Mehdi and Xu, Bing and Warde-Farley, David and Ozair, Sherjil and Courville, Aaron and Bengio, Yoshua},
  journal={Advances in neural information processing systems},
  volume={27},
  year={2014},
  doi = {10.1007/978-3-658-40442-0_9}
}

@article{mirza2014conditional,
  title={Conditional generative adversarial nets},
  author={Mirza, Mehdi and Osindero, Simon},
  journal={arXiv preprint arXiv:1411.1784},
  year={2014},
  eprint = {1411.1784}
}

@article{gelfand2000gibbs,
  title={Gibbs sampling},
  author={Gelfand, Alan E},
  journal={Journal of the American statistical Association},
  volume={95},
  number={452},
  pages={1300--1304},
  year={2000},
  publisher={Taylor \& Francis},
  doi={10.1080/01621459.2000.10474335}
}

@article{smith2007boa,
  title={boa: an R package for MCMC output convergence assessment and posterior inference},
  author={Smith, Brian J},
  journal={Journal of statistical software},
  volume={21},
  pages={1--37},
  year={2007}
}

@article{lecun2015deep,
  title={Deep learning},
  author={LeCun, Yann and Bengio, Yoshua and Hinton, Geoffrey},
  journal={nature},
  volume={521},
  number={7553},
  pages={436--444},
  year={2015},
  publisher={Nature Publishing Group UK London},
  doi = {10.1038/nature14539}
}

@article{kingma2019introduction,
  title={An introduction to variational autoencoders},
  author={Kingma, Diederik P and Welling, Max and others},
  journal={Foundations and Trends{\textregistered} in Machine Learning},
  volume={12},
  number={4},
  pages={307--392},
  year={2019},
  publisher={Now Publishers, Inc.},
  doi = {10.1561/9781680836233}
}

@incollection{efron1992bootstrap,
  title={Bootstrap methods: another look at the jackknife},
  author={Efron, Bradley},
  booktitle={Breakthroughs in statistics: Methodology and distribution},
  pages={569--593},
  year={1992},
  publisher={Springer},
  doi = {10.1214/aos/1176344552}
}

@article{diciccio1996bootstrap,
  title={Bootstrap confidence intervals},
  author={DiCiccio, Thomas J and Efron, Bradley},
  journal={Statistical science},
  volume={11},
  number={3},
  pages={189--228},
  year={1996},
  publisher={Institute of Mathematical Statistics},
  doi = {10.1214/ss/1032280214}
}

@book{efron1982jackknife,
  title={The jackknife, the bootstrap and other resampling plans},
  author={Efron, Bradley},
  year={1982},
  publisher={SIAM},
  doi = {10.1137/1.9781611970319}
}

@book{efron1994introduction,
  title={An introduction to the bootstrap},
  author={Efron, Bradley and Tibshirani, Robert J},
  year={1994},
  publisher={Chapman and Hall/CRC},
  doi = {10.1201/9780429246593}
}

@article{hannig2016generalized,
  title={Generalized fiducial inference: A review and new results},
  author={Hannig, Jan and Iyer, Hari and Lai, Randy CS and Lee, Thomas CM},
  journal={Journal of the American Statistical Association},
  volume={111},
  number={515},
  pages={1346--1361},
  year={2016},
  publisher={Taylor \& Francis},
  doi = {10.1080/01621459.2016.1165102}
}

@techreport{athey2022smiles,
  title={Smiles in profiles: Improving fairness and efficiency using estimates of user preferences in online marketplaces},
  author={Athey, Susan and Karlan, Dean and Palikot, Emil and Yuan, Yuan},
  year={2022},
  institution={National Bureau of Economic Research},
  doi = {10.3386/w30633}
}

\begin{appendix}

\section{Bayes Risk}\label{app_bayes_risk}

The bounds in this appendix apply to the parameter-generation class where the unknown function $f$ is the posterior map and $\theta$ is the target. This appendix uses the standard nonparametric regression convention: $X_i$ denotes the input (corresponding to the data summary $S(Y_i)$ in the main text) and $Y_i$ denotes the response (corresponding to the parameter $\theta_i$ in the main text); these differ from the convention of the main text where $Y_i$ represents observed data and $\theta_i$ is the parameter. Consider the nonparametric regression model \( Y_i = f(X_i) + \varepsilon_i \), where \( X_i = (x_{1i}, \ldots, x_{di}) \in [0,1]^d \), and the goal is to estimate the unknown multivariate function \( f : [0,1]^d \to \mathbb{R} \). A natural measure of estimation accuracy is the integrated squared error, defined as
\[
R(f, \hat{f}_N) = \mathbb{E}_{X,Y} \left[ \lVert f - \hat{f}_N \rVert^2 \right],
\]
where \( \lVert \cdot \rVert \) denotes the \( L^2(P_X) \)-norm and \( \hat{f}_N \) is an estimator based on a sample of size \( N \).

A common strategy is to construct \( \hat{f}_N \) as a regularized solution to the empirical risk minimization problem:
\[
\hat{f}_N = \arg\min_{f \in \mathcal{F}} \left\{ \frac{1}{N} \sum_{i=1}^N (Y_i - f(X_i))^2 + \lambda \, \phi(f) \right\},
\]
where \( \lambda > 0 \) is a regularization parameter and \( \phi(f) \) is a penalty functional controlling model complexity. Such estimators often correspond to the mode of a posterior distribution under a Bayesian prior, resulting in a regularized maximum a posteriori estimate. Under appropriate smoothness and complexity conditions, these estimators have a posterior concentration: the risk \( R(f, \hat{f}_N) \) converges to zero at a rate determined by the regularity of \( f \) and the structure of the function class \( \mathcal{F} \).

The posterior distribution \( p(f \mid X_i, Y_i) \) can concentrate around the true function at near-optimal rates (up to a $\log N$ factor). For instance, \cite{barron1993universal} shows that Fourier-based models can achieve convergence rates of order \( O(N^{-1/2}) \) in mean squared error. More generally, for functions \( f \) belonging to a \(\beta\)-H\"{o}lder class, the minimax risk over all estimators satisfies
\[
\inf_{\hat{f}} \sup_{f} R(f, \hat{f}_N) = O_p\left(N^{-2\beta / (2\beta + d)}\right),
\]
where \( d \) is the input dimension and \( \hat{f}_N \) denotes any estimator based on \( N \) observations. These rates reflect the curse of dimensionality: estimation becomes harder as \( d \) increases. By restricting the class of functions, better rates can be obtained; when the function class has low-dimensional compositional structure, the effective dimension $d$ in the rate can be much smaller than the ambient dimension, mitigating the curse of dimensionality. For example, it is common to consider the class of linear superpositions, also called ridge functions, or projection pursuit.

Consider the nonparametric regression model \( Y_i = f(X_i) + \varepsilon_i \), where \( f \) is an unknown function to be estimated from data \( (X_i, Y_i)_{i=1}^N \). Suppose that \( f \) is a composite function,
\[
f = g_L \circ g_{L-1} \circ \cdots \circ g_0,
\]
where each component \( g_\ell \) is a H\"{o}lder-smooth function of \( d_\ell \) variables with smoothness index \( \beta_\ell \). That is, for all \( x, y \in \mathbb{R}^{d_\ell} \),
\[
|g_\ell(x) - g_\ell(y)| \leq C \|x - y\|^{\beta_\ell}.
\]
Under this assumption, recent results show that deep neural network estimators can achieve the convergence rate
\[
O\left( \max_{1 \leq \ell \leq L} N^{-2 \beta_\ell^* / (2 \beta_\ell^* + d_\ell)} \right), \quad \text{where} \quad \beta_\ell^* = \beta_\ell \prod_{k=\ell+1}^{L} \min(\beta_k, 1).
\]
This bound shows how the local smoothness and dimensionality of each layer interact to determine the overall estimation rate. As an example, consider a generalized additive model of the form
\[
f_0(x) = h\left( \sum_{p=1}^d f_{0p}(x_p) \right),
\]
where \( h \) is a Lipschitz function and each \( f_{0p} \) is a univariate smooth function. This structure can be viewed as a composition of three low-dimensional functions: $g_0(z) = h(z)$ with $z \in \mathbb{R}$; $g_1(x_1, \ldots, x_d) = (f_{01}(x_1), \ldots, f_{0d}(x_d))$; and $g_2(y_1, \ldots, y_d) = \sum_{p=1}^d y_p$.
Since each component operates over one-dimensional or additive functions and \( h \) is Lipschitz, the resulting estimator achieves the rate \( O(N^{-1/3}) \), which is independent of the dimension \( d \).
\cite{schmidt-hieber2020nonparametric} show that deep rectified-linear-unit networks achieve the optimal rate \( O(N^{-1/3}) \) under similar assumptions. \cite{coppejans2004kolmogorovs} finds a convergence rate of \( O(N^{-3/7}) \), equivalently \( O(N^{-3/(2\beta + d)}) \) with $\beta=3$ and $d=1$, for three-times differentiable cubic B-splines. \cite{igelnik2003kolmogorovs} derives a parametric rate \( O(N^{-1}) \) for Kolmogorov spline networks.

\section{Bayes Rule for Quantiles}\label{app_bayes_rule}
The compositional quantile identity developed in this appendix underlies the generative Bayesian computation framework for parameter generation (Section~\ref{sec:gbc}) and the forward-emulator quantile network for predictive inference (Section~\ref{sec:ebola}); the same  left-continuous generalized inverse \citep{parzen2004quantile} appears as $G(q, \tilde{s}, a)$ in the counterfactual generator for outcome generation in Section~\ref{sec:tasks_cat1}. \cite{parzen2004quantile} shows that quantile methods are direct alternatives to density estimation. Following the convention of this work, we use $u$ for the quantile level in this appendix (corresponding to $\tau$ in the main text). Given the conditional cumulative distribution function $F_{\theta \mid y}$, assumed to be continuous and non-decreasing, the associated quantile function is defined as
\[
Q_{\theta \mid y}(u) \coloneqq F^{-1}_{\theta \mid y}(u) = \inf\left\{ \theta : F_{\theta \mid y}(\theta) \geq u \right\},
\]
which is left-continuous and non-decreasing. A fundamental property established by \citet{parzen2004quantile} is the probability integral transform
\[
\theta \overset{d}{=} Q_{\theta \mid y}(U), \qquad U \sim \mathrm{Unif}(0,1),
\]
which shows that the quantile function generates draws from the target distribution by transforming uniform random variables: monotonicity of the quantile map aligns the order statistics of the parameter with those of the baseline distribution, so sampling reduces to evaluating $Q_{\theta \mid y}$ at uniform draws. Consider now a non-decreasing, left-continuous function \( g(y) \), with generalized inverse defined by
\[
g^{-1}(z) = \sup \{ y : g(y) \leq z \}.
\]
Then the quantile of a transformed variable satisfies the composition identity
\[
Q_{g(Y)}(u) = g(Q_Y(u)),
\]
illustrating that quantile operations are compositional.
\begin{proof}
Let \( g \) be strictly increasing and continuous. Then:
\[
F_{g(Y)}(z) = p(g(Y) \leq z) = p(Y \leq g^{-1}(z)) = F_Y(g^{-1}(z)).
\]
The quantile function of \( g(Y) \) is:
\[
Q_{g(Y)}(u) = \inf \left\{ z : F_{g(Y)}(z) \geq u \right\}
= \inf \left\{ z : F_Y(g^{-1}(z)) \geq u \right\}.
\]
Substitute \( z = g(y) \), which implies \( y = g^{-1}(z) \):
\[
Q_{g(Y)}(u) = \inf \left\{ g(y) : F_Y(y) \geq u \right\}
= g\left( \inf \left\{ y : F_Y(y) \geq u \right\} \right)
= g(Q_Y(u)).
\]
\end{proof}

This property supports the interpretation of nested quantile transformations as layered representations, akin to deep learning architectures. In Bayesian models, this compositional structure underlies the conditional quantile representation of the posterior:
\[
Q_{\theta \mid Y = y}(u) = Q_{\theta}(s), \quad \text{where} \quad s = Q_{F(\theta) \mid Y = y}(u).
\]
To determine \( s \), observe that
\[
u = p(F(\theta) \leq s \mid Y = y) = p(\theta \leq Q_{\theta}(s) \mid Y = y) = F_{\theta \mid Y = y}(Q_{\theta}(s)),
\]
which confirms that \( s = F^{-1}_{F(\theta) \mid Y = y}(u) \). This recursive identity provides a quantile-based method for posterior updating, bypassing the need for explicit density evaluation.

\section{Comparative Analysis}\label{sec:arch_cross}

The three generator classes share a common representation theorem (noise outsourcing theorem of \cite{kallenberg1997foundations} applied to different variable pairs), but condition on different objects, rely on different identification assumptions, and answer different statistical questions. Counterfactual outcome generators condition on covariates and treatment states and ask what outcome distribution would arise under a specified intervention. Parameter--outcome generators condition on observed outcomes and ask what parameter values are plausible under a forward model. Predictive generators condition on observed covariates, design points, or simulator inputs and ask what future outcomes are plausible. Methods can move across classes only when their loss function, conditioning variables, and output interpretation are changed accordingly. For example, a conditional flow can be a posterior estimator in the second class or a predictive density estimator in the third; a diffusion model can be an outcome generator in the first class or a conditional posterior sampler in the second; a quantile network can estimate counterfactual outcome quantiles, posterior quantiles, or predictive quantiles depending on whether it is trained on \((\tilde{S}(X),A,Y)\), \((Y,\theta)\), or \((X,Y)\), respectively. Table~\ref{tab:comparison} records the respective methods, training object, main output, and main limitations for each class.

\begin{table}[H]
\centering
\scriptsize
\setlength{\tabcolsep}{2pt}
\begin{tabular}{@{}>{\raggedright\arraybackslash}p{0.17\linewidth}>{\raggedright\arraybackslash}p{0.19\linewidth}>{\raggedright\arraybackslash}p{0.22\linewidth}>{\raggedright\arraybackslash}p{0.19\linewidth}>{\raggedright\arraybackslash}p{0.17\linewidth}@{}}
\hline
Generator class & Training object & Methods & Main statistical output & Main limitation \\
\hline
Counterfactual outcome distributions & Observed or simulated triples $(\tilde{S}(X),A,Y)$; more generally $(c,Y)$ & Conditional adversarial generators, diffusion models, conditional variational autoencoders, conditional flows, quantile generators & Conditional outcome distribution, potential-outcome distribution, treatment-effect functionals & Identification depends on causal assumptions; calibration and support overlap must be checked \\
Parameter--outcome generators & Simulated pairs $(\theta,Y)$ from a prior and forward simulator & Approximate Bayesian computation, fiducial inference, conditional flows, conditional diffusion models, generative Bayesian computation & Posterior draws, posterior quantiles, or posterior density approximation & Summary choice, invertibility, sampling cost, or absence of closed-form density depending on method \\
Predictive generators & Input--output pairs $(X,Y)$ from data or simulator runs & Gaussian process emulators, quantile networks, conditional flows, conditional variational autoencoders, conditional adversarial and diffusion generators & Predictive distribution, prediction intervals, conditional quantiles & Covariate shift, conditional calibration, and computational scaling \\
\hline
\end{tabular}
\caption{Three generator classes used throughout the paper. The same architecture may appear in more than one row, but its statistical meaning changes with the training object and target distribution.}
\label{tab:comparison}
\end{table}

Table~\ref{tab:method_class} additionally summarizes class membership for the eight architectures reviewed above. A primary assignment indicates that the architecture is naturally suited to that class; a secondary assignment indicates that the architecture can be adapted to that class with a change of training object, loss function, or conditioning structure; the absence of an entry indicates that the architecture does not naturally serve that class.

\begin{table}[H]
\centering
\footnotesize
\setlength{\tabcolsep}{4pt}
\begin{tabular}{lccc}
\hline
Architecture & Counterfactual outcome & Parameter--outcome & Predictive \\
\hline
Approximate Bayesian computation & -- & Primary & -- \\
Fiducial inference & -- & Primary & -- \\
Variational autoencoder & Secondary & Secondary & Secondary \\
Normalizing flow & Secondary & Primary & Secondary \\
Generative adversarial network & Primary & -- & Secondary \\
Diffusion model & Primary & Secondary & Secondary \\
Gaussian process emulator & -- & Secondary & Primary \\
Generative Bayesian computation & Secondary & Primary & Secondary \\
\hline
\end{tabular}
\caption{Membership of architectures across the three generator classes. Primary entries indicate that the architecture is naturally suited to the class; secondary entries indicate that the architecture serves the class after a change of training object, loss function, or conditioning structure. Empty entries indicate that the architecture does not naturally serve the class.}
\label{tab:method_class}
\end{table}

A third comparison records the computational properties of the individual architectures. Table~\ref{tab:arch_properties} indexes each by primary class and by whether it is likelihood-free, provides exact density evaluation, amortizes inference across observations, handles multimodality, generates samples, requires an invertible architecture, or refines sequentially, together with approximate training and sampling costs. The central trade-off is between exact density evaluation (normalizing flows) and sample-only output (approximate Bayesian computation, generative adversarial networks, generative Bayesian computation), and between amortized estimators that evaluate at any new observation in one pass (variational autoencoders, normalizing flows, generative adversarial networks, generative Bayesian computation) and methods that iterate per observation (approximate Bayesian computation, diffusion models).

\begin{table}[H]
\centering
\small
\begin{tabular}{lccccccc}
\hline
 & ABC & VAE & NF/ICA & GAN & Diffusion & Fiducial & GBC \\
\hline
Primary class & 2 & 1, 2, 3 & 1, 2, 3 & 1, 3 & 1, 2, 3 & 2 & 2 \\
\hline
Likelihood-free & \checkmark & & \checkmark$^\dagger$ & \checkmark & \checkmark & & \checkmark \\
Exact density & & & \checkmark & & & & \\
Amortized inference & & \checkmark & \checkmark & \checkmark & & & \checkmark \\
Handles multimodality & \checkmark & & $\sim$$^\ddagger$ & \checkmark & \checkmark & \checkmark & \checkmark \\
Generates samples & \checkmark & \checkmark & \checkmark & \checkmark & \checkmark & \checkmark & \checkmark \\
Invertible architecture & & & Required & & & & \\
Sequential refinement & \checkmark & & & & \checkmark & & \\
\hline
Training cost & Low & Medium & Medium & High & High & Low & Medium \\
Sampling cost & High & Low & Low & Low & High & High & Low \\
\hline
\end{tabular}
\caption{Computational properties of the generative architectures, indexed by primary class: Cat~1 (counterfactual outcome), Cat~2 (parameter--outcome), Cat~3 (predictive). See Section~\ref{sec:architectures} for a detailed summary of the algorithms. $^\dagger$Conditional normalizing flows can be trained on simulated pairs without likelihood evaluation, though the flow itself provides exact density computation. $^\ddagger$Normalizing flows can represent mildly multimodal distributions but may struggle with well-separated modes because of the continuity of the bijective map.}
\label{tab:arch_properties}
\end{table}

The pattern of primary and secondary memberships in Table~\ref{tab:method_class} reflects three architectural choices: invertibility, the form of the conditioning input, and the form of the output. Approximate Bayesian computation and fiducial inference are built on inverting a forward data-generating equation; they are tightly coupled to the parameter--outcome class because the simulated table $\{(\theta_i, Y_i)\}_{i=1}^N$ is their natural training object. They have no direct counterfactual or forward-predictive analogue. The two methods differ in interpretation rather than mechanism: approximate Bayesian computation treats the resulting distribution as the Bayesian posterior under a specified prior, while generalized fiducial inference \citep{hannig2016generalized} treats it as a confidence distribution induced by the structural equation, with no prior input; in regular models the two coincide under Jeffreys' prior and otherwise differ by a Jacobian. Generative adversarial networks and diffusion models, in contrast, are designed to match an unconditional or conditional outcome distribution; they belong primarily to the first class. Their use in the second class requires explicit conditioning on observed data and a posterior target, which has been pursued in conditional diffusion methods for simulation-based inference \citep{geffner2023compositional, sharrock2024sequential}; this is a secondary use because it changes the training object and loss. Gaussian process emulators are primarily predictive: they fit a forward map from input parameters to outcomes; their use as posterior estimators requires an additional MCMC calibration step (the quantile kriging construction of \citealt{fadikar2018calibrating} used in Section~\ref{sec:ebola}), which is a secondary use.

Normalizing flows and generative Bayesian computation move most easily across classes. A normalizing flow is invertible, so it can serve as a conditional density estimator for both the outcome distribution and the posterior distribution. Generative Bayesian computation is not invertible but trains a quantile function; by exchanging the roles of $\theta$ and $Y$ (Section~\ref{sec:gbc}), the same architecture serves either as a posterior generator or a predictive generator, and by changing the conditioning to $(\tilde{S}(X), A, Y)$ it serves as a counterfactual outcome generator. Variational autoencoders straddle all three classes through the encoder--decoder structure: the encoder approximates a conditional distribution that can be a posterior, an outcome distribution, or a predictive distribution depending on what is encoded. Read against classical statistical structure, the variational autoencoder is a nonlinear factor-analysis model: the Gaussian latent prior is a modeling convention (not the Bayesian prior of the inferential problem), the decoder is a nonlinear analogue of the factor-loading map, and the evidence lower bound trades exact likelihood for tractability. The primary point is that the architecture is not the determinant of which class a method belongs to: the training object and conditioning structure are, and any architecture that can represent a conditional distribution can in principle serve any of the three classes provided the appropriate training object and loss are used.

\section{Computational Cost Comparison}\label{app_cost}

Table~\ref{tab:cost} compares the computational requirements of four methods for the parameter-outcome generator class, illustrated with the Ebola application ($d = 5$ parameters, $N = 10{,}000$ simulated trajectories from \citealt{fadikar2018calibrating}). For brevity, we abbreviate generative Bayesian computation as GBC in the table and surrounding discussion.

\begin{table}[H]
\centering
\footnotesize
\setlength{\tabcolsep}{3pt}
\begin{tabular}{>{\raggedright}p{0.20\linewidth}>{\raggedright}p{0.14\linewidth}>{\raggedright}p{0.18\linewidth}>{\raggedright}p{0.18\linewidth}>{\centering\arraybackslash}p{0.10\linewidth}}
\hline
Method & Upfront simulator calls & Fit / calibration cost (paid once) & Per new observation (after fit) & MCMC \\
\hline
GBC (this paper) & $N$ & Neural network fit & $B$ network evaluations for $B$ posterior draws & No \\
ABC, nearest-neighbor & $N$ & None & $k$-NN query over the $N$ stored pairs & No \\
Quantile kriging (Fadikar 2018) & $N$ & Gaussian process fit & Fresh MCMC chain of length $N_\text{chain}$ & Yes \\
MCMC with simulated likelihood & 0 upfront & None & $O(N_\text{chain})$ fresh simulator calls per chain & Yes \\
\hline
\end{tabular}
\caption{Computational requirements for the parameter-outcome generator class, Ebola application ($d = 5$, $N = 10{,}000$). Columns separate the upfront simulator budget, the one-time training cost, and the marginal cost paid for each new observation $y$ after fitting. GBC, ABC, and quantile kriging share the same upfront simulator budget; GBC and ABC amortize fitting over future observations. The critical difference is the per-observation cost after fit: GBC and ABC require no further simulator calls, while quantile kriging requires a fresh Markov chain per observation and likelihood-based MCMC requires $O(N_\text{chain})$ fresh simulator calls per observation.}
\label{tab:cost}
\end{table}

The shared simulation budget $N$ is the dominant upfront cost for GBC, ABC, and quantile kriging. After fitting, each GBC posterior draw at a new $y$ requires one network forward evaluation, so producing $B = 1000$ draws costs $B$ evaluations of a fixed-size network. For quantile kriging \citep{fadikar2018calibrating}, posterior inference for a new $y$ requires a fresh Markov chain run against the fitted Gaussian process as a likelihood surrogate. The per-observation cost is therefore $O(N_\text{chain})$ steps. For standard MCMC that calls the simulator directly at each iteration, the per-observation cost is $O(N_\text{chain} \times n_\text{chains})$ simulator evaluations; in the Ebola setting, where each ABM run costs several seconds, this makes per-observation MCMC infeasible at scale. Posterior uncertainty is quantified in GBC through the trained quantile network: the marginal per-coordinate 90\% credible interval for parameter $\theta_j$ at observation $y$ is $[\hat{G}(0.05, y), \hat{G}(0.95, y)]$. The empirical marginal per-coordinate coverage reported in Section~\ref{sec:app_ebola_inverse} (88--90\% for nominal 90\%) is comparable to coverage reported for quantile kriging by \citet{fadikar2018calibrating}, with the advantage that GBC requires no covariance function specification and no per-observation Markov chain.

\section{Quantile Transport versus Adversarial Generators}\label{app_gbc_vs_gan}

This appendix develops the contrast between the generative Bayesian computation quantile-transport construction of Section~\ref{sec:gbc} and the adversarial encoder--decoder family. Generative Bayesian computation and the adversarial encoder--decoder both define a generative map that pushes a simple baseline measure forward onto the target conditional law, but they differ in what map is learned and what objective is optimized, and the differences favor the quantile-transport construction for inference. The quantile network of Section~\ref{sec:gbc} learns a \emph{deterministic} transport map, the conditional quantile (or, for vector parameters, the Brenier) map, by minimizing a fixed proper scoring rule. A generative adversarial network instead learns an implicit generator through a two-player game. Recall the adversarial construction of Section~\ref{sec:architectures}: a generator $G_{\theta_g}$ driven by an abstract latent prior $Z \sim p(Z)$ and a discriminator $D_{\theta_d}$, trained at the two-player saddle point
\[
\min_{\theta_g}\max_{\theta_d}\; \mathbb{E}_{Y \sim p_{\mathrm{obs}}}\!\bigl[\log D_{\theta_d}(Y)\bigr] + \mathbb{E}_{Z \sim p_Z}\!\bigl[\log\bigl(1 - D_{\theta_d}(G_{\theta_g}(Z))\bigr)\bigr].
\]
For a fixed generator the optimal discriminator is $D^\star(Y) = p_{\mathrm{obs}}(Y)/(p_{\mathrm{obs}}(Y) + p_G(Y))$, and substituting it shows the generator minimizes a divergence,
\[
V(G_{\theta_g}, D^\star) = 2\,\mathrm{JS}\bigl(p_{\mathrm{obs}} \,\|\, p_G\bigr) - \log 4,
\]
while the Wasserstein variant \citep{arjovsky2017wasserstein} replaces the Jensen--Shannon divergence with the Kantorovich--Rubinstein dual of $W_1(p_{\mathrm{obs}}, p_G)$. In both cases the quantity actually descended is an \emph{adversarial estimate} of a divergence: it depends on an inner maximization over $\theta_d$ that a finite-capacity discriminator, trained for finitely many steps, only approximates. Encoder--decoder variants, bidirectional adversarial inference \citep{donahue2017adversarial, dumoulin2017adversarially}, the variational-autoencoder--adversarial hybrid \citep{larsen2016autoencoding}, and adversarial autoencoders \citep{makhzani2015adversarial}, add an encoder and sometimes a reconstruction term, but the latent code remains unidentified: many distinct generator--encoder pairs induce the same pushforward $p_G$.

Generative Bayesian computation solves $\min_w L(w)$ for a fixed risk $L$ that is an expectation of a proper scoring rule and is convex in the network outputs $\hat{Q}(\tau, y)$, with a unique population minimizer equal to the true quantile map. The adversarial problem $\min_{\theta_g}\max_{\theta_d}$ has no single objective: its solution is a saddle point, and simultaneous or alternating gradient play need not converge even in convex--concave idealizations. This is the structural source of the mode collapse, oscillation, and hyperparameter sensitivity seen in adversarial training. The quantile-network gradient is the subgradient $\mathbf{1}\{\theta < q\} - \tau \in [-\tau,\, 1-\tau]$, bounded and almost surely non-vanishing, so the objective always supplies a descent signal. The adversarial generator gradient $\nabla_{\theta_g}\log\bigl(1 - D_{\theta_d}(G_{\theta_g}(z))\bigr)$ vanishes whenever the discriminator saturates, precisely when $p_G$ and $p_{\mathrm{obs}}$ have nearly disjoint support and the Jensen--Shannon divergence is locally constant at $\log 2$. Generative Bayesian computation has no regime in which the loss stops providing a gradient.

Additionally, the check loss and the energy score are strictly proper, so the quantile network is an ordinary M-estimator: as the simulation budget $N$ grows and capacity increases, the fitted map converges to the true posterior quantile map under the empirical-risk-minimization rates established in Section~\ref{sec:gbc}. Consistency of adversarial networks instead requires the discriminator to attain its supremum at every generator update. With a finite, imperfectly trained discriminator the generator descends a biased surrogate of the divergence, and no analogous guarantee holds. This is the same gap flagged for the adversarial objective in the convergence-rate discussion of Section~\ref{sec:gbc}.

In generative Bayesian computation the latent is the uniform index $U$, the level $\tau$ \emph{is} the probability level, and the target map is pinned down, by monotonicity in one dimension and by the convex Brenier potential in several, so the map is unique. In an adversarial model the generator is identified only up to its pushforward $p_G$; the encoder--decoder pair admits a large equivalence class of solutions, and the latent space carries no probabilistic ordering. This identifiability is what lets the quantile network report calibrated quantiles and credible intervals, which the adversarial latent does not support.

Because the fitted quantile map is monotone it is invertible, so generative Bayesian computation returns the conditional cumulative distribution function directly and the density through a scalar Jacobian, alongside samples, quantiles, and credible intervals from one trained object. A generative adversarial network is a pure sampler: it provides no tractable likelihood, and density or interval estimates require additional steps applied to its samples.

Generative Bayesian computation treats the data as a network input, so $\hat{Q}(\tau, y)$ returns the posterior for a new $y$ in a single forward pass, with one network and one loss. A conditional adversarial model must inject $y$ into both the generator and the discriminator, doubling the surface on which the two-player game can fail, and must balance generator and discriminator capacity and learning rates throughout training.

\begin{table}[H]
\centering
\small
\setlength{\tabcolsep}{4pt}
\begin{tabular}{@{}>{\raggedright\arraybackslash}p{0.20\linewidth}>{\raggedright\arraybackslash}p{0.38\linewidth}>{\raggedright\arraybackslash}p{0.34\linewidth}@{}}
\hline
Aspect & GBC quantile--transport & GAN encoder--decoder \\
\hline
Object learned & quantile or Brenier transport map & implicit generator, optionally with an encoder \\
Optimization & $\min_w L(w)$, fixed convex risk & $\min_{\theta_g}\max_{\theta_d}$ saddle point \\
Objective & proper scoring rule (check or energy) & adversarial estimate of a divergence \\
Generator gradient & bounded, almost surely non-vanishing & vanishes when the discriminator saturates \\
Consistency & M-estimation guarantees & needs exact inner maximization \\
Latent & uniform $U$; level $\tau$ identified & abstract $z$, unidentified \\
Density and CDF & available (scalar Jacobian) & not tractable \\
Conditioning & native input, one forward pass & injected into both $G$ and $D$ \\
Networks to train & one & two, or three with an encoder \\
Failure modes & quantile crossing (preventable) & mode collapse, non-convergence \\
\hline
\end{tabular}
\caption{Generative Bayesian computation (quantile transport) versus the adversarial encoder--decoder as estimators of a conditional distribution. The contrasts are developed in the text of this appendix; the adversarial construction is given in Section~\ref{sec:architectures}.}
\label{tab:gbc_vs_gan}
\end{table}

The well-defined one-dimensional quantile characterization relies on the total ordering of $\mathbb{R}$. In genuinely multivariate problems the relevant object is the Brenier map, monotone in the optimal-transport sense rather than coordinate-wise, and the practical objective becomes a multivariate proper scoring rule such as the energy score introduced in Section~\ref{sec:gbc} rather than the scalar check loss; the structural advantages, deterministic transport, a fixed proper objective, no discriminator, and an identified map, carry over, but the idea of directly modeling the quantile function should be read in its transport-map generalization. These guarantees are also relative to the simulator: bias in the assumed prior or likelihood propagates into the fitted map, as it does in any simulation-based inference scheme.

\section{Counterfactual outcome distributions: a synthetic example}
\label{sec:app_lalonde}

This subsection first develops the counterfactual outcome generator, its associated causal estimands, and the quantile treatment effects from the same formalism used elsewhere in the paper, and then illustrates the construction on a  synthetic example with known ground truth so that the estimates of the conditional average treatment effects can be tested against the truth. The two larger numerical comparisons of the paper, against existing methods on a real simulator, remain the inverse-inference application in Section~\ref{sec:app_ebola_inverse} and the forward-emulation application in Section~\ref{sec:app_ebola_prediction}.

To illustrate the first generator class, consider a binary treatment program evaluation. A population of eligible individuals is assigned to a treatment arm ($A = 1$) or a control arm ($A = 0$). The outcome $Y$ is a post-program labor-market measure such as annual earnings; the covariate vector $X$ records baseline characteristics. The estimand of interest is the average treatment effect $\Delta = \mathbb{E}[Y(1) - Y(0)]$ and its conditional version $\Delta(x) = \mathbb{E}[Y(1) - Y(0) \mid X = x]$. The generator here is not a posterior generator over model parameters. It is the counterfactual outcome generator
\[
G(q, \tilde{s}, a) = F^{-1}_{Y \mid \tilde{S}(X) = \tilde{s},\, A = a}(q), \qquad q \in (0, 1), \quad a \in \{0, 1\},
\]
where $\tilde{S}(X)$ is a sufficient covariate reduction, for example the estimated propensity score or a low-dimensional learned summary. The training object is the observed causal triple
\[
\{(\tilde{S}(X_i), A_i, Y_i)\}_{i=1}^N.
\]
A quantile generator estimates $G$ directly by the pinball loss. Conditional adversarial generators, conditional variational autoencoders, conditional diffusion models, and conditional flows target the same conditional outcome distribution with different numerical criteria.

Under consistency, unconfoundedness given $\tilde{S}(X)$, positivity, and the stable unit treatment value assumption, the conditional distribution of $Y$ given $(\tilde{S}(X)=\tilde{s}, A=a)$ identifies the conditional distribution of $Y(a)$ given $\tilde{S}(X)=\tilde{s}$. Causal estimands are then functionals of the trained generator:
\[
\Delta(\tilde{s}) = \int_0^1 \left[\widehat{G}(q, \tilde{s}, 1) - \widehat{G}(q, \tilde{s}, 0)\right] dq, \qquad
\Delta = \int \Delta(\tilde{s})\, dF_{\tilde{S}(X)}(\tilde{s}).
\]
Quantile treatment effects are recovered as $\Delta_q(\tilde{s}) = \widehat{G}(q, \tilde{s}, 1) - \widehat{G}(q, \tilde{s}, 0)$.

This setup illustrates the key distinction between the first and second generator classes. Bayesian approaches to program evaluation, such as Bayesian causal forest \citep{hahn2020bayesian}, parameterize the response surface and recover treatment effects through MCMC posterior sampling. That places them in the second generator class: the target is a posterior over model parameters, and draws require a Markov chain whose length scales with the convergence properties of the sampler. The counterfactual outcome generator here belongs to the first class: it targets the conditional outcome distribution directly without parameterizing the data-generating process, and each draw requires only one function evaluation after the generator has been fitted. Appendix~\ref{app_cost} compares the computational cost of generator-based and MCMC-based methods in detail for the parameter-outcome generator task. The same pattern (MCMC requires a chain per observation, generator methods require one network evaluation) applies in the counterfactual outcome generator class.

To illustrate the performance of the generative Bayesian computation, we apply the quantile-network counterfactual generator to a one-dimensional synthetic data-generating process where the true treatment effect is known. The covariate $X \sim \mathcal{N}(0, 1)$, the treatment indicator $A \mid X \sim \mathrm{Bernoulli}(\sigma(0.5\, X))$ with a smooth propensity (no extreme overlap), and the potential outcomes are $Y(0) = X + \varepsilon_0$ and $Y(1) = 1.5\, X + 1 + \varepsilon_1$ with $\varepsilon_0, \varepsilon_1 \sim \mathcal{N}(0, 0.5^2)$. The observed outcome is $Y = A\, Y(1) + (1 - A)\, Y(0)$. The true conditional average treatment effect is $\Delta(x) = 1 + 0.5\, x$ and the true average treatment effect is $\Delta = \mathbb{E}[\Delta(X)] = 1$. We fit an implicit quantile network with a cosine quantile embedding (Section~\ref{sec:gbc}) on $N = 2{,}000$ observed triples $\{(X_i, A_i, Y_i)\}$ by minimizing the pinball loss, then estimate the conditional treatment effect as $\hat{\Delta}(x) = \int_0^1 [\widehat{G}(q, x, 1) - \widehat{G}(q, x, 0)]\, dq$, approximated by an average over $200$ quantile levels, and the average treatment effect as $\hat{\Delta}$, the mean of $\hat{\Delta}(\cdot)$ over a held-out covariate test set of size $n_\text{test} = 500$.

\begin{figure}[H]
\centering
\includegraphics[width=\linewidth]{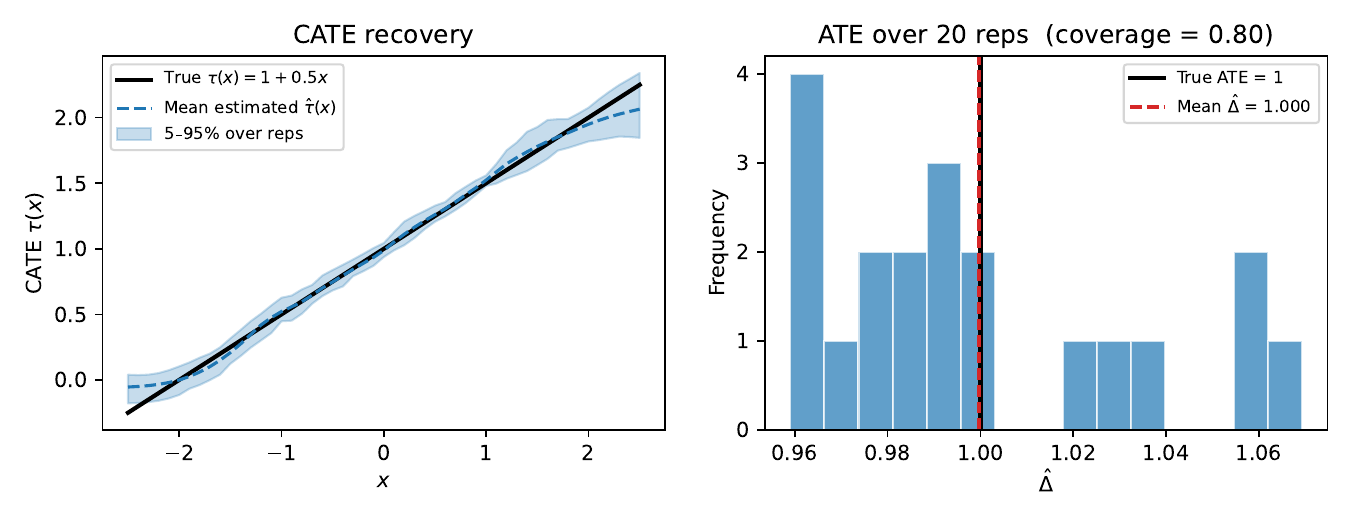}
\caption{Counterfactual outcome generator on a synthetic one-dimensional example. \emph{Left:} recovered conditional average treatment effect $\hat{\Delta}(x)$ across $20$ replicates (mean as the dashed line; $5$th--$95$th percentile band shaded) against the true $\Delta(x) = 1 + 0.5\, x$ (solid line). \emph{Right:} empirical distribution of the estimated average treatment effect $\hat{\Delta}$ across $20$ replicates; the true value $\Delta = 1$ is marked with a solid vertical line and the empirical mean $1.000$ with a dashed line. Empirical $90\%$ coverage of the training-data percentile bootstrap interval is $0.80$ ($16/20$).}
\label{fig:cat1_toy}
\end{figure}

Across $20$ Monte Carlo replicates, the estimator $\hat{\Delta}$ has mean $1.000$ against the true value $1.000$ (bias $-0.0003$, root mean squared error $0.033$). Figure~\ref{fig:cat1_toy} shows the recovered conditional treatment-effect curve, which traces the true $\Delta(x) = 1 + 0.5\, x$ across the support of $X$, and the empirical distribution of $\hat{\Delta}$ around the true ATE. A $90\%$ confidence interval on $\hat{\Delta}$ is constructed by a nonparametric bootstrap over the training data: for each replicate, the quantile network is retrained on $B = 20$ bootstrap resamples of $\{(X_i, A_i, Y_i)\}$ and $\hat{\Delta}^{(b)}$ is recomputed on the held-out test set. The $90\%$ interval is the $5$th to $95$th percentile of the resulting bootstrap distribution. Empirical coverage of the true $\Delta = 1$ across the $20$ replicates is $0.80$ ($16/20$). With $B = 20$ bootstrap retrains the percentile interval is somewhat noisy, but the interval now reflects both Monte Carlo over the test covariates and the training-data sampling variability of the generator. The headline message of the illustration is the recovery of the conditional treatment-effect map together with the smallness of the average bias and root mean squared error: the counterfactual outcome generator, fit by a single pass of pinball-loss minimization, returns an estimator $\hat{\Delta}(x)$ that is essentially unbiased for the true $\Delta(x)$ on this data-generating process.

\section{Calibration Checks for the Ebola Inverse Problem}\label{app_calibration}
\label{sec:app_ebola_inverse_calibration}
\label{sec:tractable_benchmark}

The Ebola coverage statistics of Section~\ref{sec:app_ebola_inverse} are simulation-based calibration of the average posterior. To complement them, we report two diagnostics that test the approximation more directly: a closed-form conjugate benchmark where the exact posterior is available, and per-parameter simulation-based calibration rank histograms on the Ebola test set. We also report a comparison against a flow-based neural posterior estimator on the same simulated table.

\subsubsection*{Conjugate benchmark} 

\begin{figure}[H]
\centering
\includegraphics[width=\linewidth]{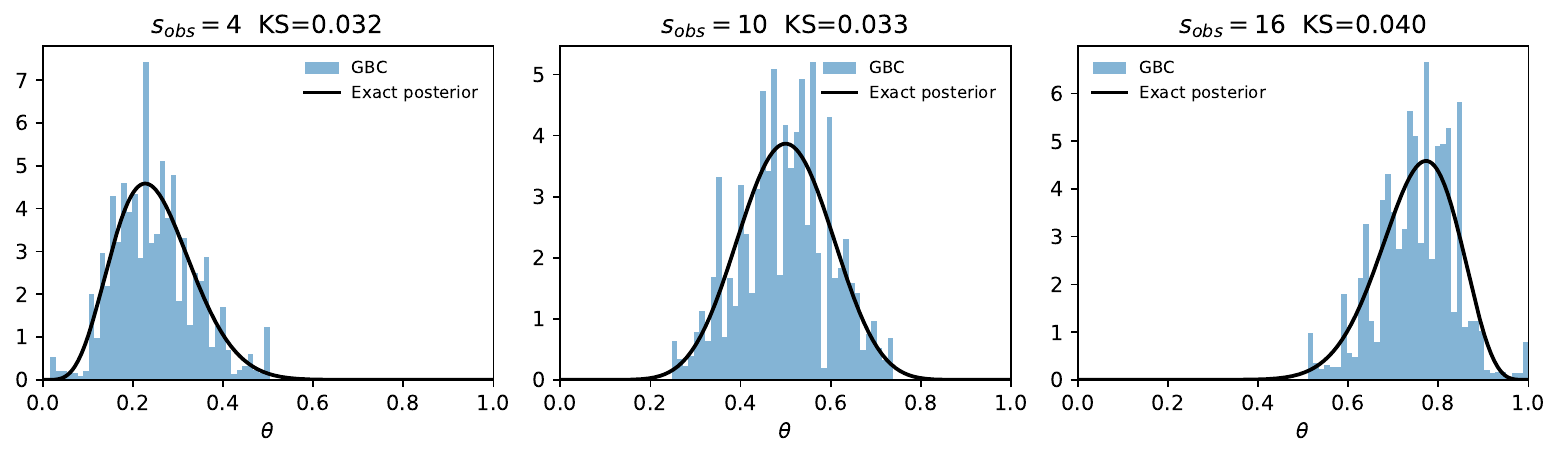}
\caption{Conjugate Beta-Binomial benchmark with exact posterior available in closed form. Each panel overlays 5000 generative Bayesian computation posterior samples (blue histogram) against the exact $\mathrm{Beta}(2 + s_\text{obs},\, 2 + 20 - s_\text{obs})$ density (black curve), at three values of the observed sufficient statistic. Kolmogorov--Smirnov statistics between the generative Bayesian computation sample and the exact posterior cumulative distribution function are 0.032--0.040.}
\label{fig:conjugate_overlay}
\end{figure}

Let $\theta \sim \mathrm{Beta}(2, 2)$ and $Y_i \mid \theta \sim \mathrm{Bernoulli}(\theta)$ for $i = 1, \ldots, 20$. The sufficient statistic $s = \sum_i Y_i$ is Bayes-sufficient and the exact posterior is $\theta \mid s \sim \mathrm{Beta}(2 + s,\, 2 + 20 - s)$ in closed form. We train an implicit quantile network on $N = 50{,}000$ simulated $(\theta, s)$ pairs by minimizing the pinball loss, then draw 5000 posterior samples at each of $s_\text{obs} \in \{4, 10, 16\}$ and compare against the exact Beta density (Figure~\ref{fig:conjugate_overlay}). Across the three values of $s_\text{obs}$, the Kolmogorov--Smirnov (KS) statistic between the generative Bayesian computation sample and the exact Beta cumulative distribution function is between $0.032$ and $0.040$; posterior means and standard deviations agree with the exact values to within $0.003$. The exact posterior is closely approximated by the generative Bayesian computation sample on this benchmark where the ground truth is available in closed form.

\subsubsection*{Simulation-based calibration on the Ebola inverse problem} 

\begin{figure}[H]
\centering
\includegraphics[width=\linewidth]{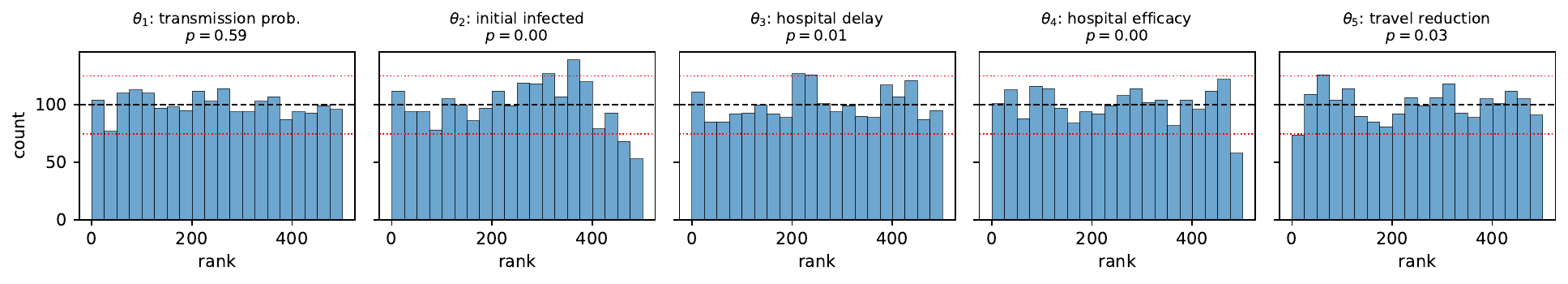}
\caption{Simulation-based calibration rank histograms on the Ebola inverse problem (2000 held-out trajectories, $B = 500$ posterior draws per trial). Dashed line is the uniform expectation; red dotted lines are the $99\%$ pointwise confidence band. Per-panel chi-square uniformity $p$-values are reported above each panel.}
\label{fig:sbc_ranks}
\end{figure}

\citet{talts2018validating} show that for a correctly calibrated posterior approximation, the ranks of the true parameter values among posterior draws across simulated $(\theta, Y)$ pairs are uniformly distributed. We retrain per-parameter quantile networks on 8000 of the 10,000 simulated Ebola trajectories using the same partial least squares summary, then for each of the 2000 held-out trajectories draw $B = 500$ posterior samples and compute the rank of the true $\theta_j$ among them. Figure~\ref{fig:sbc_ranks} shows the resulting rank histograms with a $99\%$ confidence band for uniformity. Table~\ref{tab:sbc} reports a chi-square uniformity test with 19 degrees of freedom.

\begin{table}[H]
\centering
\begin{tabular}{c|c|c|c}
parameter & $\chi^2_{19}$ & $p$-value & mean rank \\ \hline
$\theta_1$ & 17.0 & 0.59 & 247.8 \\
$\theta_2$ & 81.8 & $9\times 10^{-10}$ & 245.9 \\
$\theta_3$ & 35.1 & 0.014 & 254.2 \\
$\theta_4$ & 41.6 & 0.002 & 246.6 \\
$\theta_5$ & 32.4 & 0.028 & 251.6
\end{tabular}
\caption{Simulation-based calibration of the generative Bayesian computation posterior on the Ebola inverse problem, by parameter. Mean ranks are close to the uniform target of $B/2 = 250$ in every case, but the chi-square test rejects uniformity at the $0.05$ level for $\theta_2$, $\theta_3$, $\theta_4$, and (borderline) $\theta_5$. This refines the coverage diagnostic of Section~\ref{sec:app_ebola_inverse}: marginal coverage averages over the test set can be near nominal even when the rank distribution is non-uniform.}
\label{tab:sbc}
\end{table}

The interpretation is: marginal coverage and simulation-based calibration are complementary, not equivalent. Mean rank near $B/2$ is consistent with coverage near nominal, but the rank distribution itself probes the shape of the posterior approximation, and simulation-based calibration detects structural mismatch in $\theta_2$--$\theta_4$ that the coverage statistic alone does not.

\subsubsection*{Neural posterior estimation baseline} 

The natural comparison is a flow-based neural posterior estimator on the same simulated table. We train an amortized neural posterior estimator with a Masked Autoregressive Flow (MAF) density estimator, using the SNPE-C implementation of \citet{greenberg2019automatic} from the \texttt{sbi} package \citep{tejerocantero2020sbi} (single-round training on the full simulation budget, with no sequential refinement against the observed datum), the same 8000/2000 train/test split, and the same partial least squares summary as the quantile network; we then form $90\%$ marginal credible intervals from $1000$ posterior draws per test point. Table~\ref{tab:npe_vs_gbc} reports the comparison.

\begin{table}[H]
\centering
\begin{tabular}{l|ccccc}
method & $\theta_1$ & $\theta_2$ & $\theta_3$ & $\theta_4$ & $\theta_5$ \\ \hline
GBC (quantile network) & 0.89 & 0.88 & 0.90 & 0.90 & 0.90 \\
NPE flow (MAF, defaults) & 0.82 & 0.84 & 0.71 & 0.79 & 0.74
\end{tabular}
\caption{Empirical $90\%$ marginal credible interval coverage on the Ebola inverse problem, by parameter. Both methods are trained on the same 8000 simulated $(\theta, Y)$ pairs with the same partial least squares summary. Coverage is evaluated on 2000 held-out trajectories. Quantile-based generative Bayesian computation matches or exceeds the masked autoregressive flow at default settings on every parameter. The comparison uses sbi v0.26.1 default hyperparameters for the flow; no claim is made that the flow cannot be improved with extensive tuning.}
\label{tab:npe_vs_gbc}
\end{table}

The quantile network has between $4$ and $19$ percentage points higher marginal coverage than the flow at default settings on every parameter. The largest gaps are on $\theta_3$ ($0.90$ vs $0.71$) and $\theta_5$ ($0.90$ vs $0.74$). The likely explanation is the one in Section~\ref{sec:gbc}: the pinball training criterion directly targets the conditional quantile function, which is the object that determines credible interval endpoints, while flow training targets the log-likelihood of the posterior and only indirectly controls quantile placement. This is a single comparison on one dataset and should be read as a reference point rather than a general statement about the two architectures.

\end{appendix}

\end{document}